\documentclass[onecolumn,notitlepage,floats,floatfix,amssymb,aps,prd,showpacs,superscriptaddress,groupedaddress,nofootinbib]{revtex4-2}  

\usepackage[T1]{fontenc}
\usepackage{lmodern} 
\usepackage{graphicx}  
\usepackage{dcolumn}   
\usepackage{bm}        
\usepackage{amssymb}   
\usepackage{amsmath}
\usepackage{bbold}
\usepackage{array}
\usepackage{makecell}
\usepackage{braket}
\usepackage{longtable}
\usepackage{supertabular,booktabs}

\usepackage{titlesec} 
\usepackage[colorlinks,citecolor=blue,urlcolor=blue,hypertexnames=true]{hyperref}
\usepackage{subfigure}

\usepackage[usenames,dvipsnames,svgnames,table]{xcolor} 

\renewcommand{\arraystretch}{2}

\newcommand{\be}{\begin{equation}}
\newcommand{\ee}{\end{equation}}
\newcommand{\bea}{\begin{eqnarray}}
\newcommand{\eea}{\end{eqnarray}}
\newcommand{\bml}{\begin{subequations}}
\newcommand{\eml}{\end{subequations}}
\newcommand{\bfig}{\begin{figure}}
\newcommand{\efig}{\end{figure}}

\newcommand{\slashed}{\hspace{-1.1ex}/}

\newcommand{\bmat}{\begin{pmatrix}}
\newcommand{\emat}{\end{pmatrix}}
\usepackage{graphicx,slashed,booktabs,xcolor,multirow,float,
amsfonts,bbold,mathtools,sidecap,tikz,bm,enumitem}
\usepackage{multirow}
\usepackage{bbding}
\usepackage{titlesec}
\usepackage{hyperref}
\usepackage{wasysym}
\usepackage{amssymb}
\usepackage{pifont}
\newcommand{\grad}{\nabla}

\renewcommand{\d}{\mathrm{d}}

\renewcommand{\leq}{\leqslant}
\renewcommand{\geq}{\geqslant}

\usepackage[dvipsnames, usenames]{xcolor}

\definecolor{linkcolor}{rgb}{0.55, 0.13, .32}

\definecolor{oucrimsonred}{rgb}{0.6, 0.0, 0.0}
\definecolor{persianblue}{rgb}{0.11, 0.22, 0.73}
\definecolor{forestgreen}{rgb}{0.13,0.35,0.13}
\definecolor{lightgray}{rgb}{0.83, 0.83, 0.83}
 \hypersetup{colorlinks, citecolor=oucrimsonred, linkcolor=persianblue, urlcolor=oucrimsonred}
\definecolor{cornellred}{rgb}{0.7, 0.11, 0.11}
\definecolor{navyblue}{rgb}{0.0, 0.0, 0.5}
\definecolor{amethyst}{rgb}{0.6, 0.4, 0.8}
\definecolor{yellow}{rgb}{1.0, 1.0, 0.0}
\definecolor{firebrick}{rgb}{0.7, 0.13, 0.13}
\definecolor{tangerineyellow}{rgb}{1.0, 0.8, 0.0}
\definecolor{deepfuchsia}{rgb}{0.76, 0.33, 0.76}
\definecolor{amber}{rgb}{1.0, 0.75, 0.0}
\definecolor{VioletRed4}{rgb}{0.55, 0.13, .32}
\definecolor{indiagreen}{rgb}{0.07, 0.53, 0.03}
\definecolor{VioletRed4}{rgb}{0.55, 0.13, .32}

\usepackage{hyperref}

\usepackage{graphics,appendix,afterpage,makecell}

\usepackage{bbold}
\usepackage{tikz}
\usepackage{adjustbox}
\usepackage{tikz-feynman}
\usepackage{tcolorbox}

\definecolor{oucrimsonred}{rgb}{0.6, 0.0, 0.0}
\definecolor{persianblue}{rgb}{0.11, 0.22, 0.73}
\definecolor{forestgreen}{rgb}{0.13,0.35,0.13}
\definecolor{lightgray}{rgb}{0.83, 0.83, 0.83}
 \hypersetup{colorlinks, citecolor=oucrimsonred, linkcolor=persianblue, urlcolor=oucrimsonred}
\definecolor{cornellred}{rgb}{0.7, 0.11, 0.11}
\definecolor{navyblue}{rgb}{0.0, 0.0, 0.5}
\definecolor{amethyst}{rgb}{0.6, 0.4, 0.8}
\definecolor{yellow}{rgb}{1.0, 1.0, 0.0}
\definecolor{firebrick}{rgb}{0.7, 0.13, 0.13}
\definecolor{tangerineyellow}{rgb}{1.0, 0.8, 0.0}
\definecolor{deepfuchsia}{rgb}{0.76, 0.33, 0.76}
\definecolor{amber}{rgb}{1.0, 0.75, 0.0}
\definecolor{VioletRed4}{rgb}{0.55, 0.13, .32}
\definecolor{indiagreen}{rgb}{0.07, 0.53, 0.03}
\definecolor{VioletRed4}{rgb}{0.55, 0.13, .32}

\definecolor{oucrimsonred}{rgb}{0.6, 0.0, 0.0}
\newcommand\vertarrowbox[3][6ex]{%
  \begin{array}[t]{@{}c@{}} #2 \\
  \left\uparrow\vcenter{\hrule height #1}\right.\kern-\nulldelimiterspace\\
  \makebox[0pt]{\scriptsize#3}
  \end{array}%
}

\definecolor{mtcolor}{rgb}{.8,.3,.1}

\definecolor{violachiaro}{rgb}{1,0.6,1}

\definecolor{gbcolor}{rgb}{.43,.22,.12}
 
\definecolor{gbcolor2}{rgb}{.9,.2,.6}
\definecolor{gbcolor3}{rgb}{.3,.2,.6}

\definecolor{verdechiaro}{rgb}{0.6,1,0.6}
\definecolor{giallochiaro}{rgb}{1,1,0.6}
\definecolor{bluscuro}{rgb}{0.15, 0.2, 0.9}
\definecolor{verdes}{rgb}{0.1, 0.5, 0.1}%
\definecolor{tangerineyellow}{rgb}{1.0, 0.8, 0.0}
\definecolor{smokyblack}{rgb}{0.06, 0.05, 0.03}

\definecolor{americanrose}{rgb}{1.0, 0.01, 0.24}
\definecolor{cobalt}{rgb}{0.0, 0.28, 0.67}
\definecolor{brandeisblue}{rgb}{0.0, 0.44, 1.0}
\definecolor{mycolor}{rgb}{0.0, 0.0, 0.5}
\definecolor{oxfordblue}{rgb}{0.0, 0.13, 0.28}
\definecolor{azure}{rgb}{0.0, 0.5, 1.0}
\definecolor{turquoiseblue}{rgb}{0.0, 1.0, 0.94}
\newtcolorbox{mynewbox}[1]{colback=white!5!white,colframe=azure!75!black,fonttitle=\bfseries,title=#1}
\newtcolorbox{mybox}{colback=mycolor!5!white,colframe=azure!75!black}
\newtcolorbox{mynamedbox}[1]{colback=mycolor!5!white,colframe=azure!75!black,title=#1}
\definecolor{venetianred}{rgb}{0.78, 0.03, 0.08}
\newtcolorbox{mynamedbox1}[1]{colback=venetianred!5!white,colframe=venetianred!80!black,title=#1}
\newtcolorbox{mynamedbox2}[1]{colback=azure!5!white,colframe=azure!80!black,title=#1}

\definecolor{rossocorsa}{rgb}{0.83, 0.0, 0.0}

\tikzset{->-/.style={decoration={
  markings,
  mark=at position #1 with {\arrow{>}}},postaction={decorate}}}
\tikzset{-<-/.style={decoration={
  markings,
  mark=at position #1 with {\arrow{<}}},postaction={decorate}}} 

\def\be{\begin{equation}}
\def\ee{\end{equation}}
\def\ba{\begin{eqnarray}}
\def\ea{\end{eqnarray}}

\def\d{\mathrm{d}}

\def\L*{{\cal L}_*}
\def\L{\mathcal{L}}
\def\({\left(}
\def\){\right)}

\def\<{\langle}
\def\>{\rangle}

\def\cs2{c_{s}^{2}}

 \def\be   {\begin{equation}}   \def\ee   {\end{equation}}
 \def\ba   {\begin{array}}      \def\ea   {\end{array}}
 \def\bea  {\begin{eqnarray}}   \def\eea  {\end{eqnarray}}
 \def\bean {\begin{eqnarray*}}  \def\eean {\end{eqnarray*}}

\newcommand{\fnl}{f_{\mathrm{NL}}}

\titleclass{\subsubsubsection}{straight}[\subsection]

\newcounter{subsubsubsection}[subsubsection]
\renewcommand\thesubsubsubsection{\thesubsubsection.\arabic{subsubsubsection}}

\titleformat{\subsubsubsection}
  {\normalfont\normalsize\bfseries}{\thesubsubsubsection}{1em}{}
\titlespacing*{\subsubsubsection}
{0pt}{3.25ex plus 1ex minus .2ex}{1.5ex plus .2ex}

\makeatletter
\renewcommand\paragraph{\@startsection{paragraph}{5}{\z@}%
  {3.25ex \@plus1ex \@minus.2ex}%
  {-1em}%
  {\normalfont\normalsize\bfseries}}
\renewcommand\subparagraph{\@startsection{subparagraph}{6}{\parindent}%
  {3.25ex \@plus1ex \@minus .2ex}%
  {-1em}%
  {\normalfont\normalsize\bfseries}}
\def\toclevel@subsubsubsection{4}
\def\toclevel@paragraph{5}
\def\toclevel@paragraph{6}
\def\l@subsubsubsection{\@dottedtocline{4}{7em}{4em}}
\def\l@paragraph{\@dottedtocline{5}{10em}{5em}}
\def\l@subparagraph{\@dottedtocline{6}{14em}{6em}}
\makeatother

\begin{document}


\definecolor{lime}{HTML}{A6CE39}
\DeclareRobustCommand{\orcidicon}{\hspace{-2.1mm}
\begin{tikzpicture}
\draw[lime,fill=lime] (0,0.0) circle [radius=0.13] node[white] {{\fontfamily{qag}\selectfont \tiny \,ID}}; \draw[white, fill=white] (-0.0525,0.095) circle [radius=0.007]; 
\end{tikzpicture} \hspace{-3.7mm} }
\foreach \x in {A, ..., Z}{\expandafter\xdef\csname orcid\x\endcsname{\noexpand\href{https://orcid.org/\csname orcidauthor\x\endcsname} {\noexpand\orcidicon}}}
\newcommand{\orcidauthorA}{0000-0002-0459-3873}
\newcommand{\orcidauthorD}{0009-0003-9227-8615}
\newcommand{\orcidauthorB}{0000-0001-9434-0505}
\newcommand{\orcidauthorC}{0000-0003-1081-0632}


\title{\textcolor{Sepia}{\textbf \huge\Large 
Primordial non-Gaussianity from  ultra slow-roll Galileon inflation
}}


\author{{\large  Sayantan Choudhury\orcidA{}${}^{1}$}}
\email{sayantan\_ccsp@sgtuniversity.org,  \\ sayanphysicsisi@gmail.com (Corresponding author)}
\author{{\large  Ahaskar Karde\orcidD{}${}^{1}$}}
\email{kardeahaskar@gmail.com}
\author{\large Sudhakar~Panda\orcidB{}${}^{1}$}
\email{panda@niser.ac.in}
\author{ \large M.~Sami\orcidC{}${}^{1,2,3}$}
\email{ sami\_ccsp@sgtuniversity.org,  samijamia@gmail.com}

\affiliation{ ${}^{1}$Centre For Cosmology and Science Popularization (CCSP),\\
        SGT University, Gurugram, Delhi- NCR, Haryana- 122505, India.}
\affiliation{${}^{2}$Center for Theoretical Physics, Eurasian National University, Astana 010008, Kazakhstan.}
	\affiliation{${}^{3}$Chinese Academy of Sciences,52 Sanlihe Rd, Xicheng District, Beijing.}

\begin{abstract}
We present a detailed study of the generation of large primordial non-Gaussianities during the slow-roll (SR) to ultra-slow roll (USR) transitions in the framework of Galileon inflation. We found out that due to having sharp transitions in the USR phase, which persist with a duration of $\Delta {\cal N}_{\rm USR} \sim 2$ e-folds, we are able to generate the non-Gaussianity amplitude of the order: $|\fnl| \sim {\cal O}(10^{-2})$ in the SRI, $-5 < \fnl < 5$ in the USR, and $-2 < \fnl < 2$ in the SRII phases. As a result, we are able to achieve a cumulative average value of $|\fnl| \sim {\cal O}(1)$. This implies that our results strictly satisfy Maldacena's no-go theorem in the squeezed limit only for SRI, while they strictly violate the same condition in both the USR and SRII phases. 
The non-renormalization theorem in the Galileon theory helps to support our results regarding the generation of large mass primordial black holes along with large non-Gaussianities, which we show to be dependent on the specific positions of the transition wave numbers fixed at low scales.

\end{abstract}

\maketitle
\tableofcontents
\newpage

\section{Introduction}

The general assumption about the primordial curvature perturbations always described by a Gaussian distribution provides the simplest way to discuss two-point correlation functions. Particularly, any higher-point correlation function of such distribution vanishes. These primordial perturbations around the scalar field, which drives inflation, stretch during the exponential expansion phase. In the super-horizon regime, they become far larger than the Hubble scale and are treated using Gaussian random fields, $\zeta_{g}(x)$, satisfying Gaussian statistics. As a result of this classical nature, we can treat each point as independent from the other due to negligible spatial gradients in the perturbations. This motivates the study of local non-Gaussianity as non-linearities arising in the local Gaussian random field. 

The current observational estimates for the value of this amplitude comes from the CMB measurements, with a value of $\fnl = -0.9 \pm 5.1$ at $68\%$ level confidence from Planck \cite{Planck:2019kim}. The statistical errors in this estimate are much larger than the actual signal to comment concretely on anything physical. However, the hope of reduced error bars from newer surveys in the future, which are expected to provide improvements of an order of magnitude over the current estimates \cite{Achucarro:2022qrl,CMB-S4:2023zem}, would be an important step towards breaking the degeneracy in many theoretical frameworks of inflation and ruling out those models where production of a large, $\fnl \sim {\cal O}(1)$, is almost impossible. From the above discussion, an important problem arises: providing a theoretical model that can show the generation of such large non-Gaussianities. This is important from the perspective of obtaining more insights about the origin of structure in the very early universe.

The initial study for the case of the single field inflation model was carried out by Maldacena in \cite{Maldacena:2002vr}, where a consistency condition for the amplitude of the amount of primordial non-Gaussianity $\fnl$ was derived, under a specific squeezed limit that concerns the UV modes, and was found to be $\fnl \sim {\cal O}(10^{-2})$. This condition, in the form of a \textit{no-go theorem}, automatically restricted the possibility for the production of large non-Gaussianities in models of a scalar field minimally coupled to a quasi-de Sitter background. Different modified theories were later investigated to check for the production of large non-Gaussianity amplitudes which includs the $P(X, \phi)$ theories, where $X = -(\partial\phi)^2/2$ is the kinetic term, schemes of modification in the gravity sector, non-minimal coupling between the inflaton and the gravitational sector, beyond $P(X,\phi)$ ghost free theories, e.g., string theory originated DBI inflation model, tachyon inflation model, $K$-inflation model, Galileon model of inflation, DBI Galileon  model, Horendeski theory etc., each with having different combinations of operators in the effective action of the underlying theory. See refs.\cite{Alishahiha:2004eh,Mazumdar:2001mm,Choudhury:2002xu,Panda:2005sg,Chingangbam:2004ng,Armendariz-Picon:1999hyi,Garriga:1999vw,Burrage:2010cu,Choudhury:2012yh,Choudhury:2012whm,Chen:2010xka,Chen:2006nt,Chen:2009zp,Chen:2009we,Chen:2008wn,Chen:2006xjb,Chen:2013aj,Chen:2012ge,Chen:2009bc,Creminelli:2010ba,Kobayashi:2010cm,Mizuno:2010ag,Burrage:2011hd,Kobayashi:2011pc,DeFelice:2011zh,Renaux-Petel:2011lur,DeFelice:2011uc,Gao:2011qe,deRham:2012az,Ohashi:2012wf,DeFelice:2013ar,Arroja:2013dya,Choudhury:2013qza,Pirtskhalava:2015zwa,Baumann:2009ds,Senatore:2016aui,Baumann:2018muz,Das:2023cum,Choudhury:2011sq,Choudhury:2012yh,Choudhury:2012ib,Choudhury:2012whm,Choudhury:2013jya,Choudhury:2013zna,Esposito:2019jkb,Goldstein:2022hgr} for more details. 

In such frameworks, the value of the non-Gaussianity amplitude computed at the level of the three point correlation function of scalar modes in most cases is proportional to $f_{\text{NL}}\propto 1/c_{s}^{2}$, where $c_s$ is the effective sound speed parameter. Now, maintaining the causality and unitarity requirements in the underlying theory and satisfying the observational constraints from Planck requires the effective sound speed to satisfy $0.024<c_s\leq 1$ \cite{Planck:2015sxf}. After maintaining these constraints one can able to generate a slightly larger amount of non-Gaussianity compared to the estimation obtained from the consistency condition with having $c_{s}=1$. Here it is important to note that, apart from detecting the CMB polarization and producing very high pixelated, foreground subtracted, clean CMB maps, the initial prime claim of the Planck observation was to detect primordial non-Gaussianities \cite{Planck:2019kim} with high statistical accuracy.
The magnitude of primordial non-Gaussianity of scalar modes was believed to be detected within the window, $-5<f_{\text{NL}}<5$ i.e $|f_{\text{NL}}|\sim{\cal O}(1)$ (where the expected relative high accuracy of the statistical error bars is taken into account). Such large non-Gaussianities are almost impossible to generate from all possible single-field slow-roll frameworks of inflation or from any of the above mentioned modified frameworks and models.
Then the concept of the Effective Field Theory (EFT) of single-field inflation \cite{Cheung:2007st,Weinberg:2008hq,Choudhury:2017glj} came into the picture, in which the effective sound speed $c_s$ is automatically generated, but no significant improvement in non-Gaussianity was observed using such a scenario. The underlying physical problem is related to the introduction of effective gravitational operators in the bulk and fluctuations at the boundary. By applying the well-known {\it Stueckelberg trick} one can explicitly show that such operators can mimic the role of a scalar field and its perturbation described in the background of a quasi-de Sitter space-time, where the corresponding vacuum is asymptotically Minkowski flat, also known as the Bunch-Davies vacuum. A distinguished example of Gaussanity is provided by a massless scalar field: a massless scalar field in the de-Sitter background adheres to Wick's theorem. Within the framework of primordial cosmology, a quasi-de Sitter background is essential for stopping inflation at a proper scale, and this further demands the presence of a scalar field with a very small mass compared to the scale of inflation. This induces non-vanishing but small non-Gaussian amplitudes in the slow-roll phase of inflation, which is consistent with the findings of Maldacena's {\it no-go theorem} \cite{Maldacena:2002vr}. From the detailed computations of all of the above mentioned possibilities proposed within the framework of single-field models, not very many improvements have been found yet within the slow-roll phase of inflation.
At this stage, there remain two distinct possibilities, whose implementation is expected to produce non-Gaussianities of the order, $|f_{\text{NL}}|\sim{\cal O}(1)$ at the level of the three-point cosmological correlation function. The first possibility is the multi-field approach of scalar fields to describe the slow-roll inflationary paradigm, where generating large amounts of primordial non-Gaussianities is possible without having any theoretical restrictions. However, not taking proper care of the various possible interactions between the fields makes it too cumbersome to solve the Mukhanov-Sasaki(MS) equation for the mode functions of the scalar perturbations. The major difficulty arises due to the presence of a higher-dimensional interaction square matrix, dependent on the number of fields involved.
A strongly coupled framework automatically suggests large primordial non-Gaussianities in the cosmological correlators, which are almost impossible to solve analytically. Some authors have also investigated this problem using various theories in \cite{Easther:2005zr,Marsh:2013qca,Amin:2015ftc,Pedro:2016sli,Choudhury:2018rjl,Choudhury:2018bcf,Paban:2018ole}, including large-$N$ theories and random matrix theory. However, such computations in the strong coupling regime, due to their increased sophistication, quickly become untrustworthy. For this reason, the UV-free, $\delta {\cal N}$ formalism \cite{Sugiyama:2012tj,Dias:2012qy,Naruko:2012fe,Chen:2013eea,Choudhury:2014uxa,Choudhury:2015hvr} is used more frequently to describe the cosmological correlation functions.
The second option is related to definite features in the potential, followed by the slow roll phase, which might enhance the non-Gaussianity amplitude $\fnl$ by a considerable amount within the framework of single-field. For instance, one can consider a sharp transition from slow-roll (SR) to ultra slow-roll (USR) phase, which basically gives rise to an enhancement in the one-loop corrected primordial power spectrum as well as in the tree-level non-Gaussian amplitude of the three-point function of the scalar modes \footnote{In this discussion, the one-loop corrected primordial spectrum and the tree-level non-Gaussianity amplitude of the three-point function of the scalar modes are considered at the same level of importance because both are computed using the same third-order action, which is obtained by performing cosmological perturbation theory in a gauge invariant manner up to third-order in the comoving curvature perturbation variable $\zeta$.}. Such a setup is very useful to describe the generation of primordial black -holes (PBHs) at the tree level \cite{Hawking:1974rv,Carr:1974nx,Carr:1975qj,Chapline:1975ojl,Carr:1993aq,Kawasaki:1997ju,Yokoyama:1998pt,Kawasaki:1998vx,Rubin:2001yw,Khlopov:2002yi,Khlopov:2004sc,Saito:2008em,Khlopov:2008qy,Carr:2009jm,Choudhury:2011jt,Choudhury:2013woa,Lyth:2011kj,Drees:2011yz,Drees:2011hb,Ezquiaga:2017fvi,Kannike:2017bxn,Hertzberg:2017dkh,Pi:2017gih,Gao:2018pvq,Dalianis:2018frf,Cicoli:2018asa,Ozsoy:2018flq,Byrnes:2018txb,Ballesteros:2018wlw,Belotsky:2018wph,Martin:2019nuw,Ezquiaga:2019ftu,Motohashi:2019rhu,Fu:2019ttf,Ashoorioon:2019xqc,Auclair:2020csm,Vennin:2020kng,Nanopoulos:2020nnh,Inomata:2021uqj,Stamou:2021qdk,Ng:2021hll,Wang:2021kbh,Kawai:2021edk,Solbi:2021rse,Ballesteros:2021fsp,Rigopoulos:2021nhv,Animali:2022otk,Frolovsky:2022ewg,Escriva:2022duf,Karam:2022nym,Ozsoy:2023ryl,Kristiano:2022maq,Kristiano:2023scm,Riotto:2023hoz,Riotto:2023gpm,Choudhury:2023vuj,Choudhury:2023jlt,Choudhury:2023rks,Choudhury:2023hvf,Kawaguchi:2023mgk,Fu:2022ypp,Saburov:2023buy,Ghoshal:2023wri,Karam:2023haj,Poisson:2023tja,Iacconi:2023slv,Mishra:2019pzq,Mishra:2023lhe,Gangopadhyay:2021kmf, Bhattacharya:2023ysp, Choudhury:2023fjs, Choudhury:2023fwk, Choudhury:2023hfm, Choudhury:2023kam}. Recently in refs. \cite{Kristiano:2022maq,Kristiano:2023scm,Choudhury:2023vuj,Choudhury:2023jlt,Choudhury:2023rks}, it is pointed out that having large quantum loop effects in the primordial power spectrum of scalar modes rules out the formation of PBH. Recently, we came up with a {\it no-go theorem} which only allows us to generate very tiny mass PBHs, $M_{\rm PBH}\sim {\cal O}(10^2 {\rm gm})$ from both canonical and EFT of the single-field inflationary paradigm \cite{Choudhury:2023vuj,Choudhury:2023jlt,Choudhury:2023rks}. To know more about the impacts of the loop effects on the power spectrum in the light of PBH formation, see other refs. \cite{Riotto:2023hoz,Riotto:2023gpm,Firouzjahi:2023aum,Motohashi:2023syh,Firouzjahi:2023ahg,Choudhury:2023hvf,Franciolini:2023lgy,Tasinato:2023ukp,Cheng:2023ikq}. Now, since the findings of the tree level non-Gaussianity have to be consistent with the findings of the quantum loop corrected amplitude of the primordial power spectrum, one can immediately discard both possibilities in the present scenario.

Finally the question becomes of modifying the single-field theories to produce large non-Gaussianities, and if possible then formation of large mass PBHs. For the case of the scalar-tensor theories, where the ghost-free propagator picks up a correct sign, the effective sound speed should satisfy the causality and unitarity constraints which gives us the Horendeski theory \cite{Horndeski:1974wa} having second order equation of motion. The subclass of the Horendeski theory is the theory that respects the Galilean shift symmetry, which is sufficient enough to address the ghost-free properties of the underlying theory. See refs. \cite{Kobayashi:2010wa,Jain:2010ka,Gannouji:2010au,Ali:2010gr,deRham:2011by,Tsujikawa:2010sc,Burrage:2010rs,DeFelice:2010jn,DeFelice:2010gb,Babichev:2010jd,DeFelice:2010pv,DeFelice:2010nf,Hinterbichler:2010xn,Kobayashi:2010cm,Deffayet:2010qz,Burrage:2010cu,Mizuno:2010ag,Nesseris:2010pc,Khoury:2010xi,DeFelice:2010as,Kimura:2010di,Zhou:2010di,Hirano:2010yf,Kamada:2010qe,VanAcoleyen:2011mj,Hirano:2011wj,Li:2011sd,Pujolas:2011he,Kobayashi:2011pc,DeFelice:2011zh,Khoury:2011da,Trodden:2011xh,Burrage:2011bt,Liu:2011ns,Kobayashi:2011nu,PerreaultLevasseur:2011wto,deRham:2011by,Clifton:2011jh,Endlich:2011vg,Brax:2011sv,Gao:2011mz,DeFelice:2011uc,Gao:2011qe,Babichev:2011iz,DeFelice:2011hq,Khoury:2011ay,Qiu:2011cy,Renaux-Petel:2011rmu,DeFelice:2011bh,Kimura:2011td,Wang:2011dt,Kimura:2011dc,DeFelice:2011th,Appleby:2011aa,DeFelice:2011aa,Zhou:2011ix,Goon:2012mu,Shirai:2012iw,Goon:2012dy,deRham:2012az,Ali:2012cv,Liu:2012ww,Choudhury:2012yh,Choudhury:2012whm,Barreira:2012kk,Gubitosi:2012hu,Barreira:2013jma,deFromont:2013iwa,Deffayet:2013lga,Arroja:2013dya,Li:2013tda,Sami:2013ssa,Khoury:2013tda,Burrage:2015lla,Koyama:2015vza,Brax:2015dma,Saltas:2016nkg,Ishak:2018his} for more details in this direction.

In view of the aforementioned attractive features of Galileon field, we shall compute the non-Gaussianity amplitude from the bispectrum which is expected to be large in this case as quantum loop effects are insignificant thanks to the non-renormalizability property of the underlying framework.
The paper is organized as follows: In section \ref{s2}, we have reviewed the general framework of covariantized galileon theory with a scalar field in a de-Sitter background. Section \ref{s3} presents a semi-classical treatment using the mode functions in terms of the comoving curvature perturbation to compute the tree-level power spectrum in the underlying CGT framework. In section \ref{s4}, we discuss primordial non-Gaussianities in general, including their theoretical motivation and current observational status. In Section \ref{s5}, we perform a detailed study on the evaluation of the three-point function and the associated bispectrum for all the three phases individually and cumulatively using the well-known in-in formalism and information about the scalar modes discussed in the previous sections. The numerical results are shown in section \ref{s6}. Then, in section \ref{s7} we summarise our findings. Finally, in Appendix \ref{App:A}, \ref{App:B} and \ref{App:C} we provide the detailed computations of the bispectrum and the associated non-Gaussianity amplitudes for each of the three regions respectively.

\section{General Framework of Covariantized Galileon in de Sitter Background} \label{s2}
The Galileon action was first introduced in \cite{Nicolis:2008in}. It is a framework where one can obtain equations of motion of second-order from a scalar field theory with higher-derivative terms in a Minkowski space-time. In \cite{Deffayet:2009wt}, the authors presented a construction of the Galileon theory which was ghost-free and preserved unitarity in a dynamical space-time by introducing non-minimal coupling with the background gravity. The Galileon theory is equipped with a Galilean symmetry, a modified version of the shift symmetry which has relation with the slow-roll feature of the inflationary potential. This symmetry transforms a scalar field $\phi$ as follows:
\bea \label{IIa}\phi \rightarrow \phi+ c + b_{\mu}x^{\mu}.\eea
where $c$ is a scalar constant, $b_{\mu}$ is a vector constant, and $x^{\mu}$ represents the $3+1$ dimensions space-time coordinates. The last term in the above equation represents space-time translations since it resembles coordinate transformation between non-relativistic inertial frames. 
Now, having an inflationary solution requires the soft breaking of the exact Galilean symmetry. This manner of breaking ensures the fact that our underlying theory, coupled with the gravitational sector, does not receive any significant correction due to those being suppressed by the factor $\displaystyle{\Lambda/M_{pl}}$ where $M_{pl}$ is the Planck mass.
In \cite{Deffayet:2009wt}, the authors also introduced a ghost-free version of the theory from Ostrogradski instability in a curved space background in the classical regime. This is known as the Covariantized Galileon Theory (CGT).
Starting with a five-dimensional covering theory in a curved background, the action for this theory is written to be as:
\vspace{-.15cm}
\bea
S = \int d^{4}x\sqrt{-g}\left[\frac{M^{2}_{pl}}{2}R - V_{0} + \sum^{5}_{i}c_{i}{\cal L}_{i}\right]
\eea
\vspace{-.18cm}
The explicit form of the terms above, ${\cal L}_{i}$, $\forall i =1,2,3,4$, are given by the following expressions:
\bea
{\cal L}_1 & = & \phi, \quad
 {\cal L}_2=-\frac{1}{2} (\grad \phi)^2 ,\quad
	{\cal L}_3=\frac{c_3}{\Lambda^3} (\grad \phi)^2 \Box \phi ,\quad
	{\cal L}_4= -\frac{c_4}{\Lambda^6} (\grad \phi)^2 \Big\{
					(\Box \phi)^2 - (\grad_\mu \grad_\nu \phi)
					(\grad^\mu \grad^\nu \phi)
					- \frac{1}{4} R (\grad \phi)^2
				\Big\},\\
	{\cal L}_5&=& \frac{c_5}{\Lambda^9} (\grad \phi)^2 \Big\{
					(\Box \phi)^3 - 3 (\Box \phi)( \grad_\mu \grad_\nu \phi)
					(\grad^\mu	 \grad^\nu \phi)
     + 2 ( \grad_\mu  \grad_\nu \phi)
					(\grad^\nu	 \grad^\alpha \phi)
					(\grad_\alpha \grad^\mu \phi)
					- 6 G_{\mu \nu} \grad^\mu \grad^\alpha \phi
					\grad^\nu \phi \grad_\alpha \phi
				\Big\}.	\nonumber	
	\eea
where $R$ is the Ricci scalar and $G_{\mu\nu}$ represents the Einstein tensor for the gravitational background. This CGT, with a curved background, breaks the Galilean symmetry softly. Here the coefficients $c_{i}$ are adjusted to appear in a dimensionless manner. Also, the parameter $\Lambda$ represents the physical energy scale cut-off for the theory. It must be kept in mind that given our representation, the theory given is not valid above the cut-off scale. However, it is possible for the quantum fluctuations to go beyond the energy scale $\Lambda$ if the Vainshtein effect is active. In the terms mentioned above, the lagrangians ${\cal L}_{4}$ and ${\cal L}_{5}$ do contain terms with non-minimal coupling to gravity, but they are later suppressed through powers of $\displaystyle{H/{\Lambda}}$. We have kept these terms for the completeness of the covariantization description even though we mentioned they remain insignificant during inflation, where galileon self-interactions dominates the non-linearities. A careful examination of the above expressions shows that for specific values of the coefficients, $c_4 = 0 = c_5$, one is able to recover the covariantized version of the DGP model. However, if the Galileon field $\phi$ is important only while inflation persists, then the coefficients $c_i$ must be determined only through cosmological observations. Such an analysis is performed in ref.\cite{Ali:2010gr} for the coefficients $c_2, c_3$, and $c_4$.

\textcolor{black}{Terms like a constant $V_{0}$, and those linear in the scalar field $\phi$ are also part of the only allowed possibilities that are known to respect the non-renormalization theorem \cite{Burrage:2010cu}, which is going to be the prime highlighting component of this work. To successfully implement inflation requires softly breaking the modified shift symmetry, which is possible in the presence of such terms. Beyond this, one can look into other theories, such as the Horendeski theory \cite{Horndeski:1974wa}, where hard symmetry breaking would render the non-renormalization theorem inapplicable to implement inflation further. We localize our present discussion towards the Galileon theory, and hence, the terms mentioned above are the only ones necessary to consider to implement inflation in the present discussion successfully.}

We now begin the discussion of the inflationary solution in a quasi-de Sitter background. The effective inflationary potential is required to satisfy the respective constraint given by $|\Delta V/V| \ll 1$. This leads to the Galileon theory in a quasi-de Sitter background where the scale factor satisfies $a(t) = \exp{(Ht)}$, with $H$ representing the Hubble parameter which does not remain a constant. Now focusing on the Galileon part of the action, we find that upon performing integration by parts and discarding the boundary terms together gives us the action of a time-dependent, background Galileon field $\bar{\phi}_0(t)$ in the following form:
\bea\label{Z}  S_0=\int d^4x\,a^3 \,\Bigg\{\dot{\bar{\phi}}^2_0\Bigg(\frac{c_2}{2}+2c_3Z+\frac{9c_4}{2}Z^2+6c_5Z^3\Bigg)+\lambda^3\bar{\phi}_0\Bigg\}\quad\quad{\rm where}\quad\quad Z\equiv \frac{H\dot{\bar{\phi}}_0}{\Lambda^3}.\eea
From this we obtain solutions under specific conditions as:
\bea \dot{\bar{\phi}}_0=\frac{\Lambda^3}{12H}\frac{c_2}{c_3}\Bigg[-1+\sqrt{1+\frac{8c_3}{c^2_2}\frac{\lambda^3}{\Lambda^3}}\Bigg]=
\left\{
	\begin{array}{ll}
		\displaystyle \frac{\lambda^3}{3c_2H}\quad\quad\quad & \mbox{when}\quad  Z\ll 1    \\ 
			\displaystyle 
			\displaystyle \sqrt{\frac{\Lambda^3}{18c_3}\frac{\lambda^3}{H^2}}\quad\quad\quad & \mbox{when }\quad  Z\gg 1 
	\end{array}
\right. \eea
When considering the theory in a weak-coupling regime, i.e., $Z \ll 1$, it eventually resembles the canonical slow-roll inflation model. On the other hand, in a strong-coupling regime, i.e., $Z \gg 1$, we encounter the DGP model. However, if $Z \simeq 1 $, then the theory lies in between two extremes of the strong and weak regimes. The relative contributions become controlled if we compare the higher-derivative and lower-derivative terms as a result of the coupling parameter $Z$ having positive powers in the construction. This is because the terms with non-minimal couplings to gravity will become insignificant due to there being no interactions with the background gravity. The non-linearities from the Galileon sector will still be present from various derivative terms. However, under the condition $Z \ll 1$,  we have to consider the mixing contributions of the Galileon and the non-minimal couplings from the gravitation sector since changes due to them in the canonical slow-roll inflation will be significant in nature. In this paper, we are concerned with the intermediate regime, i.e., $Z \simeq 1 $. 

\section{Semi-Classical modes from Cosmological Perturbation}
\label{s3}	
In this section, we discuss the second-order perturbation under the framework of the Covariantized Galileon Theory (CGT). We begin by constructing the classical equation of motion for the generalized curvature perturbation modes in the Fourier space. This is the well-known Mukhanov-Sasaki equation. After this, in the subsequent sections, we solve this equation in the three regions of interest, namely the first slow-roll (SRI), Ultra slow-roll (USR), and the second slow-roll (SRII) regions. An analytic approach would require us to establish a quantum initial boundary condition, known as the Bunch-Davies vacuum state, through the use of the Bogoliubov coefficients of the region of interest. This is required to fix the mode functions in the SRI region, and using these along with the continuity conditions at the transition points between SRI to USR and USR to SRII also determines the mode functions separately for the other regions.

The action second-order in the curvature perturbation modes is written as a function of conformal time in the following way:
\bea \label{IIIa} S^{(2)}_{\zeta}=\displaystyle \int d\tau\;  d^3x\;  a^2\; \frac{{\cal A}}{H^2}\big(\zeta^{'2}-c^2_s\left(\partial_i\zeta\right)^2\big)=\displaystyle \int d\tau\;  d^3x\;  a^2\; \frac{{\cal B}}{c^2_sH^2}\big(\zeta^{'2}-c^2_s\left(\partial_i\zeta\right)^2\big).\eea  
where a derivative with respect to the conformal time is performed. Here the time-dependent quantities ${\cal A}$ and ${\cal B}$ are represented as follows:
\bea \label{A}  {\cal A}:&\equiv& \frac{\dot{\bar{\phi}}^2_0}{2}\Bigg(c_2+12c_3Z+54c_4Z^2+120c_5Z^3\Bigg),\\
    \label{B} {\cal B}:&\equiv& 
   \frac{\dot{\bar{\phi}}^2_0}{2}\Bigg\{c_2+4c_3\Bigg(2Z-\frac{H\dot{\bar{\phi}}_0}{\Lambda^3}\eta\Bigg)+2c_4\Bigg[13Z^2-\frac{6}{\Lambda^6}\dot{\bar{\phi}}^2_0H^2\big(\epsilon+2\eta\big)\Bigg]-\frac{24c_5}{\Lambda^9}H^3\dot{\bar{\phi}}^3_0\big(2\epsilon+1\big)\Bigg\}.\eea
where the coupling constant $Z$ is introduced in the CGT action in the previous section. Here we introduce the first and second slow-roll parameters,
$\epsilon=-\frac{\dot{H}}{H^2},\eta=-\frac{\ddot{\bar{\phi}}_0}{H\dot{\bar{\phi}}_0}.$
The parameter $c_{s}$ in the second-order action is the effective sound speed which is defined as, 
$c_{s} = \sqrt{\frac{{\cal B}}{{\cal A}}}.$
Now that we have the second-order action with us, we move on to introduce a new variable which redefines the curvature perturbation field and is written as,  $v(\tau, {\bf x})=z(\tau)\zeta(\tau, {\bf x})$ with $ z(\tau)=\frac{a\sqrt{2{\cal A}}}{H^2}=\frac{a\sqrt{2{\cal B}}}{c_sH^2},$
this is also known as the Mukhanov-Sasaki variable. Using this variable in Eq.(\ref{IIIa}) gives us its different form whose canonically normalized version is written as:
\bea
S^{(2)}_{\zeta}=\frac{1}{2}\int d\tau\;  d^3x\;  \bigg(v^{'2}(\tau, {\bf x})-c^2_s\left(\partial_iv(\tau, {\bf x})\right)^2+\frac{z^{''}(\tau)}{z(\tau)}v^{2}(\tau, {\bf x})\bigg).
   \eea
Now we move towards the Fourier space for this action which leads to the aforementioned second order action being transformed in the following manner: 
 \bea
S^{(2)}_{\zeta}&=&\frac{1}{2}\int \frac{d^3{\bf k}}{(2\pi)^3}\;d\tau\;\bigg(|v^{'}_{\bf k}(\tau)|^{2}-\omega^2(k,c_s,\tau)|v_{\bf k}(\tau)|^{2}\bigg).
   \eea
   where the effective conformal time dependent frequency in the present context is defined as:
   \bea \omega^2(k,c_s,\tau)=\bigg(c^2_sk^2-\frac{z^{''}(\tau)}{z(\tau)}\bigg)\quad\quad{\rm where}\quad\quad \frac{z^{''}(\tau)}{z(\tau)}\approx \frac{2}{\tau^2}.\eea
After varying the above mentioned action in Fourier space we get the following equation of motion, which is frequently referred as the {\it Mukhanov-Sasaki equation} (MS) and given by the following expression:
\bea v^{''}_{\bf k}(\tau)+\omega^2(k,c_s,\tau)v_{\bf k}(\tau)=0.\eea

\textcolor{black}{In order to implement the three phases, SR, USR, and SRII, the shape of the potential is not altered here as it would disturb the symmetry-breaking feature needed to perform inflation. We can, therefore, play with the coupling coefficients $c_{i}\;\forall\;i=1,\cdots,5$. These coefficients are significant in implementing the three phases. Since we are working here in an EFT framework, one can, in general, develop a large class of possibilities to generate the conditions for the slow-roll parameters for each of the three phases. Finally, we join these three initially disconnected phases to explain the phenomenon of PBH production within the Galileon framework;  the said phases  may be joined sharply or smoothly. We  have not analyzed the smooth transition case and refer the reader to refs.\cite{Riotto:2023gpm,Firouzjahi:2023ahg,Firouzjahi:2023aum}. In this discussion, we focus on having a sharp transition feature, which we have implemented by using the Heaviside Theta function at the position of the transition, $k_{s}$ for the SRI to the USR, and $k_{e}$ for the USR to the SRII. Apart from this construction, the choice of those coupling coefficients $c_{i}$ also assures the formation of PBH and allows for a sufficient number of e-foldings by maintaining the necessary perturbative approximations. A similar effect can be implemented by our choice of the effective sound speed parameter $c_{s}$. The definition of the sound speed involves the time-dependent quantities ${\cal A}$ and ${\cal B}$. These quantities are written down explicitly in Eqs.(\ref{A},\ref{B}) and the use of the coupling coefficients $c_{i}$ and the other new coupling constant $Z$, in terms of the time-dependent background galileon field as in Eqn(\ref{Z}), is evident in their definition. Hence, to parameterize the sound speed exactly mimics the role to parameterize the couplings $c_{i}$.}

\textcolor{black}{Though we have not mentioned the explicit parameterization of these couplings, we do mention the specific parameterization of the effective sound speed. Its value is labelled as $c_{s}= c_{s,*}$ at the scale of horizon crossing at conformal time $\tau=\tau_{*}$. Throughout the SRI phase, this value remains constant. As we approach the transition moment, at $\tau=\tau_{s}$ the value it takes is of the form $c_{s}= \tilde{c_{s}}=1\pm \delta$, where $\delta \ll 1$. This value of $\tilde{c_{s}}$ is also the sound speed when the other transition moment at $\tau=\tau_{e}$ is encountered. Between the two transition moments, that is, during the USR and continuing till the end of inflation, after USR, the effective sound speed is the same as its value at horizon crossing, $c_{s}= c_{s,*}$.} 

\subsection{Region I: First Slow Roll (SRI) region}
\label{s3a}
We now discuss the general solution of the MS equation, in the SRI region ($\tau \leq \tau_{s}$), for the perturbed scalar mode. This is given as follows:
\bea
   v_{\bf k}(\tau)&=&\frac{\alpha^{(1)}_{\bf k}}{\sqrt{2c_sk}}\left(1-\frac{i}{kc_s\tau}\right)\; e^{-ikc_s\tau}+\frac{\beta^{(1)}_{\bf k}}{\sqrt{2c_sk}}\left(1+\frac{i}{kc_s\tau}\right)\; e^{ikc_s\tau}.
   \eea
where $\alpha^{(1)}_{\bf{k}}$ and $\beta^{(1)}_{\bf{k}}$ are the respective Bogoliubov coefficients for this region. We also choose the well-known, Bunch-Davies quantum vacuum state, which is obtained by fixing the mode function using the following choices for the coefficients:
\bea \label{bG1a}&&\alpha^{(1)}_{\bf k}=1, \\
 \label{bG1b}&&\beta^{(1)}_{\bf k}=0.\eea
these initial conditions helps in defining the necessary physical inflationary vacuum state.
After implementing the said initial conditions we obtain the following expression for the curvature perturbation in the SRI region ($\tau < \tau_{s}$):
\bea \zeta_{\bf{k}}(\tau) = \frac{v_{\bf{k}}(\tau)}{z} = \left(\frac{iH^{2}}{2\sqrt{\cal A}}\right)\frac{1}{(c_{s}k)^{3/2}}\left(1+ikc_{s}\tau\right)e^{-ikc_{s}\tau}.
\eea
The first slow-roll parameter, $\epsilon$, is roughly of a constant value during this phase and changes very slowly with time. The second slow-roll parameter, $\eta$, is almost zero for this phase. 

\subsection{Region II: Ultra Slow Roll (USR) region}
\label{s3b}
The USR phase is denoted by the following conformal time region $\tau_{s}<\tau<\tau_{e}$, where $\tau_{s}$ marks the beginning of the USR phase after the end from SRI phase and the conformal time $\tau_{e}$ denotes the end of the USR phase. The parameter $\epsilon$ in this phase can be explicitly written using the same parameter in the SRI component as follows,
$\epsilon(\tau)= \left(\tau/\tau_{s}\right)^{6}.$ This form clearly depicts the fact that this parameter is almost of a constant value at the moment the transition from SRI to USR phase occurs, i.e., $\epsilon(\tau = \tau_{s}) = \epsilon$. After this transition, when $\tau > \tau_{s}$, the same slow-roll parameter is no longer a constant. This fact is crucial for the interpretation of our further analysis; hence, it is worth remembering this fact at this stage. The general solution of the MS equation in the USR phase is written as follows:
\bea \zeta_{\bf{k}}(\tau) = \frac{v_{\bf{k}}(\tau)}{z} = \left(\frac{iH^{2}}{2\sqrt{\cal A}}\right)\left(\frac{\tau_{s}}{\tau}\right)^{3}\frac{1}{(c_{s}k)^{3/2}}\times \left[\alpha^{(2)}_{\bf k}\left(1+ikc_{s}\tau\right)e^{-ikc_{s}\tau} - \beta^{(2)}_{\bf k}\left(1-ikc_{s}\tau\right)e^{ikc_{s}\tau} \right].
\eea
An important thing to consider is the fact that the Bogoliubov coefficients in this region, i.e., $\alpha^{(2)}_{\bf k}$ and $\beta^{(2)}_{\bf k}$, can be expressed in terms of the initial conditions required to fix the initial Bunch-Davies vacuum state during the SRI phase. It is achieved through a Bogoliubov transformation which ultimately suggests that the underlying vacuum now differs in structure from the initial Bunch-Davies state. By imposing the continuity and differentiability conditions on the modes computed from SRI and USR phases, the Bogoliubov coefficients for the USR phase can then be obtained through the use of the Israel junction condition applied at the transition time $\tau_{s}$, which are given by: 
\bea \label{bG2a}\alpha^{(2)}_{\bf k}&=&1-\frac{3}{2ik^{3}c^{3}_s\tau^{3}_s}\left(1+k^{2}c^{2}_s\tau^{2}_s\right),\\
\label{bG2b}\beta^{(2)}_{\bf k}&=&-\frac{3}{2ik^{3}c^{3}_s\tau^{3}_s}\left(1+ikc_s\tau_s\right)^{2}\; e^{-2ikc_s\tau_s}.\eea
These values will be fundamental in further analysis of the correlation functions of the modes into this region. The parameter $\eta$ has a large magnitude in this phase, $\eta \sim -6$, and this is responsible for the enhancement of the perturbations in the this phase.

\subsection{Region III: Second Slow Roll (SRII) region}
\label{s3c}
The final slow-roll phase, SRII, is denoted by the following conformal time region $\tau_{e}<\tau<\tau_{\text{end}}$, where $\tau_{e}$ denotes the exit from USR and entry into the SRII region while $\tau_{\text{end}}$ denotes the conclusion of the inflationary paradigm. The slow-roll parameter $\epsilon$ for SRII can be written using the same parameter in the SRI region in the following manner, 
$\epsilon(\tau)= \left(\tau_{e}/\tau_{s}\right)^{6}.$
This parameter now possesses a non-constant value throughout this region, when crossing from USR to SRII and until the end of SRII region. The general solution of the MS equation also changes when taken into account this fact along with the specific time-dependent nature. This is written as follows:
\bea \zeta_{\bf{k}}(\tau) = \frac{v_{\bf{k}}(\tau)}{z} = \left(\frac{iH^{2}}{2\sqrt{\cal A}}\right)\left(\frac{\tau_{s}}{\tau_{e}}\right)^{3}\frac{1}{(c_{s}k)^{3/2}}\times \left[\alpha^{(3)}_{\bf k}\left(1+ikc_{s}\tau\right)e^{-ikc_{s}\tau} - \beta^{(3)}_{\bf k}\left(1-ikc_{s}\tau\right)e^{ikc_{s}\tau} \right].
\eea
To obtain this solution which includes the presence of a new set of Bogoliubov coefficients, we use the boundary conditions fixed using the vacuum of the USR phase. The underlying vacuum structure for the SRII phase is also now completely different and the use of the new boundary conditions, equivalently the Israel junction conditions, at the transition from USR to SRII phase gives us the following explicit form of these new coefficients:
\bea \label{bG3a}\alpha^{(3)}_{\bf k}&=&-\frac{1}{4k^6c^6_s\tau^3_s\tau^3_e}\Bigg[9\left(kc_s\tau_s-i\right)^2\left(kc_s\tau_e+i\right)^2 e^{2ikc_s(\tau_e-\tau_s)}\nonumber\\
&&\quad\quad\quad\quad\quad\quad\quad\quad\quad\quad\quad\quad\quad\quad\quad\quad-
\left\{k^2c^2_s\tau^2_e\left(2kc_s\tau_e-3i\right)-3i\right\}\left\{k^2c^2_s\tau^2_s\left(2kc_s\tau_s+3i\right)+3i\right\}\Bigg],\\
\label{bG3b}\beta^{(3)}_{\bf k}&=&\frac{3}{4k^6c^6_s\tau^3_s\tau^3_e}\Bigg[\left(kc_s\tau_s-i\right)^2\left\{k^2c^2_s\tau^2_e\left(3-2ikc_s\tau_e\right)+3\right\}e^{-2ikc_s\tau_s}\nonumber\\
&&\quad\quad\quad\quad\quad\quad\quad\quad\quad\quad\quad\quad\quad\quad\quad\quad+i\left(kc_s\tau_e-i\right)^2\left\{3i+k^2c^2_s\tau^2_s\left(2kc_s\tau_s+3i\right)\right\}e^{-2ikc_s\tau_e}\Bigg].\eea
These values will be crucial in further analysis of the correlation functions from this phase. The parameter $\eta$ for this phase is almost zero, much like what was the condition for this parameter in the first slow-roll phase.

\subsection{Tree level power spectrum from comoving curvature perturbation }
\label{s3d}

The quantization of the scalar curvature perturbations modes $\zeta_{\bold{k}}$ require the introduction of an annihilation $\hat{a}_{\bf{k}}$ and a creation operator $\hat{a}^{\dagger}_{\bf{k}}$, whose action on the initial Bunch-Davies vacuum will in turn create an excited state or annihilate an existing state. The important restriction in order to define a Bunch-Davies vacuum state is, $\hat{a}_{\bold{k}}\ket{0} = 0 \quad \forall \bold{k}$.
The following quantization conditions, at equal times ($\tau$), must also be satisfied in order to perform the quantization procedure:
\bea
\left[\hat{\zeta}_{\bold{k}}(\tau) , \hat{\Pi}_{\bold{k}^{\text{'}}}(\tau)\right] = i\delta^{3}(\bold{k} + \bold{k}^{\text{'}}), \quad \left[\hat{\zeta}_{\bold{k}}(\tau) , \hat{\zeta}_{\bold{k}^{\text{'}}}(\tau)\right] = 0, \quad \left[\hat{\Pi}_{\bold{k}}(\tau) , \hat{\Pi}_{\bold{k}^{\text{'}}}(\tau)\right] = 0
\eea
where $\Pi_{\bold{k}} = \partial_{\tau}\zeta_{\bold{k}}$ is the conjugate momentum variable of the scalar curvature mode. Once these classical modes and its respective conjugate momenta are promoted to become quantum operators, we can then mention the following expressions for the same operators:
\bea
\hat{\zeta}_{\bold{k}}(\tau) =  \zeta_{\bold{k}}a_{\bold{k}} + \zeta_{\bold{k}}^{*}a^{\dagger}_{\bold{-k}}, \quad\quad  \hat{\Pi}_{\bold{k}}(\tau) =  \Pi_{\bold{k}}a_{\bold{k}} + \Pi_{\bold{k}}^{*}a^{\dagger}_{\bold{-k}}.
\eea
Upon considering the late time limit $\tau\rightarrow 0$ with the co-moving curvature disturbance, we can write the tree-level version of the two-point cosmological correlation function as follows:
\bea \langle \hat{\zeta}_{\bf k}\hat{\zeta}_{{\bf k}^{'}}\rangle_{{\bf Tree}} =(2\pi)^{3}\;\delta^{3}\left({\bf k}+{\bf k}^{'}\right)\frac{2\pi^2}{k^3}\Delta^{2}_{\zeta,{\bf Tree}}(k),\quad\quad \text{where} \quad \Delta^{2}_{\zeta,{\bf Tree}}(k) = \frac{k^{3}}{2\pi^{2}}\langle\langle \hat{\zeta}_{\bf k}\hat{\zeta}_{-{\bf k}}\rangle\rangle_{(0,0)} = \frac{k^{3}}{2\pi^{2}}|{\zeta}_{\bf k}(\tau)|^{2}_{\tau\rightarrow 0}.
\eea

\textcolor{black}{Currently, from our analysis done in the previous section for each of the individual phases, SRI, USR, and SRII, we can use their solutions for the cosmological scalar perturbation modes to write down the tree-level version of the dimensionless power spectrum depending on the particular interval of consideration, in terms of the wavenumber, after calculation as follows:} 
\bea  \Delta^{2}_{\zeta,{\bf Tree}}(k)
&=& \displaystyle
\displaystyle\left\{
	\begin{array}{ll}
		\displaystyle\Delta_{\bf {Tree}}^{\text{SRI}}(k)=\left(\frac{H^{4}}{8\pi^{2}{\cal A} c^3_s}\right)_* \Bigg\{1+\Bigg(\frac{k}{k_s}\Bigg)^2\Bigg\}& \mbox{when}\quad  k\leq k_s  \;(\rm SRI)  \\  
			\displaystyle 
			\displaystyle\Delta_{\bf {Tree}}^{\text{USR}}(k)=\left(\frac{H^{4}}{8\pi^{2}{\cal A} c^3_s}\right)_* \left(\frac{k_e}{k_s }\right)^{6}\left|\alpha^{(2)}_{\bf k}-\beta^{(2)}_{\bf k}\right|^2 & \mbox{when }  k_s\leq k\leq k_e  \;(\rm USR)\\ 
   \displaystyle 
			\displaystyle\Delta_{\bf {Tree}}^{\text{SRII}}(k)=\left(\frac{H^{4}}{8\pi^{2}{\cal A} c^3_s}\right)_* \left(\frac{k_e}{k_s }\right)^{6}\left|\alpha^{(3)}_{\bf k}-\beta^{(3)}_{\bf k}\right|^2 & \mbox{when }  k_e\leq k\leq k_{\rm end}  \;(\rm SRII) 
	\end{array}
\right. \eea
\textcolor{black}{where the expressions follow upon considering to work in the super-horizon regime. To obtain the cumulative contribution to the primordial tree-level scalar power spectrum, we require summing over all the individual contributions, as mentioned in the equation above. The final form of this total contribution is a result of using the following equation: }
\textcolor{black}{\bea \label{totpower} \left[\Delta^{2}_{\zeta,{\bf Tree}}(k)\right]_{\bf Total} &=&\Delta_{\bf {Tree}}^{\text{SRI}}(k)+\Delta_{\bf {Tree}}^{\text{USR}}(k)\Theta(k-k_s)+\Delta_{\bf {Tree}}^{\text{SRII}}(k)\Theta(k-k_e).
\eea}

\textcolor{black}{The above equation represents the total tree-level primordial power spectrum where we have incorporated the use of two distinct Heaviside Theta functions to smoothly connect the individual contributions to the total tree-level power spectrum from the regions SRI and USR at the transition wavenumber $k=k_{s}$ and from the regions USR and SRII at the transition wavenumber $k=k_{e}$. } 

\section{General introduction to Primordial Non-Gaussianity from comoving scalar perturbation modes}
\label{s4}

In this section, we present a general discussion on primordial non-Gaussianities. 
They represent the deviations from Gaussian distribution of the primordial density perturbations locally and described by the Gaussian random fields $\zeta_{g}(x)$ which satisfies the Gaussian statistics. The original vacuum fluctuations during inflation get promoted to classical perturbations when they exit the Horizon. In the super-horizon regime, locally each position can be treated separately, and their evolution will be determined by the initial conditions in the form of Gaussian distributions. Now, the observed curvature perturbations, up to first order, can be represented as a term linear in the perturbations described by the Gaussian random fields. Hence, the non-linearities will give the presence of non-Gaussianities in observations. This would require that higher-order correlation functions should not become zero. In the present work, we have evaluated non-Gaussianities with the help of the three-point correlation function or the bispectrum for the gauge invariant scalar modes. Their non-linear nature is represented mathematically through the following expression:
\bea 
\zeta(\bold{x}) = \zeta_{g}(\bold{x}) + \frac{3}{5}f_{\text{NL}}\big(\zeta^{2}_{g}(\bold{x}) - \langle \zeta^{2}_{g}(\bold{x}) \rangle\big)+\cdots\cdots
\eea 
where $\zeta(\bold{x})$ is the primordial comoving curvature perturbation. Here $\fnl$ denotes the local amplitude of the non-Gaussianity which is a model-dependent quantity and the current observational estimates for this quantity comes from the CMB measurements with the value of $\fnl = -0.9 \pm 5.1$ at $68\%$ level confidence from Planck \cite{Planck:2019kim}. This measurement contains significant error bars and at present do not allow for a physically concrete statement to be made. However, with data from the future surveys, the new observations are estimated to provide improvements of an order of magnitude over the current estimates \cite{Achucarro:2022qrl,CMB-S4:2023zem}.
An interesting fact about the understanding of primordial non-Gaussianities also comes through the fact that one can split those density perturbations into a background part consisting of a large-scale component, in which one observes appreciable changes when looking into scales comparable to the wavelength of the long modes and a small-scale component which already provides significant deviations on scales smaller than the long-modes. This can also be written using the said decomposition, $\zeta_{g} = \zeta_{g,L} + \zeta_{g,S}$ and $\zeta = \zeta_{L} + \zeta_{S}$, in the following way: 
\bea \zeta_{S}(\bold{x}) &=& \zeta_{g,S}(\bold{x})\left(1 + \frac{6}{5}\fnl\zeta_{g,L}(\bold{x})\right) + \frac{3}{5}f_{\text{NL}}\zeta_{g,S}(\bold{x})^{2},\\
\zeta_{L}(\bold{x}) &=& \zeta_{g,L}(\bold{x}) + \frac{3}{5}\fnl\zeta_{g,L}(\bold{x})^{2}.
\eea 
where the first equation provides a non-trivial connection between the small-scale and large-scale cosmological fluctuations. These discussions lead to an important question concerned with finding a theoretical model which can show the generation of non-Gaussianities of the order $|f_{\text{NL}}| \sim {\cal O}(1)$. Since, these features arrive while studying the cosmological perturbations generated in the very early universe and, remarkably, their observational imprints are also visible in the form of the CMB data, this question becomes all the more important as it can lead to some exciting insights into the origin of structure in the very early universe.

According to the canonical single field inflation model, Maldacena derived a consistency condition \cite{Maldacena:2002vr}:
\bea f_{\rm NL}=\frac{5}{12}\left(1-n_s\right),\eea 
which is sometimes referred as a {\it no-go theorem}, stating that in the squeezed limit, the value of the amplitude of non-Gaussianity, $\fnl$, computed using the three-point cosmological correlation function of scalar modes for single field models during slow-roll is connected to the spectral index of the primordial power spectrum of scalar modes, and is estimated as ${\cal O}(10^{-2})$, which is a very small number when considering the detection of primordial non-Gaussianities in cosmological observations.

This is obviously a challenging task to generate large amount of non-Gaussianities from single field models of inflation. To achieve such an interesting goal in this paper, we completely devote ourselves to study the generation of large non-Gaussianities by explicitly breaking the Maldacena's {\it no-go theorem} in the squeezed limit of the three-point cosmological correlation function for the scalar modes using the underlying theory of Galileon inflation. In the next section we are going to provide a mechanism which allows us to produce large amount of non-Gaussian amplitude out of this computation. To fulfill the purpose as discussed before, we will introduce the three phases SRI, USR and SRII along with sharp transitions at the phase changing boundaries near the beginning and end of the USR phase. With the help of our computation we will explicitly show that such sharp transitions along with the transition scale positions described in terms of the small wave numbers is able to generate a large non-Gaussianity amplitude from the mentioned cosmological three-point function by breaking the previously mentioned {\it no-go theorem} in the squeezed limit, particularly, in the USR and SRII phases.

\section{Computing the three point function and the associated bispectrum from scalar modes}
\label{s5}

In this section, we use the information about the modes obtained in the previous section to calculate the tree level three-point correlation function and similarly the associated bispectrum for all the three SRI, USR, and SRII regions. We also focus on the validity of the consistency condition for single-field inflation in present Galileon inflation theory for all three regions, especially in the USR region, which will let us know about the scope of this condition while considering the formation of PBH. Hence, we explicitly examine the squeezed limit behavior of the bispectrum. The {\it Schwinger-Keldysh formalism}, also known as the {\it in-in formalism}, is the well-known method with which we will begin to introduce the expectation value for the required three-point correlation function, and that is used together with the third-order action of CGT to calculate the results for the bispectrum and corresponding associated non-Gaussianity amplitude. 
 
\subsection{The Schwinger-Keldysh formalism}
\label{s5a}

The method from {\it in-in formalism} is better used to evaluate the expectation value of the required three-point correlation function since we are interested in the expectation of operators at a specific instant of time, and the respective boundary conditions for the scalar modes are imposed only at very early times. This is the primary reason for this formalism to be named as {\it in-in}. 

The time-dependent expectation value for a given operator in Interaction picture is then given by in this formalism as:
\bea \bra{\Omega(\tau)}Q^{I}(\tau)\ket{\Omega(\tau)}=\bra{0}\bigg[\overline{T}\exp\Bigg(\displaystyle{i\int_{-\infty(1-i\epsilon)}^{\tau}H_{\text{Int}}^{I}(\tau')d\tau'}\Bigg)\bigg]Q^{I}(\tau)\bigg[T\exp\Bigg(\displaystyle{-i\int_{-\infty(1+i\epsilon)}^{\tau}H_{\text{Int}}^{I}(\tau')d\tau'}\Bigg)\bigg]\ket{0}. \eea
where $Q^{I}(\tau)$ is the operator in  the Heisenberg picture, $\ket{\Omega}$ and $\ket{0}$ are the interacting and free theory vacuum in the far past, described by:
\bea \ket{\Omega(\tau)}=\bigg[T\exp\Bigg(\displaystyle{-i\int_{-\infty(1+i\epsilon)}^{\tau}H_{\text{Int}}^{I}(\tau')d\tau'}\Bigg)\bigg]\ket{0},\eea
and $T$ and $\overline{T}$ tells us that the operators are either time or anti-time ordered. The early time limit is taken by $\tau_{0} \rightarrow -\infty$ and then regularizing the integral by the $i\epsilon$ prescription. Finally, the superscript ${I}$ implies the fields making up the interaction Hamiltonian and the operators are in the interaction picture. This picture is introduced to deal with the non-linearities arriving in the equations of motion as a result of the interactions in the total Hamiltonian. The expectation value of any operator is computed by first evolving the fields from the early past to the point in time of interest and then going from that moment back to the initial time.   

Now we expand this expression and take the leading order term when expanding in $H_{\text{Int}}^{I}$ since this will be responsible for the tree-level contribution which is what we are interested with for our further computations. The leading order term is then given by:
\bea \label{a3} \bra{\Omega(\tau)}Q^{I}(\tau)\ket{\Omega(\tau)} = -i\int_{-\infty}^{\tau}d\tau'\bra{0}[Q^{I}(\tau),H_{\text{Int}}^{I}(\tau')]\ket{0} = 2\times\text{Im}\left[\int_{-\infty}^{\tau}d\tau'\bra{0}Q^{I}(\tau)H^{I}_{\text{Int}}(\tau')\ket{0}\right].
\eea
where in the last equality the Hermiticity property is used to further simplify the commutator form. For the concern of this paper, $Q^{I}(\tau)$ would be the three-point correlation of the curvature perturbation written as $\hat{\zeta}(\bf{k}_1,\tau)\hat{\zeta}(\bf{k}_2,\tau)\hat{\zeta}(\bf{k}_3,\tau)$.

To begin with the evaluation of the correlation function, we must use the interaction Hamiltonian, also formed by the interaction picture fields, and start by performing contractions using the commutation relations for the fields for the quantized curvature perturbation, into a product of green's functions. The critical difference between the standard {\it in-out} and the {\it in-in formalism} here is that there is no analog of the Feynman propagator in an inflationary background. So we have to keep that in mind while performing contractions.

While performing the Wick contraction method to evaluate correlation functions, there is first the normal ordering of the fields where the positive frequency modes are kept to the right of the product to annihilate the vacuum, keeping all the negative frequency operators to the left. This arrangement will be similar under normal ordering for both the in-out and in-in cases. When we next begin to perform the contraction, we encounter the previously mentioned expression for the two-point cosmological correlation of the quantized curvature perturbation $\langle\hat{\zeta}(\bf{k},\tau)\hat{\zeta}(\bf{k'},\tau)\rangle$ which turns out to be a real quantity, more equivalently an absolute value squared of the mode functions, contrary to the imaginary quantity of the Feynman propagator in case of the {\it in-out formalism}. Combining both normal ordering and what we learned from the behavior of contractions from above, we can say that the expectation value of a string of field operators evaluated using {\it Wick's theorem} is written as:
\bea \bra{0}\hat{\zeta_1}\hat{\zeta_2}\cdots\hat{\zeta_n}\ket{0} = \bra{0}:\hat{\zeta_1}\hat{\zeta_2}\cdots\hat{\zeta_n}:\ket{0} + \textit{all possible contractions}.
\eea
where $\zeta_i = \zeta(\bold{k}_i,\tau_i)$ and \textit{all possible contractions} means including one contraction term for each way of contacting the $n$ operators into pairs. Using the methods learned here, we will explicitly compute the three-point correlation function for the different SRI, USR, and SRII regions. However, before that we have to understand the third-order action of the Covariantized Galileon Theory, which is necessary for the future computations to be possible. Hence in the next section, we discuss details of this action for the CGT properly.
\subsection{Third order action of comoving curvature perturbation}
\label{s5b}

We now analyze how the third-order action for the curvature perturbations computed from the CGT is constructed, which is used to calculate the required correlation functions. Here we must consider what changes the Galilean shift symmetry can bring to the curvature perturbation. Careful analysis of these changes leads us to understand what terms could be allowed by such symmetry and what others will not be necessary to include in the action.

We know from our earlier discussions that the Galilean shift symmetry must be broken softly to get a de-Sitter like solution in the inflationary phase. The curvature perturbation and its spatial and temporal derivatives transform differently under the Galilean shift symmetry Eq.(\ref{IIa}). Only terms breaking the shift symmetry will be of importance while there will be additional terms which will either be absorbed through field redefinition or they vanish at the boundary. There is one term of special importance when considering the USR period, which is of the form $\zeta^{'}\zeta^{2}$. It is one of those terms which vanishes at the boundary after the use of the shift transformation property but its effect at the transition points, before and after the USR phase, is much significant due to its coefficient having the form $\partial_{\tau}(\eta/{c_{s}^{2}})$. To understand how this term is actually redundant we take a look into how this term transforms under the softly broken Galilean shift symmetry:
\bea
\zeta^{'}\zeta^{2} \rightarrow \left(\zeta^{'} - \frac{H}{\dot{\bar{\phi_{0}}}}b_{0}\right)\left(\zeta^{'} - \frac{H}{\dot{\bar{\phi_{0}}}}(b.\delta x)\right)^{2} \sim \frac{1}{3}(b.\delta x)\partial_{\tau}(\zeta^{2}) \sim 0
\eea
this is the result at the boundary. Remaining terms coming from this such as, $\left(\frac{H}{\dot{\bar{\phi}}_0}\right)^2\left(b\cdot \delta x\right)\zeta^{'}$, $\left(\frac{H}{\dot{\bar{\phi}}_0}\right)^3 b_0\left(b\cdot \delta x\right)^2$, $\left(\frac{H}{\dot{\bar{\phi}}_0}\right)^2 b_0\left(b\cdot \delta x\right)$, $\left(\frac{H}{\dot{\bar{\phi}}_0}\right)\left(b\cdot \delta x\right)\zeta^{'}$ do not participate in the third-order action.
As a result of the transformation properties and the corresponding analysis done in ref.\cite{Choudhury:2023hvf} regarding the inclusion of several possible terms based on their soft symmetry-breaking behavior, we get the following final expression for the action which is of third-order in the curvature perturbation as a result of the remaining bulk self-interaction terms:
\bea &&S^{(3)}_{\zeta}=\int d\tau\;  d^3x\;  \frac{a^2}{H^3}\; \bigg[\frac{{\cal G}_1}{a}\zeta^{'3}+\frac{{\cal G}_2}{a^2}\zeta^{'2}\left(\partial^2\zeta\right)+\frac{{\cal G}_3}{a}\zeta^{'}\left(\partial_i\zeta\right)^2+\frac{{\cal G}_4}{a^2}\left(\partial_i\zeta\right)^2\left(\partial^2\zeta\right)\bigg].\eea
The coupling parameters ${\cal G}_i$, $\forall i=1,2,3,4$ appearing in the above action are are obtained to be of the following form:
\bea {\cal G}_1:&\equiv& \frac{2H\dot{\bar{\phi}}^3_0}{\Lambda^3}\Bigg(c_3+9c_4Z+30c_5Z^2\Bigg),\\
      {\cal G}_2:&\equiv& -\frac{2\dot{\bar{\phi}}^3_0}{\Lambda^3}\Bigg(c_3+6c_4Z+18c_5Z^2\Bigg),\\ 
       {\cal G}_3:&\equiv& -
       -\frac{2H\dot{\bar{\phi}}^3_0}{\Lambda^3}\Bigg(c_3+7c_4Z+18c_5Z^2\Bigg)-\frac{2\dot{\bar{\phi}}^3_0H\eta}{\Lambda^3}\Bigg(c_3+6c_4Z+18c_5Z^2\Bigg),\\
        {\cal G}_4:&\equiv& 
        \frac{\dot{\bar{\phi}}^3_0}{\Lambda^3}\bigg\{c_3+3c_4Z+6c_5\bigg[Z^2+\frac{\dot{H}\dot{\bar{\phi}}^2_0}{\Lambda^6}\bigg]\bigg\}-\frac{3\dot{\bar{\phi}}^4_0H\eta}{\Lambda^6}\bigg\{c_4+4c_5Z\bigg\},\eea
        
where factor $Z$ is the same as in Eq.(\ref{Z}). The exact details for the construction of such an action is performed by the authors in \cite{Burrage:2010cu}, where they have detailed the construction of the third-order action considering the soft breaking of Galilean symmetry.

Now, from our analysis about the slow-roll parameters $\epsilon$ and $\eta$ in the three, SRI, USR, and SRII regions, we know that parameter $\epsilon$ exhibits a smooth behavior when transitioning between the SRI to USR and USR to SRII phases at their respective conformal times $\tau=\tau_s$ and $\tau=\tau_e$. However, when we look at the behavior of the parameter $\eta$ around the exact transition times, we realise that it needs extra attention due to its value being changing abruptly in between the three phases. Consider using the following parametrization:
\bea \eta(\tau)=-6-\Delta\eta\left[\Theta(\tau-\tau_s)-\Theta(\tau-\tau_e)\right].\eea
The benefit of such a parameterization is found when differentiating with respect to the conformal time, where it gives us, 
$\eta^{'}(\tau)=-\Delta\eta\left[\delta(\tau-\tau_s)-\delta(\tau-\tau_e)\right],$
which is sharply peaked at the two consecutive transition points. However, terms like these are forbidden in the perturbative action obtained above for the curvature perturbations due to having softly broken Galilean shift symmetry. This particular term appears in refs.\cite{Kristiano:2022maq,Riotto:2023hoz,Choudhury:2023vuj,Choudhury:2023jlt,Kristiano:2023scm,Riotto:2023gpm,Choudhury:2023rks,Firouzjahi:2023aum,Motohashi:2023syh}, where large quantum fluctuation from the short range UV modes are present due to the absence of such a symmetry. So in our paper, we do not worry about the derivative of the parameter $\eta$ at the transition points. Nevertheless, the behavior of this parameter at the transition points is already made clear and further we will show in the rest of our analysis that such a behaviour of $\eta$ will be helpful to generate large amount of non-Gaussianity in the USR and SRII regions by violating Maldacena's {\it no-go theorem} in the squeezed limit of our calculated three-point functions.

We end the discussion by highlighting some important facts about the strength of the cubic interaction terms in the three regions. In the couplings ${\cal G}_3$ and ${\cal G}_4$, the second slow-roll parameter $\eta$ is present, which enhances their contributions in the USR region compared to the ones coming from the last two coupling terms, i.e., ${\cal G}_1$ and ${\cal G}_2$. Also, all of the interactions in the cubic action give suppressed contributions in the phases of SRI and SRII since $\eta$ vanishes in these regions. These observations will help us in the end while considering the final result from all three phases.

In the following subsection, we will explicitly calculate the tree-level three-point function and its associated bispectrum using the third-order action. We show the contributions due to all the operators in the three regions and later use them to derive the bispectrum in the three regions. Since we do not have a time derivative of parameter $\eta$ in any of the terms, the results which be more suppressed in regions SRI and SRII than during USR due to the presence of a finite and large $\eta$ in that region.

\subsection{Local non-Gaussianity from three point function and the associated Bispectrum computation}
\label{s5c}

From this point onward we  begin detailed analysis of the calculation of the tree-level three point cosmological correlation function which is also known as the Bispectrum. This function is relevant as the least order measure of the deviation from standard Gaussian statistics. From the third-order action in the curvature perturbations discussed using the Covariantized Galileon Theory in the previous section and working with the {\it in-in formalism} which is also discussed before the cubic action, we begin this section by introducing the general three-point function as:
 \bea \label{tpt} \langle\hat{\zeta}_{\bf k_{1}}\hat{\zeta}_{\bf k_{2}}\hat{\zeta}_{\bf k_{3}}\rangle:&&=\left\langle\bigg[\overline{T}\exp\bigg(i\int^{\tau}_{-\infty(1-i\epsilon)}d\tau^{'}\;H_{\rm int}(\tau^{'})\bigg)\bigg]\;\;\hat{\zeta}_{\bf k_{1}}(\tau)\hat{\zeta}_{\bf k_{2}}(\tau)\hat{\zeta}_{\bf k_{3}}(\tau)\right.\nonumber\\
&& \left.
\;\;\quad\quad\quad\quad\quad\quad\quad\quad\quad\quad\quad\quad\quad\quad\times\bigg[{T}\exp\bigg(-i\int^{\tau}_{-\infty(1+i\epsilon)}d\tau^{''}\;H_{\rm int}(\tau^{''})\bigg)\bigg]\right\rangle_{\tau\rightarrow 0}.\eea
where $\bar{T}$ and $T$ represent the anti-time and time ordering of the unitary operators which are made up from the time integral of the interacting Hamiltonian, which is described in this case as follows:
\bea && H_{\rm int}(\tau)=-\int d^3x\; \frac{a(\tau)^2}{H^3}\; \bigg[\frac{{\cal G}_1}{a}\zeta^{'3}+\frac{{\cal G}_2}{a^2}\zeta^{'2}\left(\partial^2\zeta\right)+\frac{{\cal G}_3}{a}\zeta^{'}\left(\partial_i\zeta\right)^2+\frac{{\cal G}_4}{a^2}\left(\partial_i\zeta\right)^2\left(\partial^2\zeta\right)\bigg].\eea
where the coefficients ${\cal G}_i, \forall i=1,2,3,4$ are the same as before in the perturbed cubic action.
Next, the contribution to the three point function coming from all the diagrams due to interactions present in the Hamiltonian is written as:
\bea &&\label{g}\langle\zeta_{\bold{k}_{1}}\zeta_{\bold{k}_{2}}\zeta_{\bold{k}_{3}}\rangle = \langle\zeta_{\bold{k}_{1}}\zeta_{\bold{k}_{2}}\zeta_{\bold{k}_{3}}\rangle_{\zeta'^{3}} + \langle\zeta_{\bold{k}_{1}}\zeta_{\bold{k}_{2}}\zeta_{\bold{k}_{3}}\rangle_{\zeta'^{2}(\partial^{2}\zeta)} + \langle\zeta_{\bold{k}_{1}}\zeta_{\bold{k}_{2}}\zeta_{\bold{k}_{3}}\rangle_{\zeta'(\partial_{i}\zeta)^{2}} + \langle\zeta_{\bold{k}_{1}}\zeta_{\bold{k}_{2}}\zeta_{\bold{k}_{3}}\rangle_{(\partial_{i}\zeta)^{2}(\partial^{2}\zeta)}.
\eea
Our main task would be explicitly calculating these contributions in all three phases, SRI, USR, and SRII respectively. To calculate them, we use the formula from the equation developed after perturbatively expanding the formula for the expectation of any operator up to the leading order in $H_{\text{Int}}(\tau)$. To keep track of formulas that we are going to use further, we mention the general expression from which the correlation functions are then derived for each interaction operator:
\bea \langle\hat{\zeta}_{\bf k_{1}}\hat{\zeta}_{\bf k_{2}}\hat{\zeta}_{\bold{k}_{3}}\rangle&=&\text{2 $\times$ Im}\left[\zeta_{\bold{k}_{1}}(\tau)\zeta_{\bold{k}_{2}}(\tau)\zeta_{\bold{k}_{3}}(\tau)\int^{\tau}_{-\infty}d\tau_{1}d^{3}x\frac{a(\tau_1)^2}{H^3}Q(x,\tau_{1})\right]\nonumber\\
&=&\text{2 $\times$ Im}\left[\zeta_{\bold{k}_{1}}(\tau)\zeta_{\bold{k}_{2}}(\tau)\zeta_{\bold{k}_{3}}(\tau)\int\frac{d^{3}\bold{q}_{1}}{(2\pi)^3}\frac{d^{3}\bold{q}_{2}}{(2\pi)^3}\frac{d^{3}\bold{q}_{3}}{(2\pi)^3}\int^{\tau}_{-\infty}d\tau_{1}\frac{a(\tau_{1})^2}{H^3}Q_{\bold{q}}(\tau_{1})\exp(-i(\bold{q}_{1}+\bold{q}_{2}+\bold{q}_{3}).\bold{x})\right].\quad\quad
\eea
where a Fourier transform:
\be Q(x, \tau_{1}) = \int^{\tau}_{-\infty}d\tau_{1}Q_{\bold{q}}(\tau_{1})\exp(-i(\bold{q}_{1}+\bold{q}_{2}+\bold{q}_{3}).\bold{x}),\ee is used and $Q_{q}( \tau_{1})$ represents the operators $\displaystyle{\frac{{\cal G}_1}{a}\zeta^{'3}}$,$\displaystyle{\frac{{\cal G}_2}{a^2}\zeta^{'2}\left(\bold{q}_{i}^{2}\zeta\right)}$,$\displaystyle{\frac{{\cal G}_3}{a}\zeta^{'}\left(\bold{q}_{i}.\bold{q}_{j}\zeta^2\right)}$, and $\displaystyle{\frac{{\cal G}_4}{a^2}((\bold{q}_{i}.\bold{q}_{j})\bold{q}_{k}^{2})\left(\zeta\right)^2\left(\partial^2\zeta\right)}$ where $a \equiv a(\tau_{1})$, and $\zeta \equiv \zeta_{\bold{q}}(\tau_{1})$ are the mode functions for the concerned regions. This expression will be used for each operator in all three regions. To perform further calculations, we would need information about the behavior of mode functions in the three regions, which has already been evaluated in the previous sections, and we can then contract operators who are outside the time integral with the interactions ones inside the integral as part of the wick contraction method. It will be explicitly shown in the following subsections, where we will evaluate the contribution from the interaction operators for each SRI, USR, and SRII region.

Also, while calculating in each said region, the integral over conformal time will be tackled by dividing it with respect to the region of concern:
\bea {\bf Conformal\;time\;integral:}\quad\quad\quad\quad\lim_{\tau\rightarrow 0}\int^{\tau}_{-\infty}:\equiv \underbrace{\Bigg(\int^{\tau_s}_{-\infty}\Bigg)}_{\bf SRI}+\underbrace{\Bigg(\int^{\tau_e}_{\tau_s}\Bigg)}_{\bf USR}+\underbrace{\Bigg(\int^{\tau_{\rm end}\rightarrow 0}_{\tau_e}\Bigg)}_{\bf SRII}.\eea

\subsubsection{Bispectrum and associated non-Gaussian amplitude for region I: SRI}
\label{s5c1}

We now begin the bispectrum calculation for each operator in the SRI region. This region is defined for the conformal time interval $-\infty < \tau < \tau_{s}$ where at $\tau_{s}$ a sharp transition occurs from SRI to USR region. Only first slow-roll parameter $\epsilon$ is finite (constant to be precise), and the second parameter $\eta$ is approximately zero for this region. From our analysis performed in this section we will show that contribution of the non-Gaussian amplitude obtained in this section for the SRI phase is going to be extremely small and in the squeezed limiting case it will be consistent with the Maldacena's {\it no-go theorem}.
Before we proceed, we mention that in \cite{Burrage:2010cu} the authors found non-zero results for the non-Gaussianity amplitude, $\fnl$, under the equilateral limit while for the squeezed limit they concluded that the value of $\fnl$ decays to zero. Our results in this section for SRI shows that under the limit $\tau_{s} \rightarrow 0$, in the absence of any USR and SRII phases, the value of $\fnl$ also tends to zero. Hence, our results agree with the findings of the authors in \cite{Burrage:2010cu}.  

We mention the detailed analysis of the contributions from all operators and their combined contribution to the tree-level scalar  three point correlation function in Appendix \ref{App:A}. The combined contribution as a result of the contributions coming from the  Eqs.(\ref{c1r1},\ref{c1r2},\ref{c1r3},\ref{c1r4}) is written as follows:
\bea \langle\hat{\zeta}_{\bold{k}_{1}}\hat{\zeta}_{\bold{k}_{2}}\hat{\zeta}_{\bold{k}_{3}}\rangle_{\text{SRI}} = (2\pi)^{3}\delta^{3}(\bold{k}_{1}+\bold{k}_{2}+\bold{k}_{3}){B}^{\text{SRI}}_{\zeta\zeta\zeta}(k_{1},k_{2},k_{3}).
\eea
where, the RHS includes the sum of the  individual contributions from each operators as:
\bea \label{c1BI} {B}^{\text{SRI}}_{\zeta\zeta\zeta}(k_{1},k_{2},k_{3}) = \sum_{i=1}^{4}{B}_{Q}^{\text{SRI}}(k_1,k_2,k_3).
\eea
where $Q$ represents the four interaction operators such that for each of them we will have the following explicit contributions:
\bea &&{B}^{\text{SRI}}_{\zeta^{'3}} = \frac{H^{12}}{(4{\cal A})^{3}}\frac{{\cal G}_{1}}{H^4}\frac{6}{(k_{1}^{3}k_{2}^{3}k_{3}^{3})}\frac{k_{1}^2 k_{2}^2 k_{3}^2}{c_{s}^{6} K^3}\left\{\left(\frac{K^{2}}{k_{s}^{2}}-2\right)\cos\left(\frac{K}{k_{s}}\right)-2\frac{K}{k_{s}}\sin\left(\frac{K}{k_{s}}\right)\right\}. \\&&{B}^{\text{SRI}}_{\zeta^{'2}(\partial^{2}\zeta)} = \frac{H^{12}}{(4{\cal A})^{3}}\frac{{\cal G}_{2}}{H^3}\frac{4}{(k_{1}^{3}k_{2}^{3}k_{3}^{3})}\displaystyle{\frac{k_{1}^2 k_{2}^2 k_{3}^2}{c_{s}^{8} K^3}\left\{\left(12-6\frac{K^{2}}{k_{s}^2}\right)\cos\left(\frac{K}{k_{s}}\right) - \left(\frac{K^{3}}{k_{s}^3}-12\frac{K}{k_{s}}\right)\sin\left(\frac{K}{k_{s}}\right)\right\}}. \eea\bea &&{B}^{\text{SRI}}_{\zeta^{'}(\partial_{i}\zeta)^{2}} = \frac{H^{12}}{(4{\cal A})^{3}}\frac{{\cal G}_{3}}{H^4}\frac{4}{k_{1}^{3}k_{2}^{3}k_{3}^{3}}\frac{k_{1}k_{2}k_{3}}{c_{s}^{8}K^{3}k_{s}^{2}}\Bigg\{\bigg(18 k_1 k_2 k_3 K+ 2 k_1^3\left(k_2+k_3\right) +2k_2^3\left(k_3+k_1\right) + 2k_3^3\left(k_1+k_2\right)\nonumber\\
   &&\quad\quad\quad\quad\quad\quad + 2\left(k_1^2 k_2^2+k_2^2 k_3^2+k_3^2 k_1^2\right)\bigg) k_s \sin
   \left(\frac{K}{k_s}\right)-3 k_1 k_2 k_3 K^{2} \cos
   \left(\frac{K}{k_s}\right)+\bigg(k_1^3 + k_2^3+k_3^3 \nonumber\\
   &&\quad\quad\quad\quad\quad\quad  + 5k_1^2\left(k_2+k_3\right) +5 k_2^2\left(k_3+k_1\right)+5k_3^2 \left(k_1+k_2\right) + 18 k_1 k_2 k_3\bigg) k_s^2 \cos
   \left(\frac{K}{k_s}\right)\Bigg\}.\\
   && {B}^{\text{SRI}}_{(\partial_{i}\zeta)^{2}(\partial^{2}\zeta)} = \frac{H^{12}}{(4{\cal A})^{3}}\frac{{\cal G}_{4}}{H^3}\frac{6}{k_{1}^{3}k_{2}^{3}k_{3}^{3}}
\frac{k_{1}k_{2}k_{3}}{c_{s}^{10}K^{3}}\left\{k_{1}k_{2}k_{3}\frac{K^{3}}{k_{s}^{3}}\left(\sin{\frac{K}{k_s}}\right)+\frac{K^{2}}{k_{s}^{2}}\bigg(k_{1}^{2}(k_{2}+k_{3})+k_{2}^{2}(k_{3}+k_{1}) +k_{3}^{2}(k_{2}+k_{1}) \right.\nonumber\\
&&\left. \quad\quad\quad\quad\quad + 6k_{1}k_{2}k_{3}\bigg)\left(\cos{\frac{K}{k_s}}\right) -\frac{1}{k_{s}}\bigg(k_{1}^{4}+ k_{2}^{4}+k_{3}^{4}+6k_{1}^{3}(k_{2}+k_{3})+6k_{2}^{3}(k_{1}+k_{3}) + 6k_{3}^{3}(k_{1}+k_{2}) + 28k_{1}k_{2}k_{3}K \right. \nonumber\\
&& \left. \quad\quad\quad\quad\quad  + 10k_{1}^{2}k_{2}^{2}+8k_{2}^{2}k_{3}^{2}+10k_{1}^{2}k_{3}^{2} \bigg)\left(\sin{\frac{K}{k_{s}}}\right)-2\bigg(k_{1}^{3}+k_{2}^{3}+k_{3}^{3}+4k_{1}^{2}(k_{2}+k_{3})+ 4k_{2}^{2}(k_{3}+k_{1}) +4k_{3}^{2}(k_{1}+k_{2}) \right.\nonumber\\
&& \left. \quad\quad\quad\quad\quad + 12k_{1}k_{2}k_{3}\bigg)\left(\cos{\frac{K}{k_{s}}}\right) \right\}. \eea

After mentioning the individual contributions towards the tree-level three-point function, coming from each interaction operator, we further evaluate the corresponding values of the non-Gaussian amplitude $\fnl$ using the following expression:
\bea \label{c1Bf} {B}^{\text{SRI}}(k_1,k_2,k_3) = \frac{6}{5}f^{\text{SRI}}_{\text{NL}}{\times}(2\pi^{2})^{2}\bigg[\frac{\Delta_{\bf {Tree}}^{\text{SRI}}(k_1)\Delta_{\bf {Tree}}^{\text{SRI}}(k_2)}{k_1^{3}k_2^{3}}+\frac{\Delta_{\bf {Tree}}^{\text{SRI}}(k_2)\Delta_{\bf {Tree}}^{\text{SRI}}(k_3)}{k_2^{3}k_3^{3}}+\frac{\Delta_{\bf {Tree}}^{\text{SRI}}(k_3)\Delta_{\bf {Tree}}^{\text{SRI}}(k_1)}{k_3^{3}k_1^{3}}\bigg].
\eea
The same factorization is also used in the USR and SRII regions to extract the information regarding the non-Gaussianity amplitude $\fnl$.
Also it is important to note that, the above expression is written using the dimensionless power spectrum in the SRI region.

\subsubsection{Bispectrum and associated non-Gaussian amplitude computation for region II: USR}
\label{s5c2}

In this section, we continue our analysis of the bispectrum for the scalar modes by working out its explicit expression in the USR region. This region is defined for the conformal time interval $\tau_{s} < \tau < \tau_{e}$, where we have sharp transitions between phases SRI and USR at $\tau_{s}$ and between phases USR to SRII at $\tau_{e}$. In the USR phase, the parameter $\epsilon$ is not a constant. However, it depends on the conformal time through the relation as defined earlier when discussing the modes for USR, and the parameter $\eta$ is also not a constant but takes the value $\eta \sim -6$. These facts will have implications on the behavior of the strength of the bispectrum and the way the Bogoliubov coefficients depend on conformal time, which is visible in the mode expansion in the USR phase from Eq.(\ref{bG2a}).

We mention the detailed analysis of all the different contributions coming from the individual operators and their combined results in Appendix \ref{App:B}. Here we mention the results for the tree-level scalar three-point correlation function in the USR region as the combined contribution coming from the individual operators using the specific functions and their details in the appendix.  

The tree-level contribution to the three point function due to all the operators in the USR region can be written as follows: 
\bea \langle\hat{\zeta}_{\bold{k}_{1}}\hat{\zeta}_{\bold{k}_{2}}\hat{\zeta}_{\bold{k}_{3}}\rangle_{\text{USR}} = (2\pi)^{3}\delta^{3}(\bold{k}_{1}+\bold{k}_{2}+\bold{k}_{3}){B}^{\text{USR}}_{\zeta\zeta\zeta}(k_{1},k_{2},k_{3}).
\eea
where the RHS consists of the sum of the individual contributions towards the tree-level bispectrum value:
\bea \label{c2BI} {B}^{\text{USR}}_{\zeta\zeta\zeta}(k_{1},k_{2},k_{3}) = \sum_{i=1}^{4}{B}_{Q}^{\text{USR}}(k_1,k_2,k_3).
\eea
here $Q$ represents the $4$ interaction operators and the explicit contributions from all the operators individually are written as follows.\\
From the first operator, the total tree-level contribution to the three-point correlation function is given, using the expressions in Eq.(\ref{c1a},\ref{c211f},\ref{c211aI},\ref{c22f},\ref{c212aI},\ref{c23f},\ref{c213aI},\ref{c24f},\ref{c214aI}), in the following manner:
\bea \label{c2r1}{B}^{\text{USR}}_{\zeta^{'3}} &=& \frac{H^{12}}{(4{\cal A})^{3}}\frac{{\cal G}_{1}}{H^4}\frac{-2}{(k_{1}^{3}k_{2}^{3}k_{3}^{3})}\left[\zeta_{\bold{k_{1}}}(\tau_{s})\zeta_{\bold{k_{2}}}(\tau_{s})\zeta_{\bold{k_{3}}}(\tau_{s})\right]\Bigg\{\bigg((\alpha_{k_{1}}^{(2)*}\alpha_{k_{2}}^{(2)*}\alpha_{k_{3}}^{(2)*})(\bold{I}_{1})_{1}({\cal K}_{1},k_1,k_2,k_3) - (\alpha_{k_{1}}^{(2)*}\beta_{k_{2}}^{(2)*}\alpha_{k_{3}}^{(2)*}) \nonumber\\ 
&&\times (\bold{I}_{1})_{2}({\cal K}_{2},k_1,-k_2,k_3) -(\alpha_{k_{1}}^{(2)*}\alpha_{k_{2}}^{(2)*}\beta_{k_{3}}^{(2)*})(\bold{I}_{1})_{3}({\cal K}_{1},k_1,k_2,-k_3) - (\beta_{k_{1}}^{(2)*}\alpha_{k_{2}}^{(2)*}\alpha_{k_{3}}^{(2)*})(\bold{I}_{1})_{4}({\cal K}_{4},-k_1,k_2,k_3) - \text{c.c} \bigg)\nonumber\\
&&+\bigg( (\alpha_{k_{1}}^{(2)*}\alpha_{k_{2}}^{(2)*}\alpha_{k_{3}}^{(2)*})\bigg(k_3^{2}(\bold{I}_{2})_{1}({\cal K}_{1},-k_1,-k_2)  + k_1^{2}(\bold{I}_{2})_{2}({\cal K}_{1},-k_2,-k_3) + k_2^{2}(\bold{I}_{2})_{3}({\cal K}_{1},-k_1,-k_3) \bigg)\nonumber\\
&& - (\alpha_{k_{1}}^{(2)*}\beta_{k_{2}}^{(2)*}\alpha_{k_{3}}^{(2)*})\bigg( k_3^{2}(\bold{I}_{2})_{4}({\cal K}_{3},k_1,-k_2) + k_1^{2}(\bold{I}_{2})_{5}({\cal K}_{3},k_2,-k_3)
+ k_2^{2}(\bold{I}_{2})_{6}({\cal K}_{3},k_3,-k_1) \bigg)  \nonumber\\
&& -(\alpha_{k_{1}}^{(2)*}\alpha_{k_{2}}^{(2)*}\beta_{k_{3}}^{(2)*})\bigg(k_3^{2}(\bold{I}_{2})_{7}({\cal K}_{2},-k_1,k_2) + k_1^{2}(\bold{I}_{2})_{8}({\cal K}_{2},-k_2,k_3) + k_2^{2}(\bold{I}_{2})_{9}({\cal K}_{2},-k_3,k_1)\bigg)\nonumber\\ 
&& - (\beta_{k_{1}}^{(2)*}\alpha_{k_{2}}^{(2)*}\alpha_{k_{3}}^{(2)*})\bigg(k_3^{2}(\bold{I}_{2})_{10}({\cal K}_{4},-k_1,-k_2) +  k_1^{2}(\bold{I}_{2})_{11}({\cal K}_{4},-k_2,-k_3) + k_2^{2}(\bold{I}_{2})_{12}({\cal K}_{4},-k_3,-k_1)\bigg) - \text{c.c} \bigg)\nonumber\\
&&+\bigg( (\alpha_{k_{1}}^{(2)*}\alpha_{k_{2}}^{(2)*}\alpha_{k_{3}}^{(2)*})\bigg(k_2^{2}k_3^{2}(\bold{I}_{3})_{1}({\cal K}_{1},-k_1) + k_3^{2}k_1^{2}(\bold{I}_{3})_{2}({\cal K}_{1},-k_2) +
k_1^{2}k_2^{2}(\bold{I}_{3})_{3}({\cal K}_{1},-k_3)\bigg) -
(\alpha_{k_{1}}^{(2)*}\beta_{k_{2}}^{(2)*}\alpha_{k_{3}}^{(2)*}) \nonumber\\
&&\bigg(k_2^{2}k_3^{2}(\bold{I}_{3})_{4}({\cal K}_{2},-k_1)+k_3^{2}k_1^{2}(\bold{I}_{3})_{5}({\cal K}_{2},k_2) + k_1^{2}k_2^{2}(\bold{I}_{3})_{6}({\cal K}_{2},-k_3)\bigg) -(\alpha_{k_{1}}^{(2)*}\alpha_{k_{2}}^{(2)*}\beta_{k_{3}}^{(2)*})\bigg(k_2^{2}k_3^{2}(\bold{I}_{3})_{7}({\cal K}_{3},-k_1)\nonumber\\
&&+k_1^{2}k_3^{2}(\bold{I}_{3})_{8}({\cal K}_{3},-k_2) + k_1^{2}k_2^{2}(\bold{I}_{3})_{9}({\cal K}_{3},k_3)\bigg) - (\beta_{k_{1}}^{(2)*}\alpha_{k_{2}}^{(2)*}\alpha_{k_{3}}^{(2)*})\bigg(k_3^{2}k_2^{2}(\bold{I}_{3})_{10}({\cal K}_{4},k_1) +k_1^{2}k_3^{2}(\bold{I}_{3})_{11}({\cal K}_{4},-k_2) \nonumber\\
&& +k_1^{2}k_2^{2}(\bold{I}_{3})_{12}({\cal K}_{4},-k_3)\bigg) - \text{c.c} \bigg) +k^2_1k^2_2k^2_3\bigg((\alpha_{k_{1}}^{(2)*}\alpha_{k_{2}}^{(2)*}\alpha_{k_{3}}^{(2)*})(\bold{I}_{4})_{1}({\cal K}_{1}) - (\alpha_{k_{1}}^{(2)*}\beta_{k_{2}}^{(2)*}\alpha_{k_{3}}^{(2)*})(\bold{I}_{4})_{2}({\cal K}_{2}) - \nonumber\\
&&(\alpha_{k_{1}}^{(2)*}\alpha_{k_{2}}^{(2)*}\beta_{k_{3}}^{(2)*})(\bold{I}_{4})_{3}({\cal K}_{3}) - (\beta_{k_{1}}^{(2)*}\alpha_{k_{2}}^{(2)*}\alpha_{k_{3}}^{(2)*})(\bold{I}_{4})_{4}({\cal K}_{4}) - \text{c.c}\bigg) + \text{2 Perms.}\Bigg\}.
\eea
From the second operator, the total contribution to the three-point, tree-level, correlation function is given, using the expressions in Eq.(\ref{c1b},\ref{c221f},\ref{c221aI},\ref{c222f},\ref{c222aI},\ref{c223f},\ref{c223aI}), in the following manner:
\bea \label{c2r2} {B}^{\text{USR}}_{\zeta^{'2}\partial^{2}\zeta} &=& \frac{H^{12}}{(4{\cal A})^{3}}\frac{{\cal G}_{2}}{H^3}\frac{4}{(k_{1}^{3}k_{2}^{3}k_{3}^{3})}\left[\zeta_{\bold{k_{1}}}(\tau_{s})\zeta_{\bold{k_{2}}}(\tau_{s})\zeta_{\bold{k_{3}}}(\tau_{s})\right]\Bigg\{\bigg( (\alpha_{k_{1}}^{(2)*}\alpha_{k_{2}}^{(2)*}\alpha_{k_{3}}^{(2)*})(\bold{I}_{1})_{1}({\cal K}_{1},k_1,k_2,k_3) - (\alpha_{k_{1}}^{(2)*}\beta_{k_{2}}^{(2)*}\alpha_{k_{3}}^{(2)*}) \nonumber\\ 
&&\times (\bold{I}_{1})_{2}({\cal K}_{2},k_1,-k_2,k_3)-(\alpha_{k_{1}}^{(2)*}\alpha_{k_{2}}^{(2)*}\beta_{k_{3}}^{(2)*})(\bold{I}_{1})_{3}({\cal K}_{3},k_1,k_2,-k_3) - (\beta_{k_{1}}^{(2)*}\alpha_{k_{2}}^{(2)*}\alpha_{k_{3}}^{(2)*})(\bold{I}_{1})_{4}({\cal K}_{4},-k_1,k_2,k_3)- \text{c.c} \bigg)\nonumber\\
&&+\bigg((\alpha_{k_{1}}^{(2)*}\alpha_{k_{2}}^{(2)*}\alpha_{k_{3}}^{(2)*}) \bigg(k^{2}_2(\bold{I}_{2})_{1}({\cal K}_{1},-k_1,-k_3) + k^{2}_1(\bold{I}_{2})_{2}({\cal K}_{1},-k_2,-k_3)\bigg) -
(\alpha_{k_{1}}^{(2)*}\beta_{k_{2}}^{(2)*}\alpha_{k_{3}}^{(2)*}) \bigg(k^{2}_2(\bold{I}_{2})_{3}({\cal K}_{2},-k_1,-k_3)\nonumber\\
&&+ k^{2}_1(\bold{I}_{2})_{4}({\cal K}_{2},k_2,-k_3) \bigg) -(\alpha_{k_{1}}^{(2)*}\alpha_{k_{2}}^{(2)*}\beta_{k_{3}}^{(2)*})\bigg(k^{2}_2(\bold{I}_{2})_{5}({\cal K}_{3},-k_1,k_3) + k^{2}_1(\bold{I}_{2})_{6}({\cal K}_{3},-k_2,k_3)\bigg) -(\beta_{k_{1}}^{(2)*}\alpha_{k_{2}}^{(2)*}\alpha_{k_{3}}^{(2)*})\nonumber\\
&&\bigg(k_2^{2}(\bold{I}_{2})_{7}({\cal K}_{4},k_1,-k_3) +k_{1}^{2}(\bold{I}_{2})_{8}({\cal K}_{4},-k_2,-k_3)\bigg) - \text{c.c} \bigg) + (k_1^{2}k_2^{2})\bigg((\alpha_{k_{1}}^{(2)*}\alpha_{k_{2}}^{(2)*}\alpha_{k_{3}}^{(2)*})(\bold{I}_{3})_{1}({\cal K}_{1},-k_3)\nonumber\\
&& - (\alpha_{k_{1}}^{(2)*}\beta_{k_{2}}^{(2)*}\alpha_{k_{3}}^{(2)*})(\bold{I}_{3})_{2}({\cal K}_{2},-k_3) - (\alpha_{k_{1}}^{(2)*}\alpha_{k_{2}}^{(2)*}\beta_{k_{3}}^{(2)*})(\bold{I}_{3})_{3}({\cal K}_{3},k_3) - (\beta_{k_{1}}^{(2)*}\alpha_{k_{2}}^{(2)*}\alpha_{k_{3}}^{(2)*})(\bold{I}_{3})_{4}({\cal K}_{4},-k_3)\nonumber\\
&& - \text{c.c}\bigg) + \text{2 Perms.} \Bigg\}
\eea
From the third operator, the total contribution to the three-point, tree-level, correlation function is given, using the expressions in Eq.(\ref{c1c},\ref{c221f},\ref{c221aI},\ref{c222f},\ref{c222aI}), in the following manner:
\bea \label{c2r3} {B}^{\text{USR}}_{\zeta^{'}(\partial_{i}\zeta)^{2}} &=& \frac{H^{12}}{(4{\cal A})^{3}}\frac{{\cal G}_{3}}{H^4}\frac{-4}{(k_{1}^{3}k_{2}^{3}k_{3}^{3})}\left[\zeta_{\bold{k_{1}}}(\tau_{s})\zeta_{\bold{k_{2}}}(\tau_{s})\zeta_{\bold{k_{3}}}(\tau_{s})\right]\Bigg\{\frac{-k_1 k_2}{3}\bigg( (\alpha_{k_{1}}^{(2)*}\alpha_{k_{2}}^{(2)*}\alpha_{k_{3}}^{(2)*})(\bold{I}_{1})_{1}({\cal K}_{1},k_1,k_2,k_3) - (\alpha_{k_{1}}^{(2)*}\beta_{k_{2}}^{(2)*}\alpha_{k_{3}}^{(2)*}) \nonumber\eea\bea
&&\times (\bold{I}_{1})_{2}({\cal K}_{2},k_1,-k_2,k_3)-(\alpha_{k_{1}}^{(2)*}\alpha_{k_{2}}^{(2)*}\beta_{k_{3}}^{(2)*})(\bold{I}_{1})_{3}({\cal K}_{3},k_3,k_2,-k_3) - (\beta_{k_{1}}^{(2)*}\alpha_{k_{2}}^{(2)*}\alpha_{k_{3}}^{(2)*})(\bold{I}_{1})_{4}({\cal K}_{4},-k_1,k_2,k_3) \nonumber\\
&&- \text{c.c}\bigg)-\frac{k_1 k_2 k_{3}^{2}}{3}\bigg((\alpha_{k_{1}}^{(2)*}\alpha_{k_{2}}^{(2)*}\alpha_{k_{3}}^{(2)*})  (\bold{I}_{2})_{1}({\cal K}_{1},-k_1,-k_2) -
(\alpha_{k_{1}}^{(2)*}\beta_{k_{2}}^{(2)*}\alpha_{k_{3}}^{(2)*})(\bold{I}_{2})_{2}({\cal K}_{2},-k_1,k_2)\nonumber\\
&& - (\alpha_{k_{1}}^{(2)*}\alpha_{k_{2}}^{(2)*}\beta_{k_{3}}^{(2)*})(\bold{I}_{2})_{3}({\cal K}_{3},-k_1,-k_2) - (\beta_{k_{1}}^{(2)*}\alpha_{k_{2}}^{(2)*}\alpha_{k_{3}}^{(2)*})(\bold{I}_{2})_{4}({\cal K}_{4},k_1,-k_2) - \text{c.c} \bigg) + \text{2 Perms.}\Bigg\}
\eea
From the fourth operator, the total contribution to the three-point, tree-level, correlation function is given, using the expressions in Eq.(\ref{c1d},\ref{c241f},\ref{c224aI}), in the following manner:
\bea \label{c2r4} {B}^{\text{USR}}_{\partial^{2}\zeta(\partial_{i}\zeta)^{2}} &=& \frac{H^{12}}{(2{\cal A})^{3}}\frac{{\cal G}_{4}}{H^3}\frac{2}{(k_{1}^{3}k_{2}^{3}k_{3}^{3})}\left[\zeta_{\bold{k_{1}}}(\tau_{s})\zeta_{\bold{k_{2}}}(\tau_{s})\zeta_{\bold{k_{3}}}(\tau_{s})\right]\Bigg\{ k_{1}k_{2}k_{3}^{2}\bigg( (\alpha_{k_{1}}^{(2)*}\alpha_{k_{2}}^{(2)*}\alpha_{k_{3}}^{(2)*})(\bold{I}_{1})_{1}({\cal K}_{1},k_1,k_2,k_3) - (\alpha_{k_{1}}^{(2)*}\beta_{k_{2}}^{(2)*}\alpha_{k_{3}}^{(2)*})\nonumber\\ 
&&\times (\bold{I}_{1})_{2}({\cal K}_{2},k_1,-k_2,k_3) -(\alpha_{k_{1}}^{(2)*}\alpha_{k_{2}}^{(2)*}\beta_{k_{3}}^{(2)*})(\bold{I}_{1})_{3}({\cal K}_{3},k_1,k_2,-k_3) - (\beta_{k_{1}}^{(2)*}\alpha_{k_{2}}^{(2)*}\alpha_{k_{3}}^{(2)*})(\bold{I}_{1})_{4}({\cal K}_{4},-k_1,k_2,k_3)\nonumber\\
&& - \text{c.c} \bigg) + \text{2 Perms} \Bigg\}
\eea
In the above expressions $(\text{c.c})$ indicates the complex conjugate of all the previous terms which takes into account the contributions coming from the negative exponential integrals. There are also other $2$ permutations in momentum variables which are taken care of when we present the numerical results.
From using the expression for the tree-level bispectrum in the USR region we can further evaluate $\fnl$ through the use of the relation:
\bea \label{c2Bf} {B}^{\text{USR}}(k_1,k_2,k_3) = \frac{6}{5}f^{\text{USR}}_{\text{NL}}{\times}(2\pi^{2})^{2}\bigg[\frac{\Delta_{\bf {Tree}}^{\text{USR}}(k_1)\Delta_{\bf {Tree}}^{\text{USR}}(k_2)}{k_1^{3}k_2^{3}}+\frac{\Delta_{\bf {Tree}}^{\text{USR}}(k_2)\Delta_{\bf {Tree}}^{\text{USR}}(k_3)}{k_2^{3}k_3^{3}}+\frac{\Delta_{\bf {Tree}}^{\text{USR}}(k_3)\Delta_{\bf {Tree}}^{\text{USR}}(k_1)}{k_3^{3}k_1^{3}}\bigg].
\eea
where in the USR region we have used the dimensionless power spectrum. 
The explicit calculation regarding the squeezed limit in the above equations for the bispectrum in the USR region and the related non-Gaussianity amplitude $\fnl$ is rather cumbersome to write here and would not be be very illuminating in itself. Hence, we present the results for the case of the squeezed limit by performing a numerical analysis for the amplitude $\fnl$ with respect to the wave numbers in the USR region while also considering different effective sound speed values.

\subsubsection{Bispectrum and associated non-Gaussian amplitude computation for region III: SRII} \label{s5c3}

In this section, we continue our analysis for the bispectrum by working out it's explicit expression in the final SRII region. This region is defined by the conformal time interval $\tau_{e}< \tau <\tau_{end}$, where $\tau_{e}$ marks the transition from USR to SRII region while $\tau_{end}$ is the end of the SRII phase, which will eventually be taken to zero in the late time limit.

We mention the detailed analysis of all the different contributions coming from the individual operators and their combined results in Appendix \ref{App:C}. In this section, we mention the results for the tree-level scalar three-point correlation function in the SRII region as the combined contribution coming from the individual operators and their specific results in the appendix.

The tree-level contribution to the three-point function due to all the operators in the SRII region can be written as follows:
\bea \langle\hat{\zeta}_{\bold{k}_{1}}\hat{\zeta}_{\bold{k}_{2}}\hat{\zeta}_{\bold{k}_{3}}\rangle_{\text{SRII}} = (2\pi)^{3}\delta^{3}(\bold{k}_{1}+\bold{k}_{2}+\bold{k}_{3}){B}^{\text{SRII}}_{\zeta\zeta\zeta}(k_{1},k_{2},k_{3}).
\eea
where the RHS consists of the sum of the individual contributions towards the tree-level bispectrum value:
\bea \label{c3BI} {B}^{\text{SRII}}_{\zeta\zeta\zeta}(k_{1},k_{2},k_{3}) = \sum_{i=1}^{4}{B}_{Q}^{\text{SRII}}(k_1,k_2,k_3).
\eea
here $Q$ represents the $4$ interaction operators and the explicit contributions from all the operators individually are
written in the following way. From the first operator, the contribution to the tree-level three-point correlation function is given, using the expressions in Eq.(\ref{c1a},\ref{c31aI}), in the following manner:
\bea \label{c3r1} B_{\zeta^{'3}}^{\text{SRII}} &=& \frac{H^{12}}{(4{\cal A})^{3}}\frac{{\cal G}_{1}}{H^4}\frac{-2}{(k_{1}^{3}k_{2}^{3}k_{3}^{3})}\left[\zeta_{\bold{k_{1}}}(\tau_{e})\zeta_{\bold{k_{2}}}(\tau_{e})\zeta_{\bold{k_{3}}}(\tau_{e})\right]\frac{k_{e}^{18}}{k_{s}^{18}}(k_1^{2}k_2^{2}k_3^{2})\Bigg\{(\alpha_{k_{1}}^{(2)*}\alpha_{k_{2}}^{(2)*}\alpha_{k_{3}}^{(2)*})(\bold{J}_{1})_{1}({\cal K}_{1})-(\alpha_{k_{1}}^{(2)*}\beta_{k_{2}}^{(2)*}\alpha_{k_{3}}^{(2)*})\nonumber\\
&& \times(\bold{J}_{1})_{2}({\cal K}_{2})-(\alpha_{k_{1}}^{(2)*}\alpha_{k_{2}}^{(2)*}\beta_{k_{3}}^{(2)*})(\bold{J}_{1})_{3}({\cal K}_{3})-(\beta_{k_{1}}^{(2)*}\alpha_{k_{2}}^{(2)*}\alpha_{k_{3}}^{(2)*})(\bold{J}_{1})_{4}({\cal K}_{4}) - \text{c.c} + \text{2 Perms.}\Bigg\}
\eea
For the second operator, the contribution is given by using the expressions in Eq.(\ref{c1b},\ref{c32aI}) in the following manner:
\bea  \label{c3r2} B_{\zeta^{'2}\partial^{2}\zeta}^{\text{SRII}} &=& \frac{H^{12}}{(4{\cal A})^{3}}\frac{{\cal G}_{2}}{H^3}\frac{4}{(k_{1}^{3}k_{2}^{3}k_{3}^{3})}\left[\zeta_{\bold{k_{1}}}(\tau_{e})\zeta_{\bold{k_{2}}}(\tau_{e})\zeta_{\bold{k_{3}}}(\tau_{e})\right]\frac{k_{e}^{18}}{k_{s}^{18}}\Bigg\{(\alpha_{k_{1}}^{(2)*}\alpha_{k_{2}}^{(2)*}\alpha_{k_{3}}^{(2)*})k_{1}^{2}k_{2}^{2}(\bold{J}_{2})_{1}({\cal K}_{1},-k_{3})-(\alpha_{k_{1}}^{(2)*}\beta_{k_{2}}^{(2)*}\alpha_{k_{3}}^{(2)*})\nonumber\\
&& \times k_{1}^{2}k_{2}^{2}(\bold{J}_{2})_{2}({\cal K}_{2},-k_3)-(\alpha_{k_{1}}^{(2)*}\alpha_{k_{2}}^{(2)*}\beta_{k_{3}}^{(2)*})k_{1}^{2}k_{2}^{2}(\bold{J}_{2})_{3}({\cal K}_{3},k_3)-(\beta_{k_{1}}^{(2)*}\alpha_{k_{2}}^{(2)*}\alpha_{k_{3}}^{(2)*})k_{1}^{2}k_{2}^{2}(\bold{J}_{2})_{4}({\cal K}_{4},-k_3)\nonumber\\
&& - \text{c.c} +  \text{2 Perms.} \Bigg\}
\eea
For the third operator, the contribution is given by using the expressions in Eq.(\ref{c1c},\ref{c33aI}) in the following manner:
\bea  \label{c3r3} B_{\zeta^{'}(\partial_{i}\zeta)^{2}}^{\text{SRII}} &=& \frac{H^{12}}{(4{\cal A})^{3}}\frac{{\cal G}_{3}}{H^4}\frac{-4}{(k_{1}^{3}k_{2}^{3}k_{3}^{3})}\left[\zeta_{\bold{k_{1}}}(\tau_{e})\zeta_{\bold{k_{2}}}(\tau_{e})\zeta_{\bold{k_{3}}}(\tau_{e})\right]\frac{k_{e}^{18}}{k_{s}^{18}}\Bigg\{(\alpha_{k_{1}}^{(2)*}\alpha_{k_{2}}^{(2)*}\alpha_{k_{3}}^{(2)*})k_{3}^{2}(\bold{J}_{3})_{1}({\cal K}_{1},-k_{1},-k_{2})-(\alpha_{k_{1}}^{(2)*}\beta_{k_{2}}^{(2)*}\alpha_{k_{3}}^{(2)*})\nonumber\\
&& \times k_{3}^{2}(\bold{J}_{3})_{2}({\cal K}_{2},-k_1,k_2)-(\alpha_{k_{1}}^{(2)*}\alpha_{k_{2}}^{(2)*}\beta_{k_{3}}^{(2)*})k_{3}^{2}(\bold{J}_{3})_{3}({\cal K}_{3},-k_1,-k_2)-(\beta_{k_{1}}^{(2)*}\alpha_{k_{2}}^{(2)*}\alpha_{k_{3}}^{(2)*})k_{3}^{2}(\bold{J}_{3})_{4}({\cal K}_{4},k_1,-k_2)\nonumber\\
&&  - \text{c.c} +  \text{2 Perms.} \Bigg\}
\eea
For the fourth operator, the contribution is given by using the expression in Eq.(\ref{c1d},\ref{c34aI}) in the following manner:
\bea \label{c3r4} B_{(\partial^{2}\zeta)(\partial_{i}\zeta)^{2}}^{\text{SRII}} &=& \frac{H^{12}}{(4{\cal A})^{3}}\frac{{\cal G}_{4}}{H^3}\frac{2}{(k_{1}^{3}k_{2}^{3}k_{3}^{3})}\left[\zeta_{\bold{k_{1}}}(\tau_{e})\zeta_{\bold{k_{2}}}(\tau_{e})\zeta_{\bold{k_{3}}}(\tau_{e})\right]\frac{k_{e}^{18}}{k_{s}^{18}}\Bigg\{(\alpha_{k_{1}}^{(2)*}\alpha_{k_{2}}^{(2)*}\alpha_{k_{3}}^{(2)*})(\bold{J}_{4})_{1}({\cal K}_{1},k_{1},k_{2},k_{3})-(\alpha_{k_{1}}^{(2)*}\beta_{k_{2}}^{(2)*}\alpha_{k_{3}}^{(2)*})\nonumber\\
&& \times (\bold{J}_{4})_{2}({\cal K}_{2},k_1,-k_2,k_3) -(\alpha_{k_{1}}^{(2)*}\alpha_{k_{2}}^{(2)*}\beta_{k_{3}}^{(2)*})(\bold{J}_{4})_{3}({\cal K}_{3},k_1,k_2,-k_3)-(\beta_{k_{1}}^{(2)*}\alpha_{k_{2}}^{(2)*}\alpha_{k_{3}}^{(2)*})(\bold{J}_{4})_{4}({\cal K}_{4},-k_1,k_2,k_3)\nonumber\\
&& - \text{c.c} +  \text{2 Perms.} \Bigg\}
\eea
In the above expressions (c.c) indicates the contributions from all the negative exponential integrals and the terms obtained from the other $2$ permutations in the momentum variables are also taken care of in these results. From using the expression for the tree-level bispectrum in the SRII region we can further evaluate $\fnl$ through the use of the relation:
\bea \label{c3Bf} {B}^{\text{SRII}}(k_1,k_2,k_3) = \frac{6}{5}f^{\text{SRII}}_{\text{NL}}{\times}(2\pi^{2})^{2}\bigg[\frac{\Delta_{\bf {Tree}}^{\text{SRII}}(k_1)\Delta_{\bf {Tree}}^{\text{SRII}}(k_2)}{k_1^{3}k_2^{3}}+\frac{\Delta_{\bf {Tree}}^{\text{SRII}}(k_2)\Delta_{\bf {Tree}}^{\text{SRII}}(k_3)}{k_2^{3}k_3^{3}}+\frac{\Delta_{\bf {Tree}}^{\text{SRII}}(k_3)\Delta_{\bf {Tree}}^{\text{SRII}}(k_1)}{k_3^{3}k_1^{3}}\bigg].
\eea
where in the SRII region we have used the dimensionless power spectrum. 
The explicit calculations for the squeezed limit in the above equations for the bispectrum and the related non-Gaussianity amplitude $\fnl$ in the SRII region is dealt in a similar way, as done for the USR region, by performing a numerical analysis for the amplitude $\fnl$ with respect to the wave numbers in SRII region, for different values of the effective sound speed. 

\subsection{Total bispectrum  and associated non-Gaussian amplitude }
\label{s5d}
\textcolor{black}{In this section, we present the combined version from the individual bispectrum contributions and the associated non-Gaussian amplitudes from the three-point function of the scalar modes computed for all three regions, SRI, USR, and SRII. To perform this, we would have to be extremely careful about the behavior of the amplitude at the transition points, $\tau_s$ for SRI to USR, and $\tau_e$ for USR to SRII transitions. The sharp transition features between the three regions are taken care of at their respective wavenumbers in the following manner: }
\textcolor{black}{\bea \label{c4B} 
{B}^{\text{Total}}_{\zeta\zeta\zeta}(k_{1},k_{2},k_{3}) = {B}^{\text{SRI}}_{\zeta\zeta\zeta}(k_{1},k_{2},k_{3}) + \Theta(k-k_{s}){B}^{\text{USR}}_{\zeta\zeta\zeta}(k_{1},k_{2},k_{3}) + \Theta(k-k_{e}){B}^{\text{SRII}}_{\zeta\zeta\zeta}(k_{1},k_{2},k_{3}) \eea }
\textcolor{black}{This quantity represents the total bispectrum, and here we refer to the Eqs.(\ref{c1BI},\ref{c2BI},\ref{c3BI}) while adding the individual bispectrum results. It is this quantity that will be ultimately useful to give us the total behaviour of the non-Gaussian amplitude across all wavenumbers. This expression also involves using Heaviside Theta functions to join the individual contributions carefully at their respective wavenumbers, as done similarly in the case of the total tree-level scalar power spectrum in Eqn.(\ref{totpower}).}

\textcolor{black}{For the case of the squeezed limit where one of the momenta is much shorter (long wavelength) than the others, i.e., $k_{1} \rightarrow 0$ and $k_{2} \approx k_{3} = k$, we can express the total contribution to the tree-level bispectrum as the following:
\bea \label{c4Bsq}
B^{\text{Total}}_{\zeta\zeta\zeta}(k_{1} \rightarrow 0, k, k) &=& B^{\text{SRI}}_{\zeta\zeta\zeta}(k_{1} \rightarrow 0, k, k | k \leq k_{s}) + \Theta(k-k_{s})B^{\text{USR}}_{\zeta\zeta\zeta}(k_{1} \rightarrow 0, k, k | k_{s} \leq k \leq k_{e})\nonumber\\
&& + \Theta(k-k_{e})B^{\text{SRII}}_{\zeta\zeta\zeta}(k_{1} \rightarrow 0, k, k | k_{e} \leq k \leq k_{\text{end}}).
\eea }
\textcolor{black}{Through the use of this we now mention the expression for the total non-Gaussianity amplitude $\fnl$ in the squeezed limit as:}
\textcolor{black}{\bea \label{c4fsq}
f^{\text{Total}}_{\text{NL}}(k) = f^{\text{SRI}}_{\text{NL}}(k \leq k_{s}) + \Theta(k-k_{s})f^{\text{USR}}_{\text{NL}}(k_{s} \leq k \leq k_{e}) + \Theta(k-k_{e})f^{\text{SRII}}_{\text{NL}}(k_{e} \leq k \leq k_{\text{end}}).
\eea }
\textcolor{black}{The results related to this quantity will be examined in the next section, starting with the individual contributions before presenting the final behaviour of the total non-Gaussian amplitude across all wavenumbers joined carefully using Heaviside Theta functions. }
    \begin{figure*}[htb!]
    	\centering
    \subfigure[]{
      	\includegraphics[width=8.5cm,height=7cm] {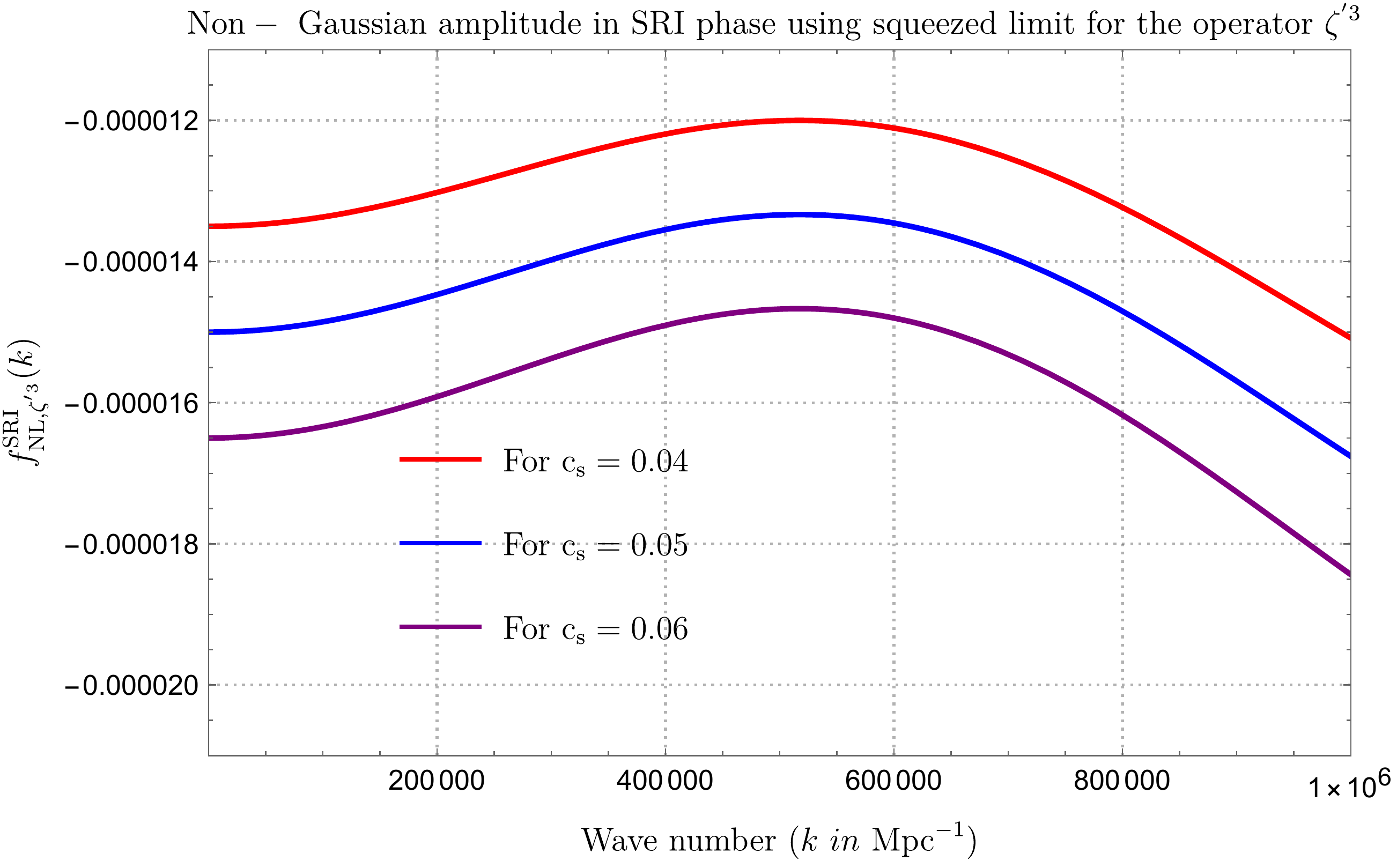}
        \label{I1}
    }
    \subfigure[]{
       \includegraphics[width=8.5cm,height=7cm] {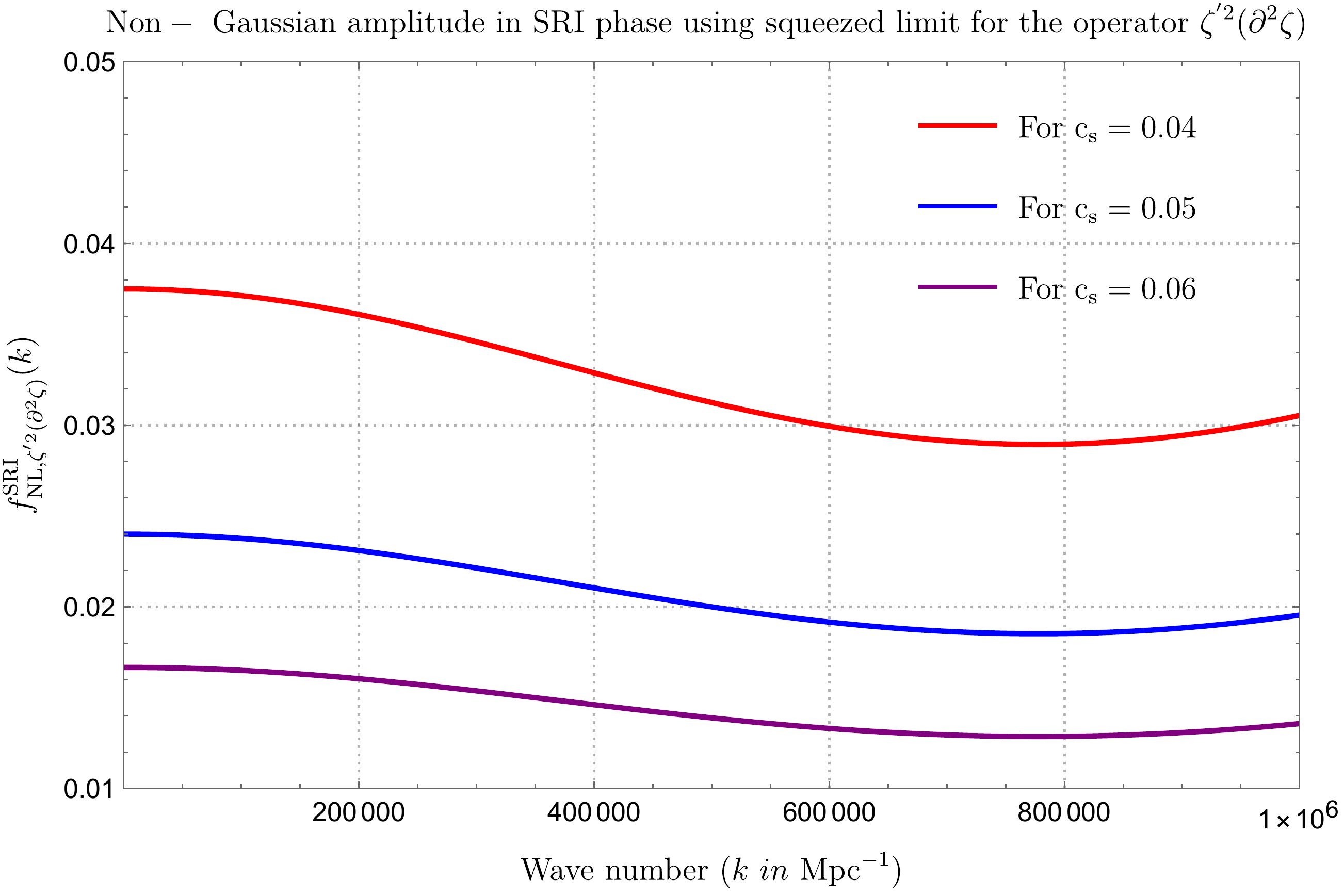}
        \label{I2}
       }
        \subfigure[]{
       \includegraphics[width=8.5cm,height=7cm] {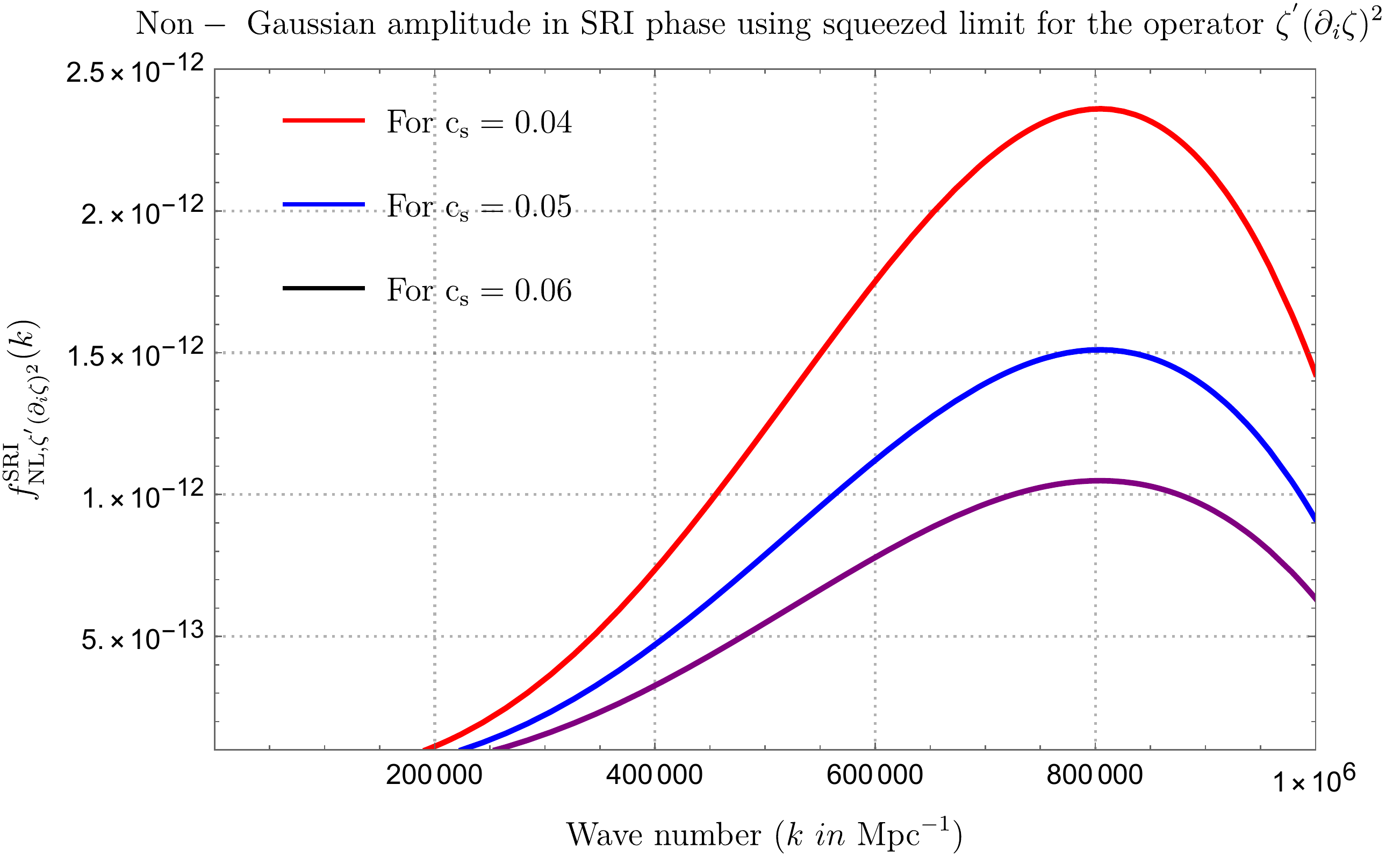}
        \label{I3}
       }
        \subfigure[]{
       \includegraphics[width=8.5cm,height=7cm] {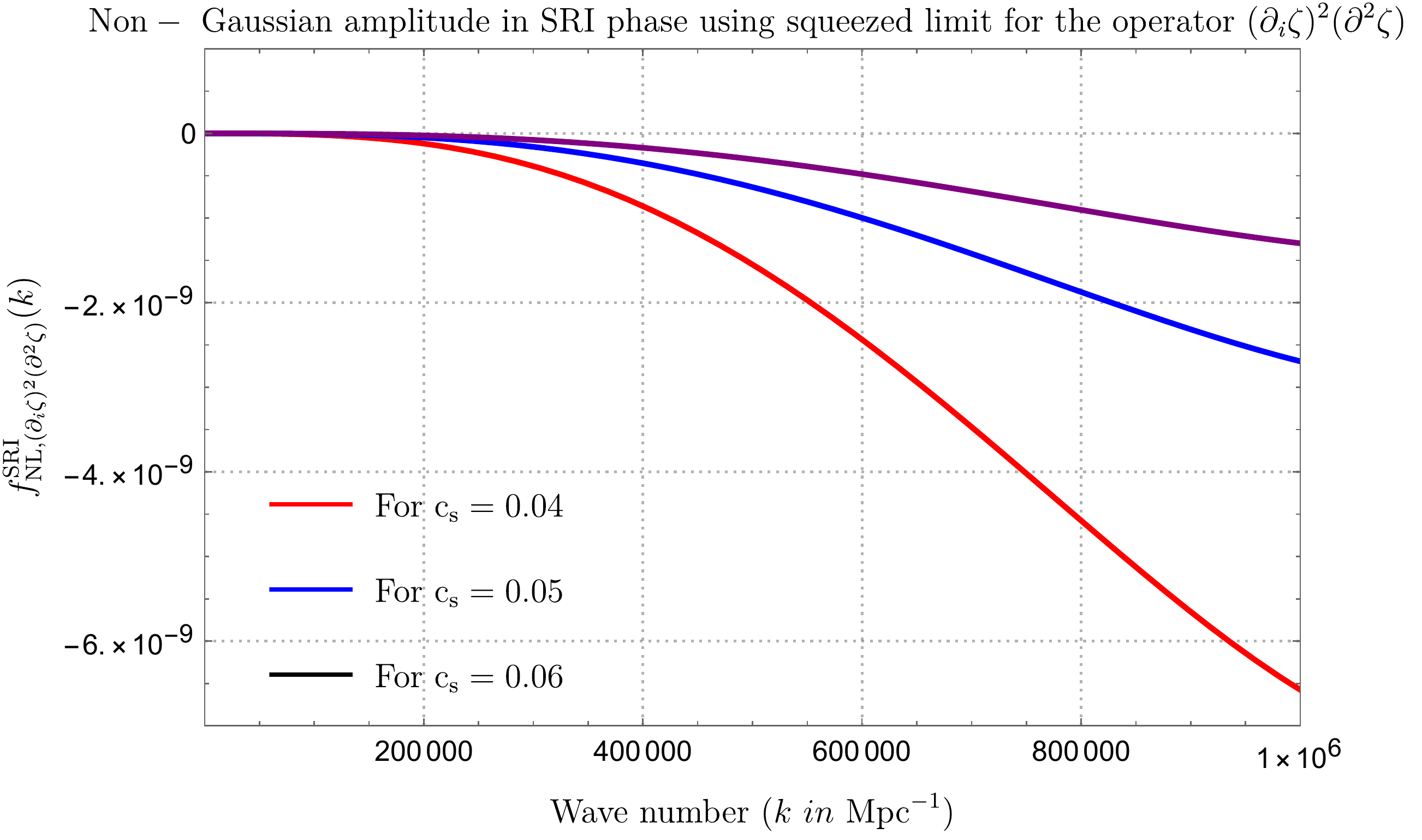}
        \label{I4}
       }
    	\caption[Optional caption for list of figures]{The plot represents the contributions of individual operators to the non-Gaussian amplitude $f_{\text{NL}}$ in the SRI region as a function of the wave number. The values of the effective sound speed parameter $c_{s}$ are chose to be as follows $c_{s} = 0.04,0.05,0.06$. The plots in the top row represent contributions from the first and second operator. The plots in the bottom row represent contributions from the third and fourth operator.} 
    	\label{fNLSRI:a}
    \end{figure*}
    \begin{figure*}[htb!]
    	\centering
{
      	\includegraphics[width=18cm,height=12.5cm] {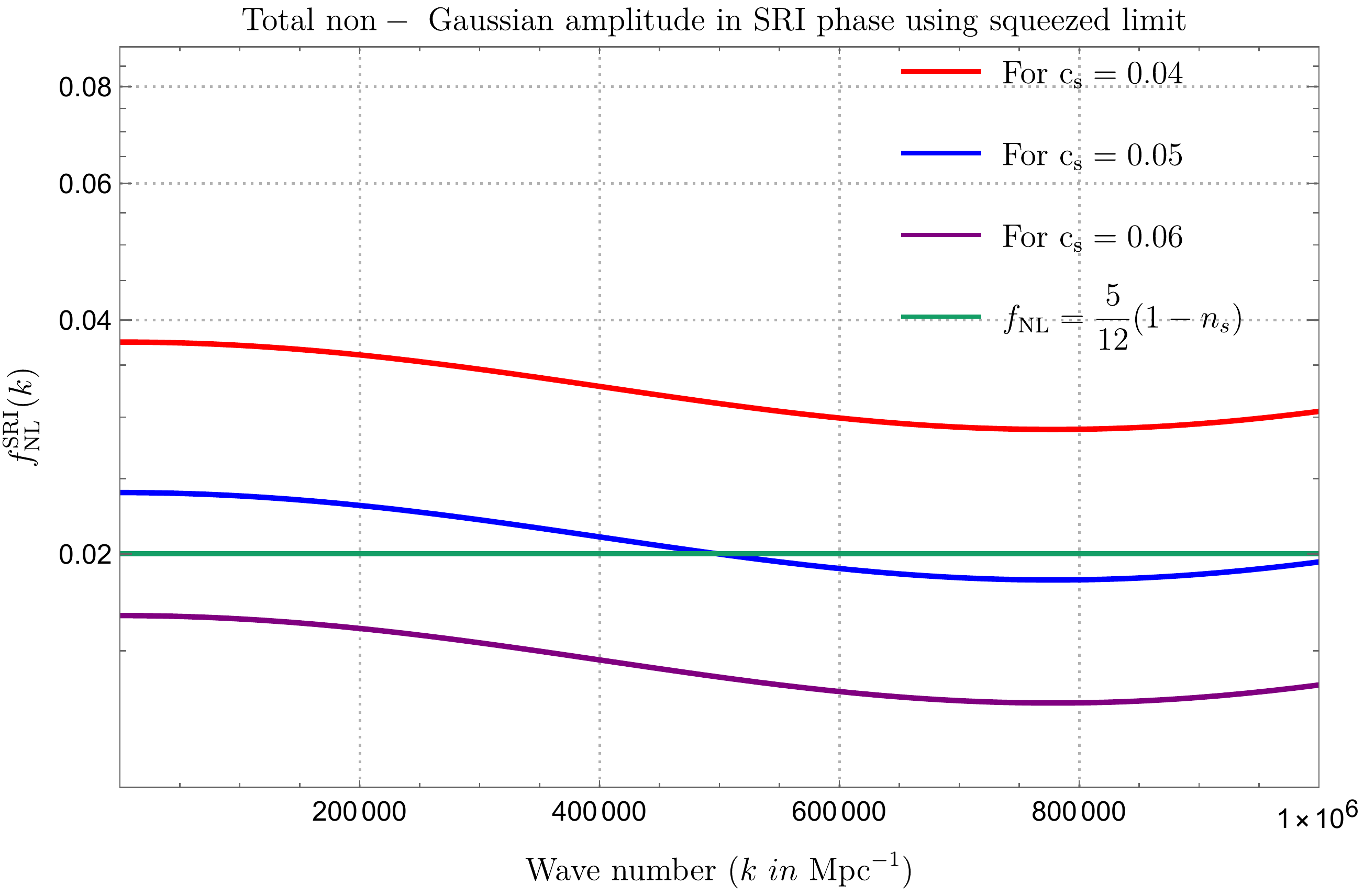}
        \label{K1}
    }
    	\caption[Optional caption for list of figures]{Plot represents the total contribution of all operators to the non-Gaussian amplitude $f_{\text{NL}}$ as a function of the wave number in the SRI phase under the squeezed limit. The values of the effective sound speed parameter $c_{s}$ are chose to be as follows $c_{s} = 0.04,0.05,0.06$. Here we find that $c_{s} = 0.05$ gives the best agreement with the value predicted from the consistency condition.} 
    	\label{fNLSRI:b}
    \end{figure*}
\section{Numerical Results}
\label{s6}
In this section, we present our results obtained from the analysis of the non-Gaussian amplitude $\fnl$ in the squeezed limit. We will begin by mentioning the plots showing the behavior of the parameter $\fnl$ with respect to the wave number ($k$), for different values of the effective sound speed ($c_{s}$), coming from all the four interaction operators individually and finally their combined contributions separately for all the three regions, SRI, USR and SRII. This gives us essential insights into the way the coefficients for the interaction operators change between different regions, since they are time-dependent, to validate the consistency condition as per Maldacena's {\it no-go theorem} in the first slow-roll region, and to also control the behaviour of the non-Gaussianities produced in the next two USR and SRII regions in the squeezed limit. Here we will find out that the consistency condition is clearly violated in both the USR and SRII regions and we will see the production of large non-Gaussianity in both the regions, with amplitude in SRII being much less than in USR but still greater when compared to SRI.

    \begin{figure*}[htb!]
    	\centering
    \subfigure[]{
      	\includegraphics[width=8.5cm,height=7.5cm] {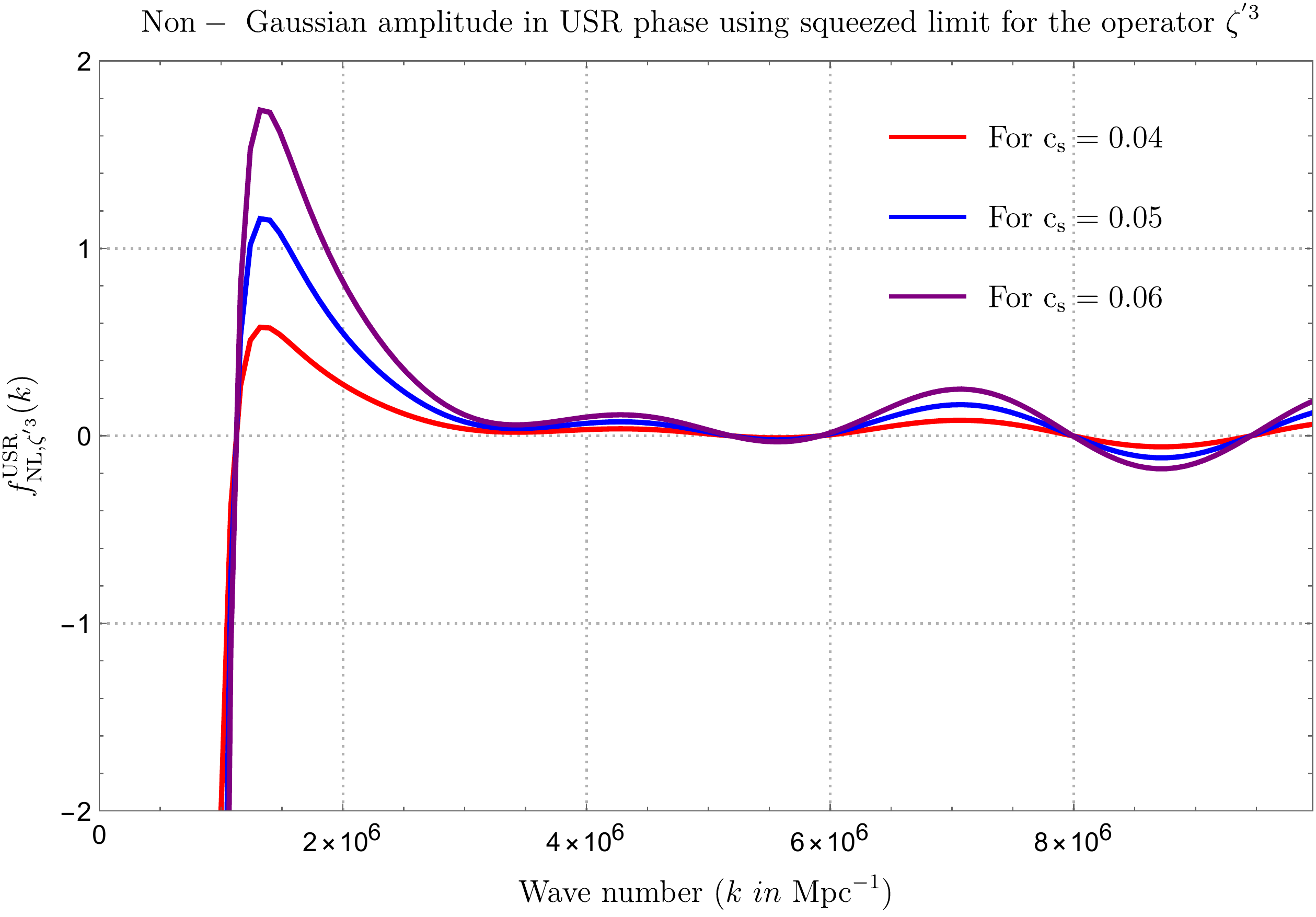}
        \label{IU1}
    }
    \subfigure[]{
       \includegraphics[width=8.5cm,height=7.5cm] {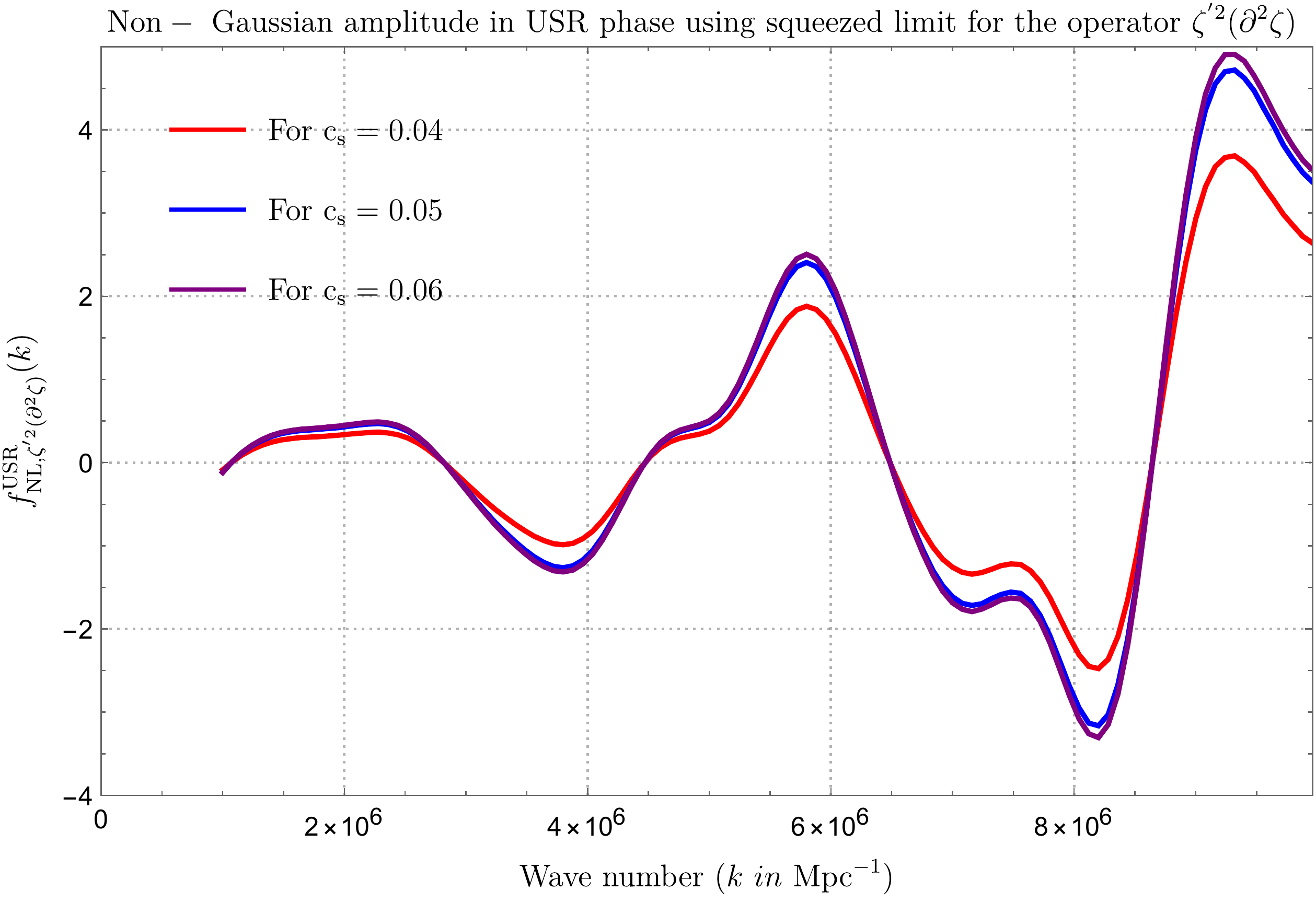}
        \label{IU2}
       }
        \subfigure[]{
       \includegraphics[width=8.5cm,height=7.5cm] {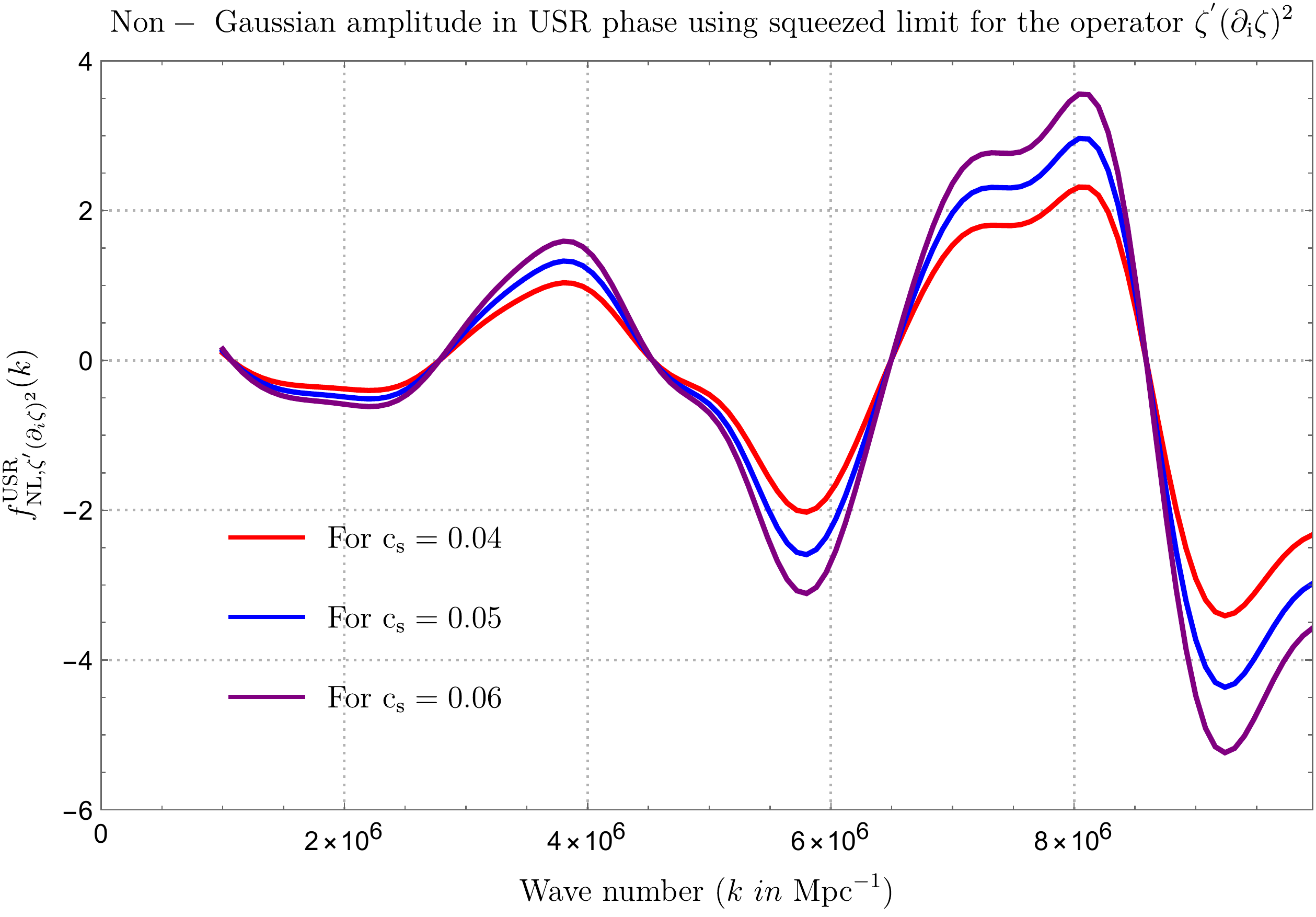}
        \label{IU3}
       }
        \subfigure[]{
       \includegraphics[width=8.5cm,height=7.5cm] {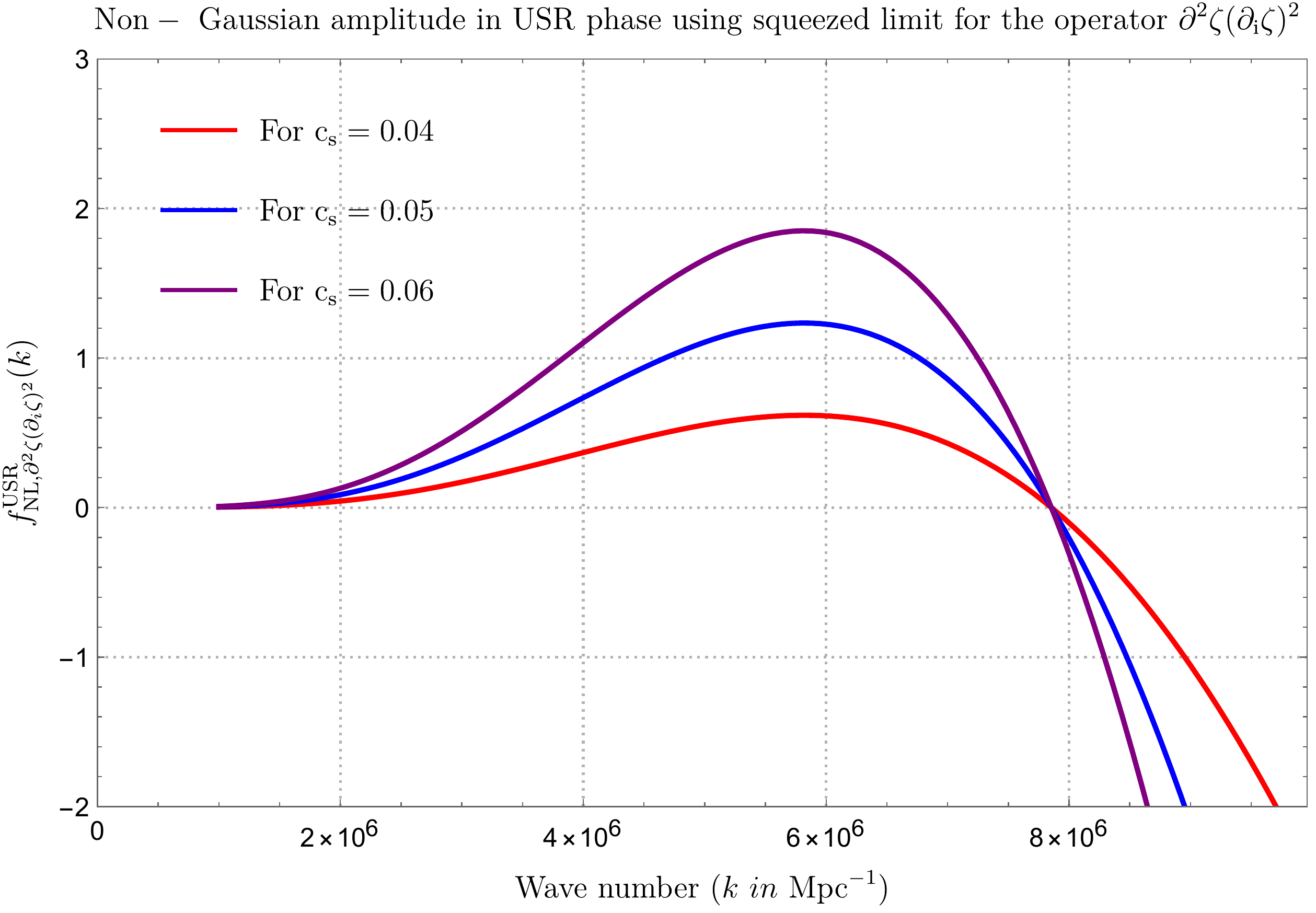}
        \label{IU4}
       }
    	\caption[Optional caption for list of figures]{The plot represents the contributions of individual operators to the non-Gaussian amplitude $f_{\text{NL}}$ in the USR region as a function of the wave number. The values of the effective sound speed parameter $c_{s}$ are chose to be as follows $c_{s} = 0.04,0.05,0.06$. The plots in the top row represent contributions from the first and second operator. The plots in the bottom row represent contributions from the third and fourth operator.} 
    	\label{fNLUSR:a}
    \end{figure*}

    \begin{figure*}[htb!]
    	\centering
   {
      	\includegraphics[width=18cm,height=12.5cm] {
      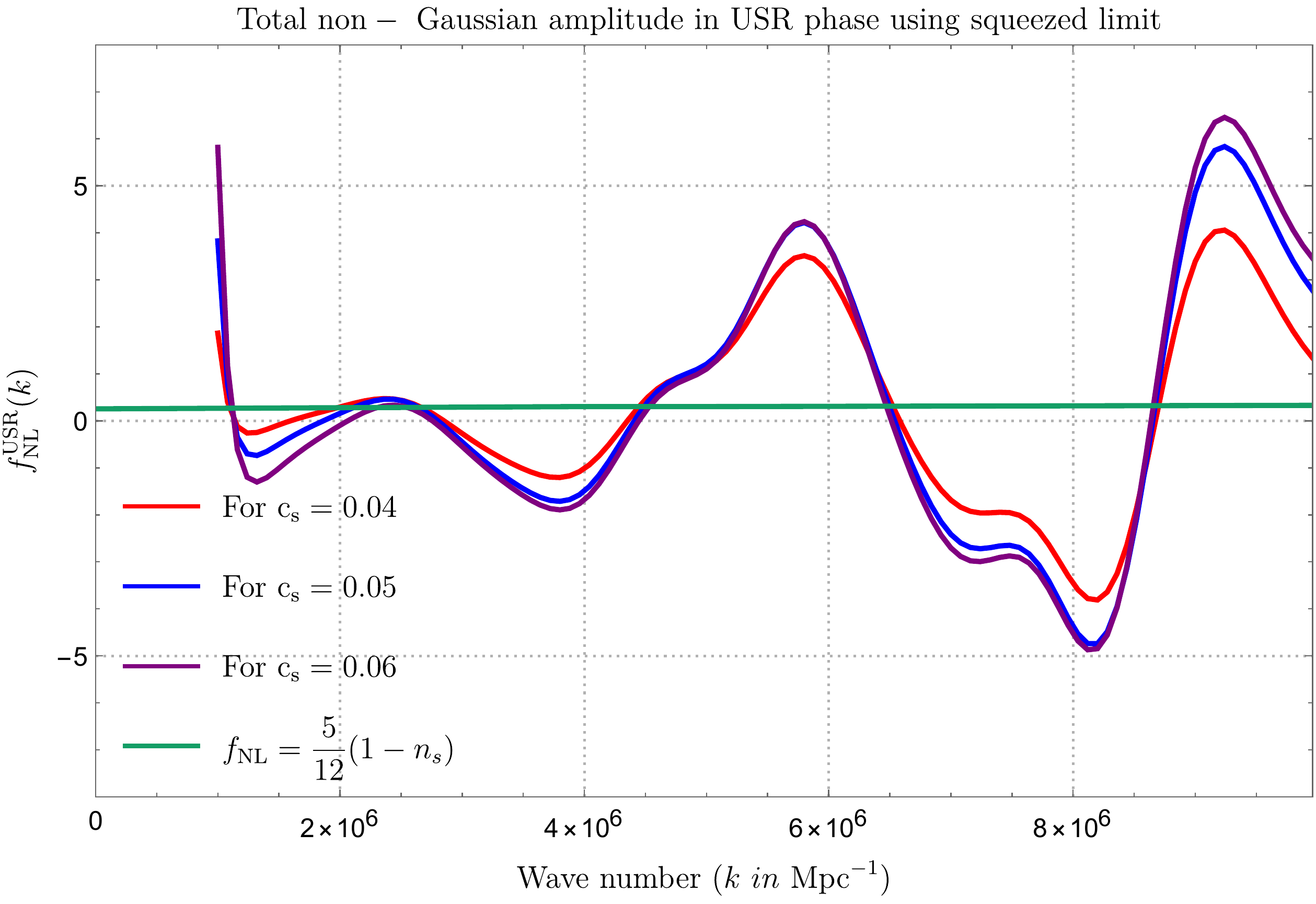}
        \label{KU1}
    }
    	\caption[Optional caption for list of figures]{Plot represents the total contribution of all operators to the non-Gaussian amplitude $f_{\text{NL}}$ as a function of the wave number in the USR phase under the squeezed limit. The values of the effective sound speed parameter $c_{s}$ are chosen to be as follows $c_{s} = 0.04,0.05,0.06$. Here we find that the value predicted from the consistency condition is being violated} 
    	\label{fNLUSR:b}
    \end{figure*}

\subsection{Results obtained from region I: SRI}
\label{s6a}
In this subsection, we present our findings through representative plots regarding the variation of the non-Gaussian amplitude $\fnl$ with respect to the wave number ($k$) in the SRI region. We first show the individual contributions of each operator towards this behavior and then give the combined behavior of all the operators together.
 
From the figure in Fig.(\ref{fNLSRI:a}), we can see that the contribution from the first operator $\zeta^{'3}$ towards $\fnl$ in plot (\ref{I1}) is much smaller than the contribution from the second operator which gives the highest contribution overall in the SRI, plot (\ref{I2}). The operators third $\zeta^{'}(\partial_{i}\zeta)^{2}$ and fourth $(\partial_{i}\zeta)^{2}\partial^{2}\zeta$ gives the least significant contribution in plots (\ref{I3},\ref{I4}). Results from the plot (\ref{I2}) are very close to what the value of $\fnl$ is obtained from the Maldacena's consistency relation in the squeezed limit, i.e., $\fnl \sim {\cal O}(10^{-2})$, and especially for $c_{s}=0.05$ the result is in exact agreement with the consistency relation. All the individual contributions tend to move towards zero as $c_{s}$ is increased except for those coming from the first operator. This specific behavior is the result of the  analytic structure of the individual terms and their dependence on $c_{s}$, where throughout the analysis the causality and unitarity constraints are perfectly maintained.
    
The combined contributions from all the operators are given in Fig.(\ref{fNLSRI:b}) where the $\fnl$ amplitude is plotted against the wave number while considering the same set of different values for the sound speed to check for the variation in results. According to the analysis for the plots in Fig.(\ref{fNLSRI:a}), the total contribution in Fig.(\ref{fNLSRI:b}) looks similar to the contributions from the second operator, with the inclusion of contributions from the first operator evident in the final results. The fact that the second operator contributes so heavily, in addition to the slight but visible contribution of the first operator and the least significant effects of the other two operators, is encoded in the coefficients ${\cal G}_{i}$, $\forall i=1,2,3,4$. In the end, from the overall results in Fig.(\ref{fNLSRI:b}), it is clear that for $c_{s}=0.05$, the consistency relation as a result of Maldacena's \textit{no-go theorem} is strictly valid, and even for small changes around this value does not lead to any violation of the said consistency relation as expected for the case of the SRI region. 

    \begin{figure*}[htb!]
    	\centering
    \subfigure[]{
      	\includegraphics[width=8.5cm,height=7.5cm] {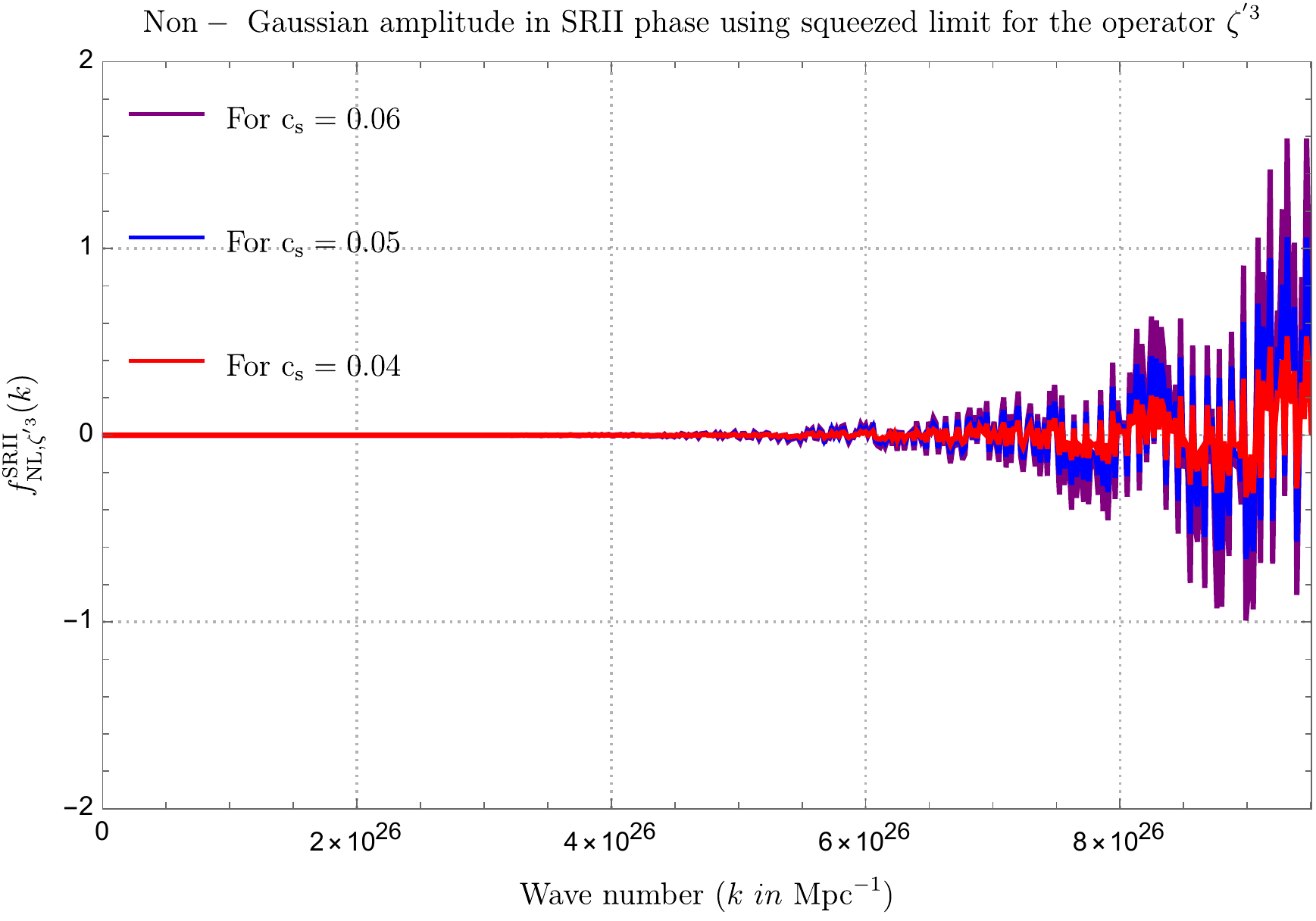}
        \label{Is21}
    }
    \subfigure[]{
       \includegraphics[width=8.5cm,height=7.5cm] {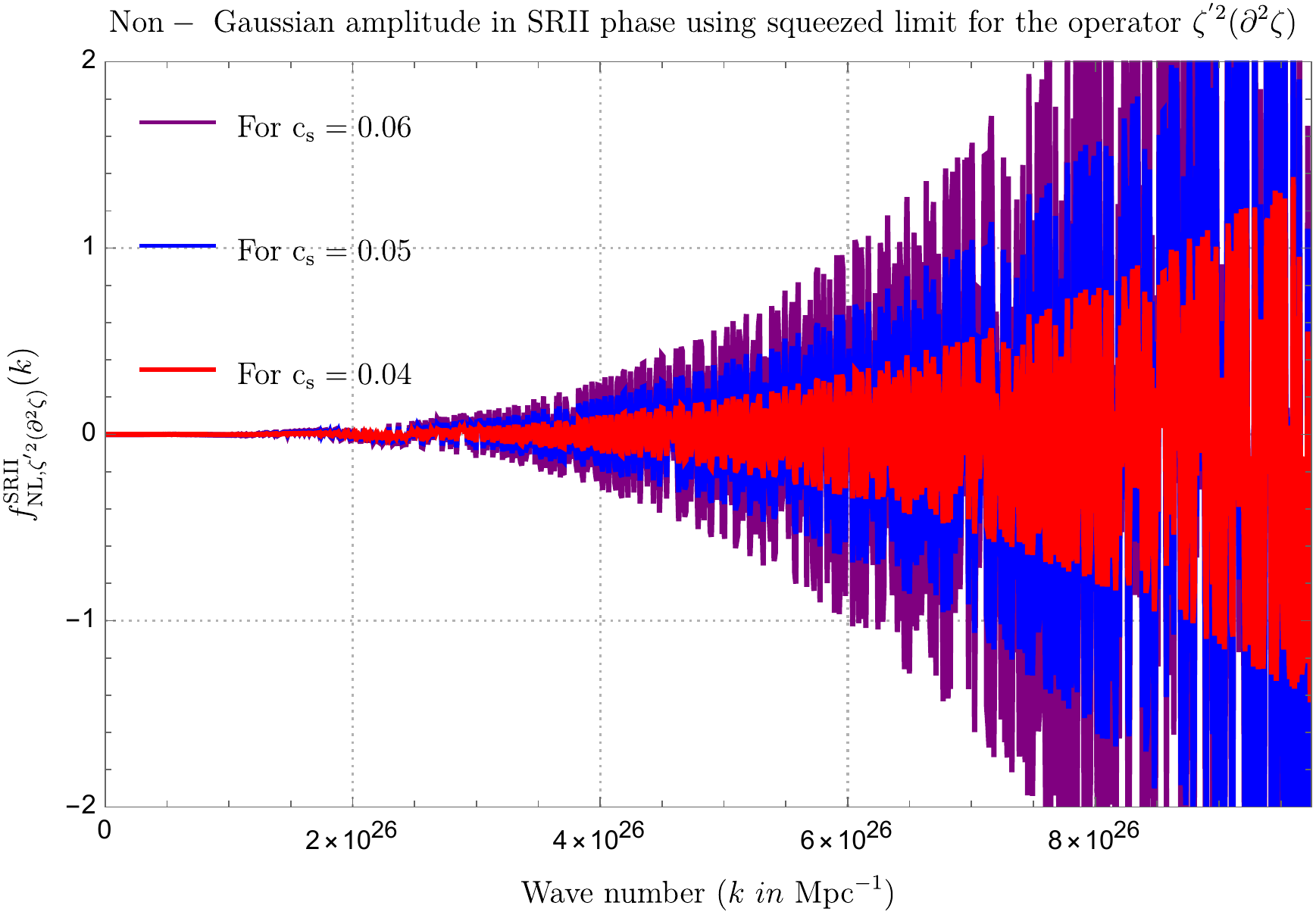}
        \label{Is22}
       }
        \subfigure[]{
       \includegraphics[width=8.5cm,height=7.5cm] {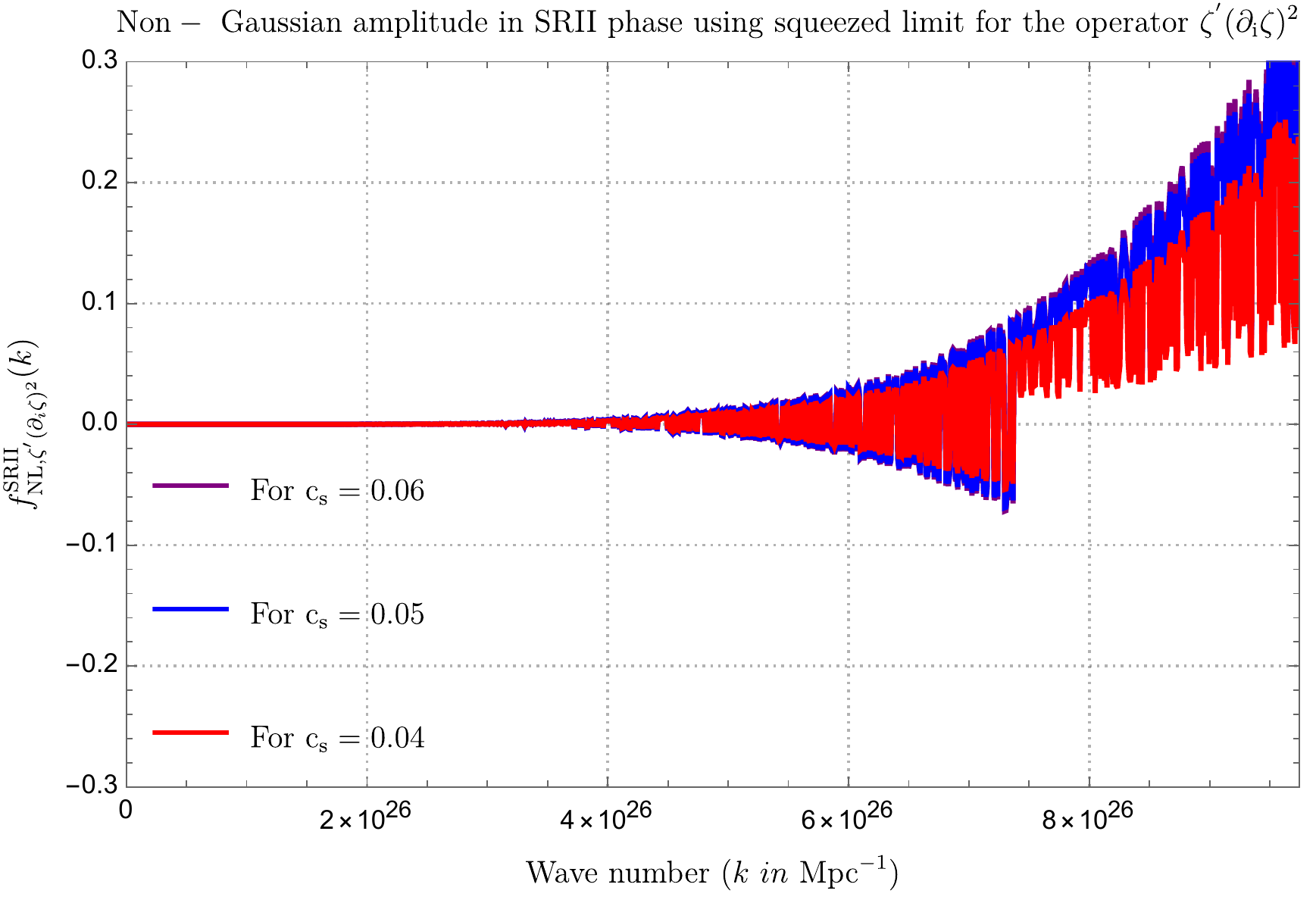}
        \label{Is23}
       }
        \subfigure[]{
       \includegraphics[width=8.5cm,height=7.5cm] {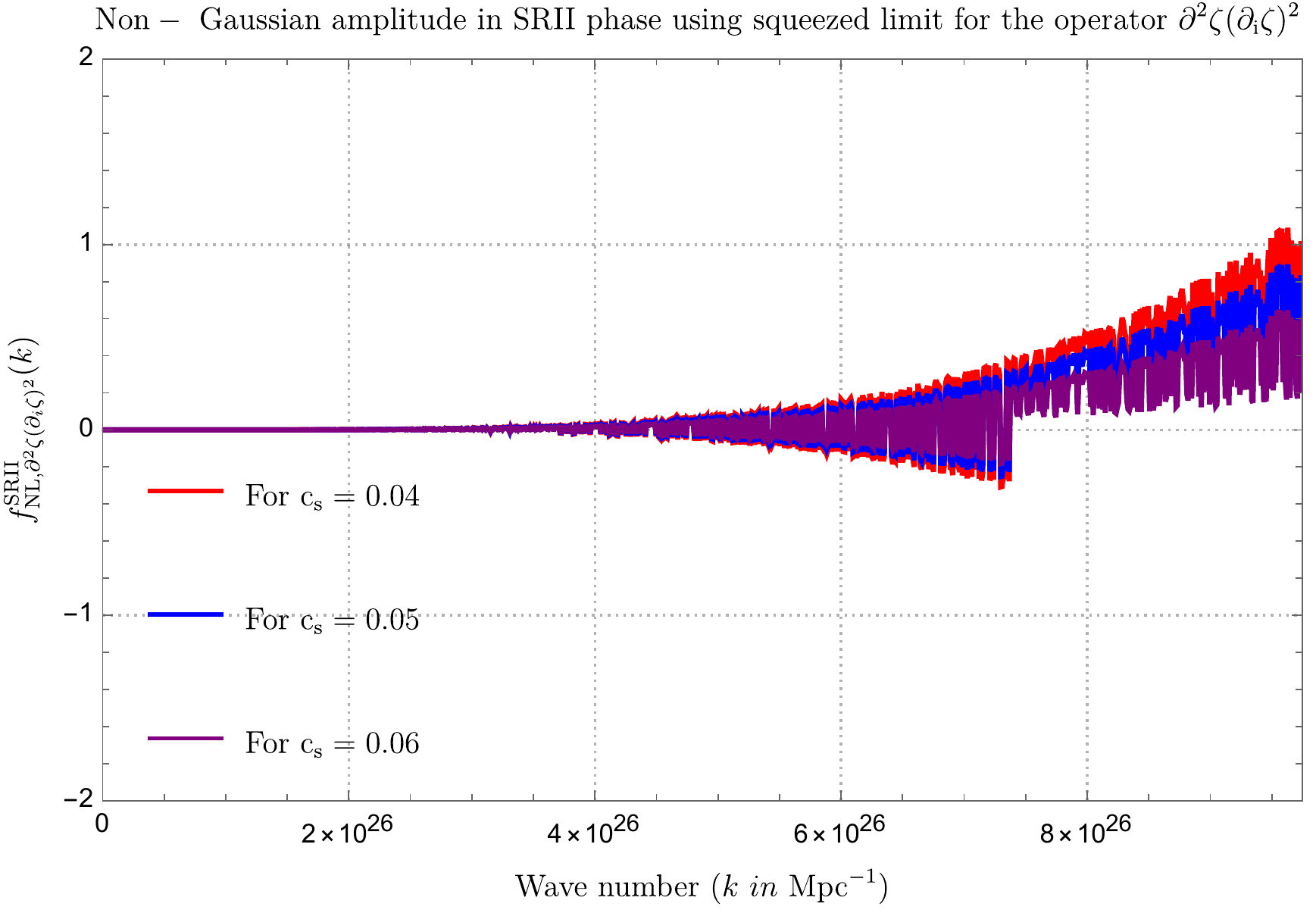}
        \label{Is24}
       }
    	\caption[Optional caption for list of figures]{The plot represents the contributions of individual operators to the non-Gaussian amplitude $f_{\text{NL}}$ in the SRII region as a function of the wave number. The values of the effective sound speed parameter $c_{s}$ are chose to be as follows $c_{s} = 0.04,0.05,0.06$. The plots in the top row represent contributions from the first and second operator. The plots in the bottom row represent contributions from the third and fourth operator.} 
    	\label{fNLSRII:a}
    \end{figure*}
    \begin{figure*}[htb!]
    	\centering
   {
      	\includegraphics[width=18cm,height=12.5cm] {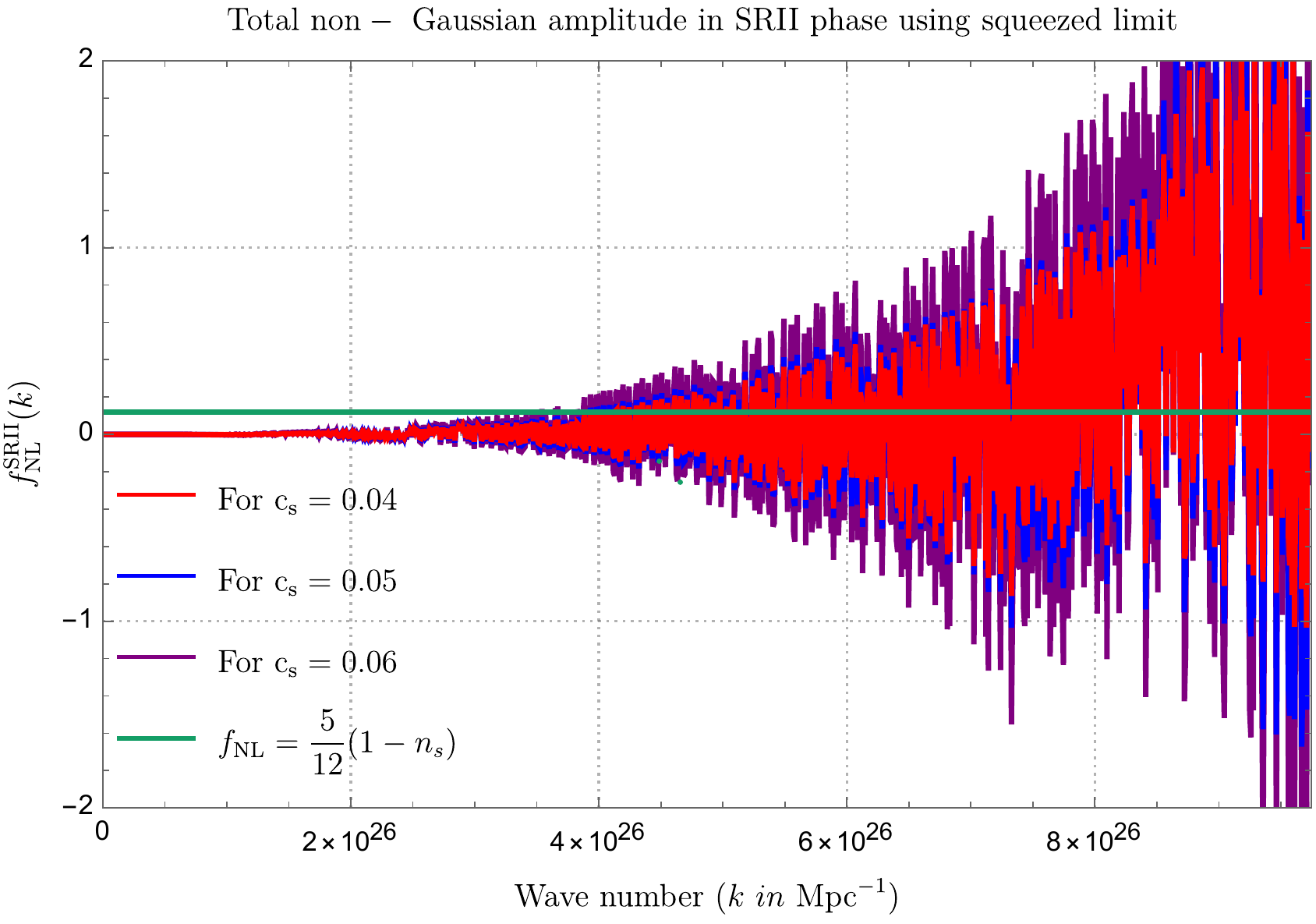}
        \label{Ks21}
    }
      	\caption[Optional caption for list of figures]{Plot represents the total contribution of all operators to the non-Gaussian amplitude $f_{\text{NL}}$ as a function of the wave number in the SRII phase under the squeezed limit. The values of the effective sound speed parameter $c_{s}$ are chose to be as follows $c_{s} = 0.04,0.05,0.06$. Here we find that the consistency condition is again being violated.} 
    	\label{fNLSRII:b}
    \end{figure*}

\subsection{Results obtained from region II: USR}
\label{s6b}
In this subsection, we present the representative plots regarding the variation of the non-Gaussian amplitude $f_{\text{NL}}$ with respect to the wave number in the USR region. Here also, we first show the individual contributions of each operator towards this behavior and then give the combined behavior of all the operators. 
 
From the figure Fig.(\ref{fNLUSR:a}), we see that the contributions from individual operators in the USR region to the non-Gaussian amplitude, $\fnl$ is significantly higher than the value produced by the consistency relation in the squeezed limit, i.e., $\fnl \sim {\cal O}(10^{-2})$. This behavior shows the clear violation of the {\it Maldacena's consistency relation} in the USR region. Contributions from all the operators are essential for getting the overall behavior in this region, unlike in the SRI region, where only a few operators had a visible impact. The values in these plots are well within the bounds for maintaining the perturbativity approximation and give us the required enhancement for the production of PBH and in the corresponding primordial power spectrum amplitude. The contribution from the first operator in plot (\ref{IU1}) consists of oscillations which are less significant when compared to the contributions from the second and third operators in plots (\ref{IU2}, \ref{IU3}), while the fourth operator in plot (\ref{IU4}) does not exhibit a clear oscillatory nature. The presence of a sharp rise or fall in the behavior of the amplitude is also visible, coming from all the operators, at either the beginning of the USR phase, for $k_{s} \approx 10^{6}$ Mpc$^{-1}$, or near the end of the phase, for $k_{e} \approx 10^{7}$ Mpc$^{-1}$. From this, we conclude that a sharp transition will be present when crossing over from the SRI to the USR region or from the USR to the SRII region. For different sound speed values, the plots do not deviate significantly relative to each other and the causality, unitarity constraints are perfectly maintained. However, as we increase the sound speed values, the behavior is more enhanced than for the previous values, which can be clearly visible in all of these representative figures.
In the plot in Fig.(\ref{fNLUSR:b}), we present the total contribution of all the operators towards the non-Gaussian amplitude $\fnl$. After their addition, we see a definite sharp behavior in the values right at the position of the transition wave numbers $k_{s}$ and $k_{e}$. The green line shows the value obtained from the consistency relation, and the values for  $\fnl$ in the USR region are significantly larger than that, indicating the violation of the consistency relation. The coefficients ${\cal G}_{i}$ $\forall i=1,2,3,4$ are adjusted accordingly to control the behaviour of the amplitude.

    \begin{figure*}[htb!]
    	\centering
   {
      	\includegraphics[width=18cm,height=12.5cm] {
      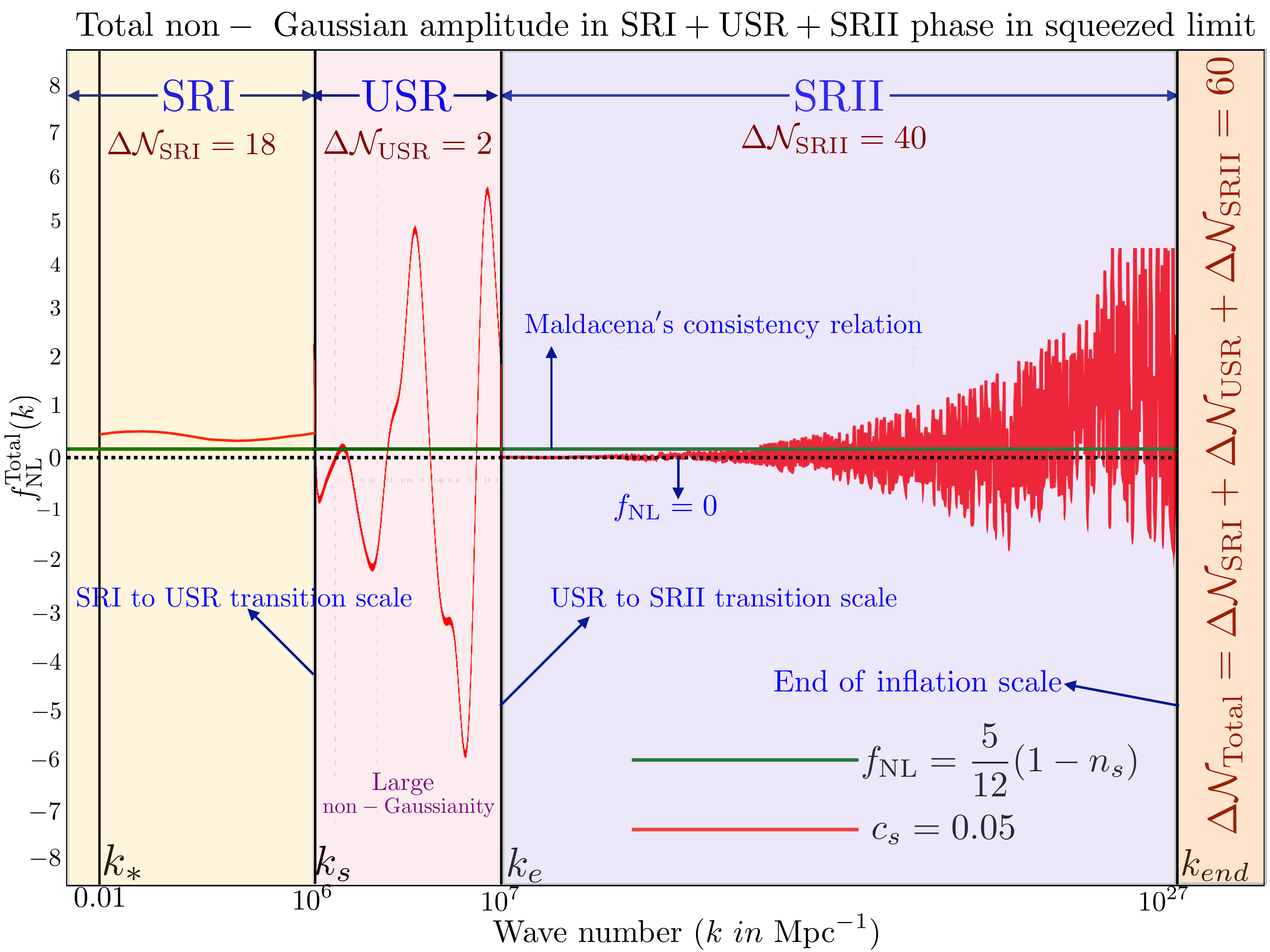}
        \label{KUPBH}
    }
    	\caption[Optional caption for list of figures]{Representative diagram of the cumulative behavior of all three phases, SRI, USR, and SRII, towards the non-Gaussian amplitude $f_{\text{NL}}$ as a function of the wave number in the squeezed limit. The effective sound speed $c_{s}$ is fixed as $c_{s} = 0.05$. The pivot scale is fixed at $k_{*} = 0.01$ Mpc$^{-1}$, the SRI to USR transition is fixed at $k_{s} = 10^{6}$ Mpc$^{-1}$,  the USR to SRII transition is fixed at $k_{e} = 10^{7}$ Mpc$^{-1}$, and the end of inflation at $k_{\text{end}} = 10^{27}$ Mpc$^{-1}$. Large non-Gaussianity is observed within the USR phase with the peak non-Gaussianity value being observed at the scale $\sim$ 8 $\times$ 10$^{6}$ Mpc$^{-1}$. Lastly, a sufficient amount of 60 e-foldings is achieved from our analysis.}
        \label{totPBH}
    \end{figure*}
    \begin{figure*}[htb!]
    	\centering
   {
      	\includegraphics[width=18cm,height=13.5cm] {
      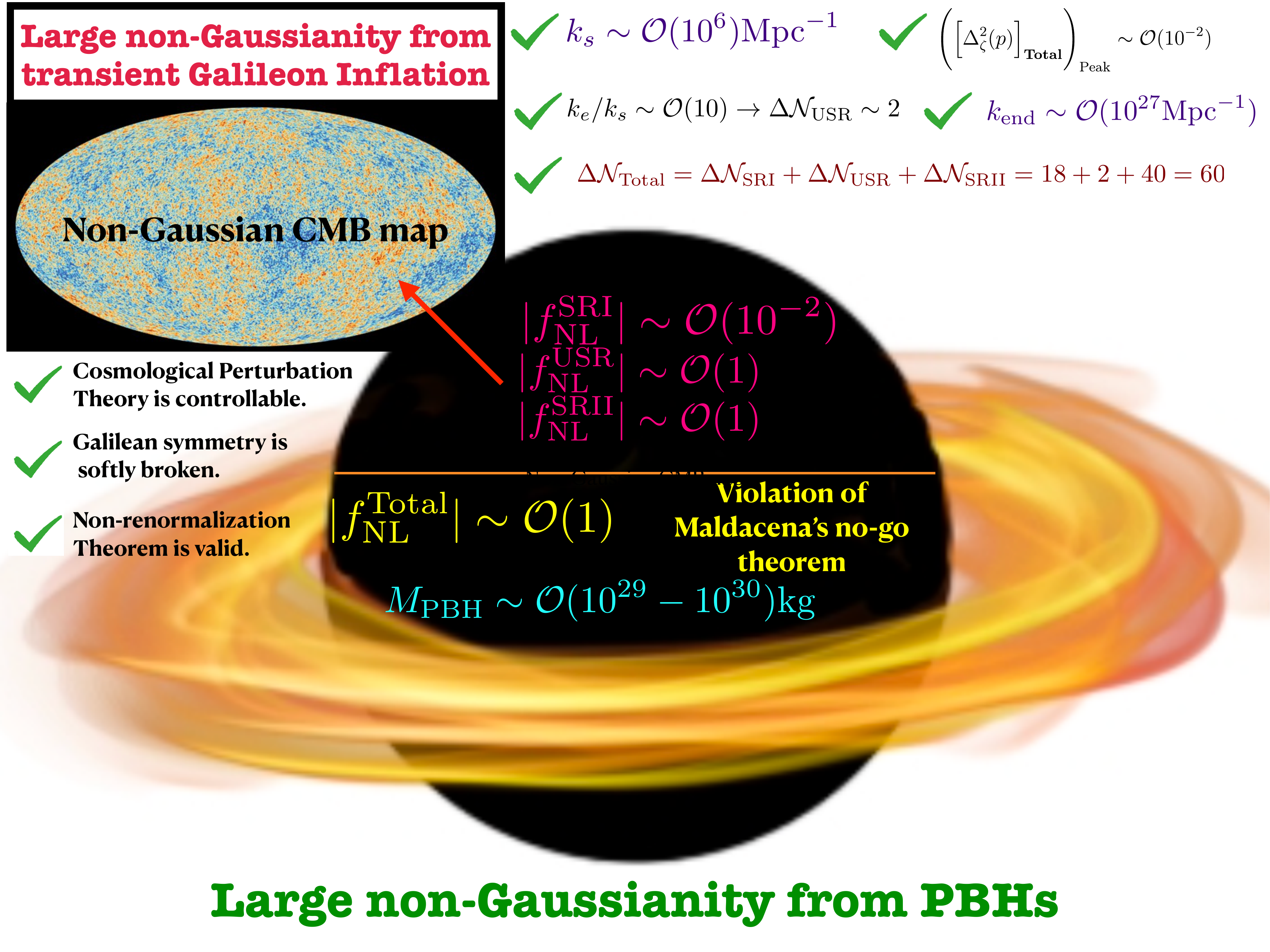}
        \label{OUTPBH}
    }
    	\caption[Optional caption for list of figures]{Representative flow chart of the underlying process of generating large primordial non-Gaussianity from transient ultra slow-roll Galileon inflationary paradigm and its connection with large mass PBHs formation.} 
        \label{totPBH2}
    \end{figure*}

\subsection{Results obtained from region III: SRII}
\label{s6c}
In this subsection, we present the plots regarding the variation of the non-Gaussian amplitude $f_{\text{NL}}$ with respect to the wave number in the second slow-roll (SRII) region. We first show the individual contributions of each operator towards this behavior and then give the combined behavior of all the operators.

From the figure Fig.(\ref{fNLSRII:a}), we see that
these contributions are much larger in amplitude when compared to their corresponding operator plots in the SRI region, and consequently display larger deviation compared to the value from the consistency condition, i.e., $\fnl \sim {\cal O}(10^{-2})$. This is expected as the fluctuations in SRII region are already enhanced after going through the USR phase and the vacuum state corresponding to this region is also a non Bunch-Davies type which supports large non-Gaussianities. The amplitude must be larger than the values in the SRI region but still smaller when compared to the USR region, which is clear from the plots. Also, the fluctuations only seem to increase drastically when the end of inflation, i.e., $k_{\text{end}} \geq 6 \times 10^{26}$ Mpc$^{-1}$ is reached. In terms of individual contributions, those coming from the plot (\ref{Is21}) contains less rapid oscillations and the fluctuations tend to increase fast as $c_{s}$ is increased, while fluctuations in the plot (\ref{Is22}) are violent in nature and increases as $c_{s}$ is increased. Similar is the case of the plots (\ref{Is23}, \ref{Is24}) where after a certain value of the wave number we only get positive oscillations. However, only for the fourth operator do the oscillations decrease when $c_{s}$ is increased. This can be understood from the analytic relation for this operator and its dependence on the parameter $c_{s}$.

The plot in Fig.(\ref{fNLSRII:b}) contains the combined behavior of all the operators, where the parameter $\fnl$ is plotted against the wave number while considering the same set of different values for the sound speed to check for the variation in results. The result eventually displays the similar violent oscillations as found in the individual contributions and the values increases as the parameter $c_{s}$ is increased. The deviation from the consistency condition is also clear from the plot.

\subsection{\textcolor{black}{Bispectrum related shape function and the corresponding numerical plots}}
\label{s6d}

\textcolor{black}{In the previous subsections, we demonstrated our results for the behaviour of the non-Gaussiantiy parameter, $f_{\rm NL}$, throughout the three regions after utilizing the explicitly calculated form for the three-point correlation functions as shown before. In this section, we elaborate on another crucial property of the bispectrum, which is the shape function, and examine its results for all three phases. }

\textcolor{black}{From its definition, we know that the bispectrum is a function of the three momenta, ${\bf k_{1}}, {\bf k_{2}}$, and ${\bf k_{3}}$, such that under the conditions imposed by translation invariance (homogeneity) and rotation invariance (isotropy), those three momenta form various triangular configurations depending on their magnitudes relative to each other. The resulting shape of the triangle contains information about the non-Gaussianity source. Thus, its qualitative analysis also helps to support the determination of the $f_{\rm NL}$ parameter \cite{Babich:2004gb, Chen:2006nt}. }

\textcolor{black}{We now introduce the other version used to define the bispectrum as follows: 
\bea \label{shapeftn}
{B}_{\zeta\zeta\zeta}(k_{1},k_{2},k_{3}) = {S}(k_{1},k_{2},k_{3})\times\frac{6}{5}f_{\rm NL}(2\pi^{2})^{2}\prod\limits_{i=1}^{3}\frac{1}{k_{i}^{3}}.
\eea 
The above expression results from the existing relation for the bispectrum that  involves the tree-level scalar power spectrum, in Eqs.(\ref{c1Bf},\ref{c2Bf},\ref{c3Bf}), and which ultimately enables the calculation of the quantity $f_{\rm NL}$ for each region.}

\textcolor{black}{The shape of the bispectrum refers to the particular form of function $S(k_{1},k_{2},k_{3})$ which treats the momentum ratios, say $x_{2} \equiv k_{2}/k_{1}$ and $x_{3} \equiv k_{3}/k_{1}$, as its variables while the total momentum, $K=k_{1}+k_{2}+k_{3}$, is kept fixed. Different names for the shape exist depending on the kind of relation the function has with the momenta. One of them is the \textit{squeezed} configuration, where one of the momenta is very small compared to the other two, which becomes equal due to momentum conservation, $k_{1} \ll k_{2} \sim k_{3}$. Another one is the \textit{equilateral} configuration, where all the three momenta remain equal, $k_{1}=k_{2}=k_{3}$. To visualize these particular shapes, we require the $3$D plot of the shape function, and there we start with the assumption that $x_{2} \leq x_{3}$, which helps to avoid considering repetition in configuration, and the arguments must obey the triangle inequality $x_{2}+x_{3}\geq 1$. The actual quantity to $3$D plot is the dimensionless quantity,  $x_{2}^{-1}x_{3}^{-1}S(1,x_{2},x_{3})$, because the definition used in Eqn.(\ref{shapeftn}) must determine the bispectrum as a homogeneous function of degree $-6$. Also, we set to zero the region outside the interval, $1-x_{2} \leq x_{3} \leq x_{2}$, so that the same configuration does not get plotted twice.
}

    \begin{figure*}[htb!]
    	\centering
   {
      	\includegraphics[width=16cm,height=12.5cm] {
    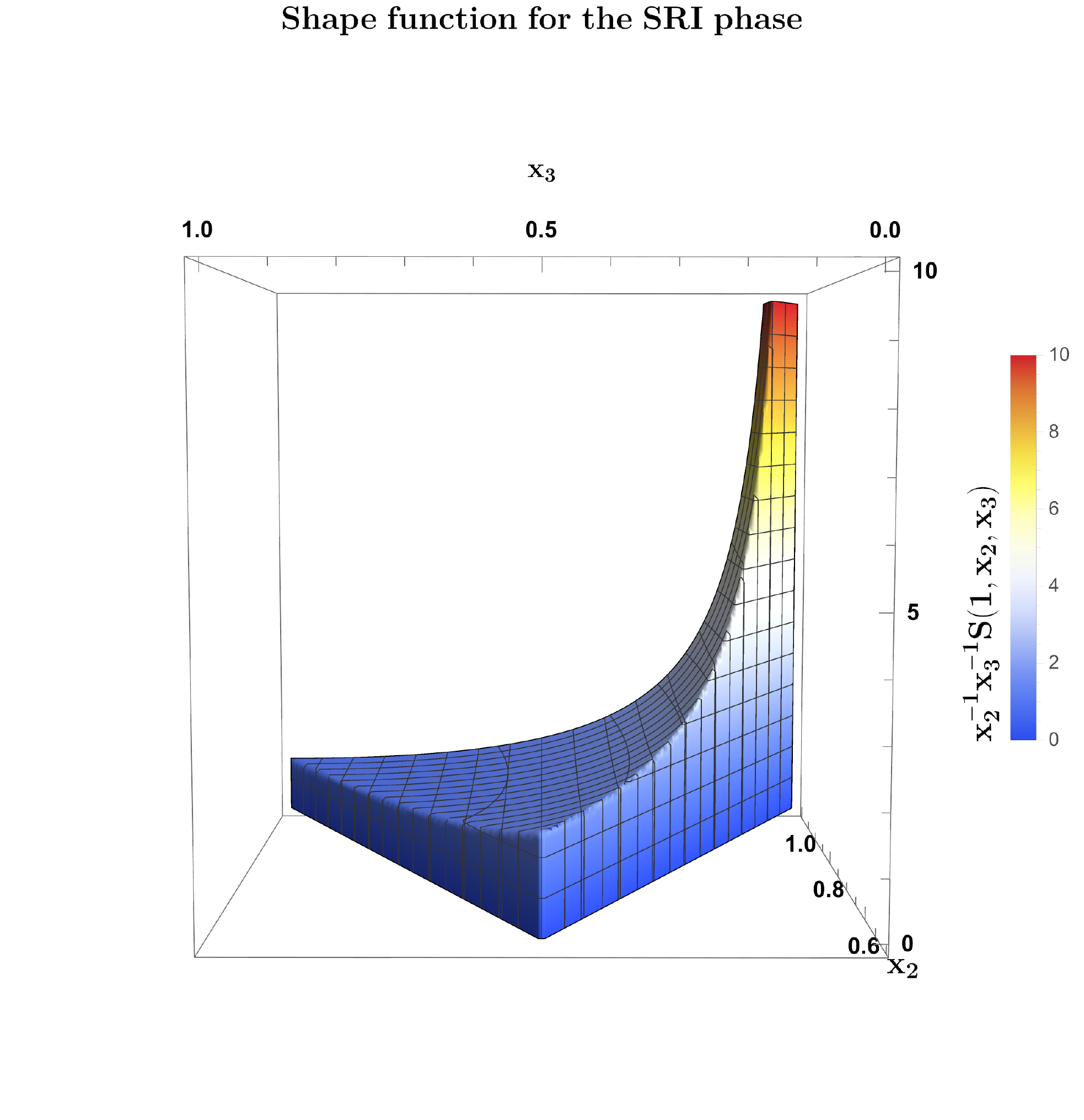}
        \label{shape1}
    }
    	\caption[Optional caption for list of figures]{\textcolor{black}{Plot of the function $x_{2}^{-1}x_{3}^{-1}S(1,x_{2},x_{3})$ for the SRI region. The shape is normalized to $1$ in the equilateral limit, $x_{2}=x_{3}=1$, and vanishe outside the region $1-x_{2} \leq x_{3} \leq x_{2}$.}}  
        \label{SRIshape}
    \end{figure*}

    \begin{figure*}[htb!]
    	\centering
   {
      	\includegraphics[width=16cm,height=12.5cm] {
    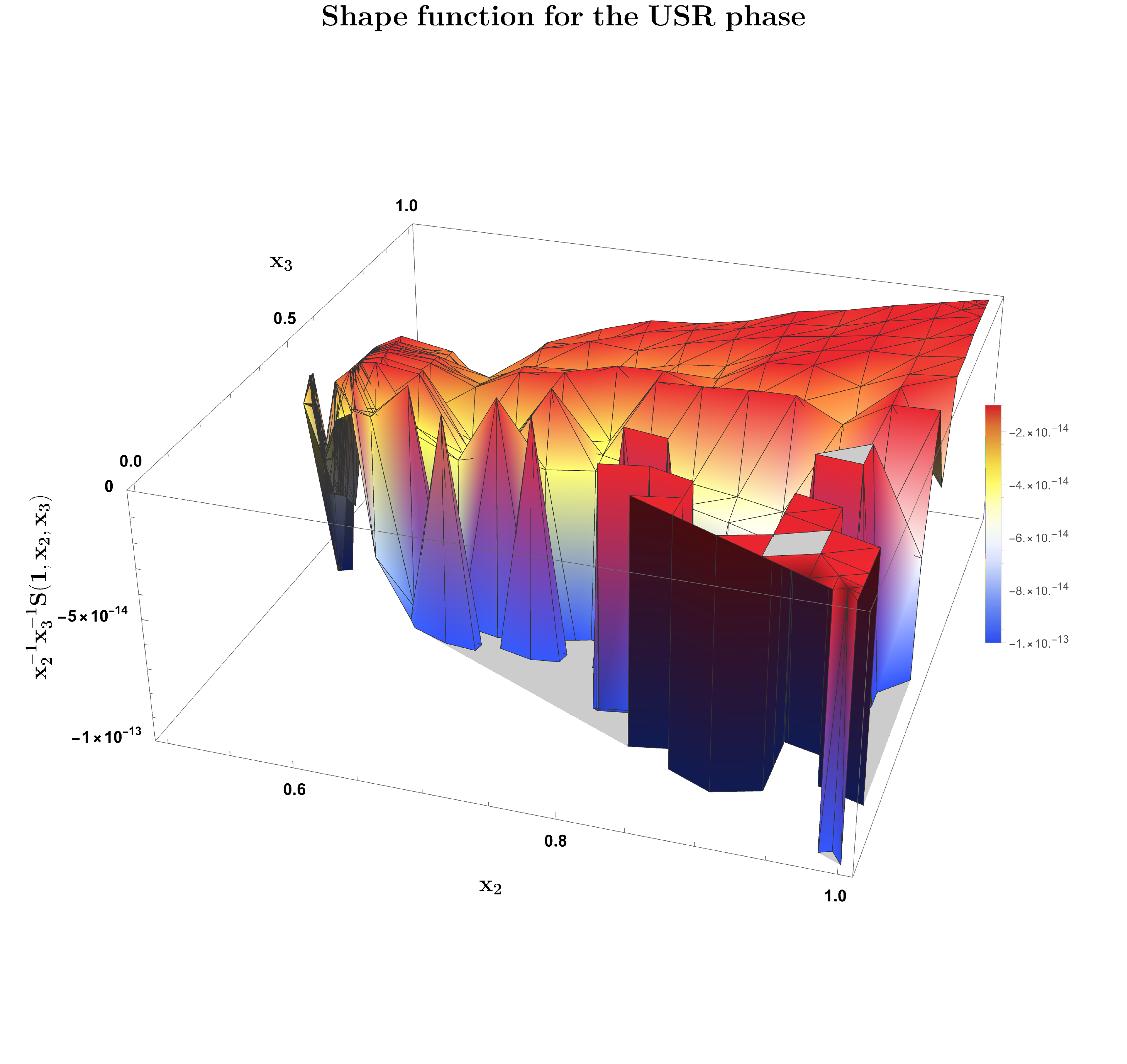}
        \label{shape2}
    }
    	\caption[Optional caption for list of figures]{\textcolor{black}{Plot of the function $x_{2}^{-1}x_{3}^{-1}S(1,x_{2},x_{3})$ for the USR region. The shape is set to vanish outside the region $1-x_{2} \leq x_{3} \leq x_{2}$.}}  
        \label{USRshape}
    \end{figure*}

    \begin{figure*}[htb!]
    	\centering
   {
      	\includegraphics[width=16cm,height=12.5cm] {
    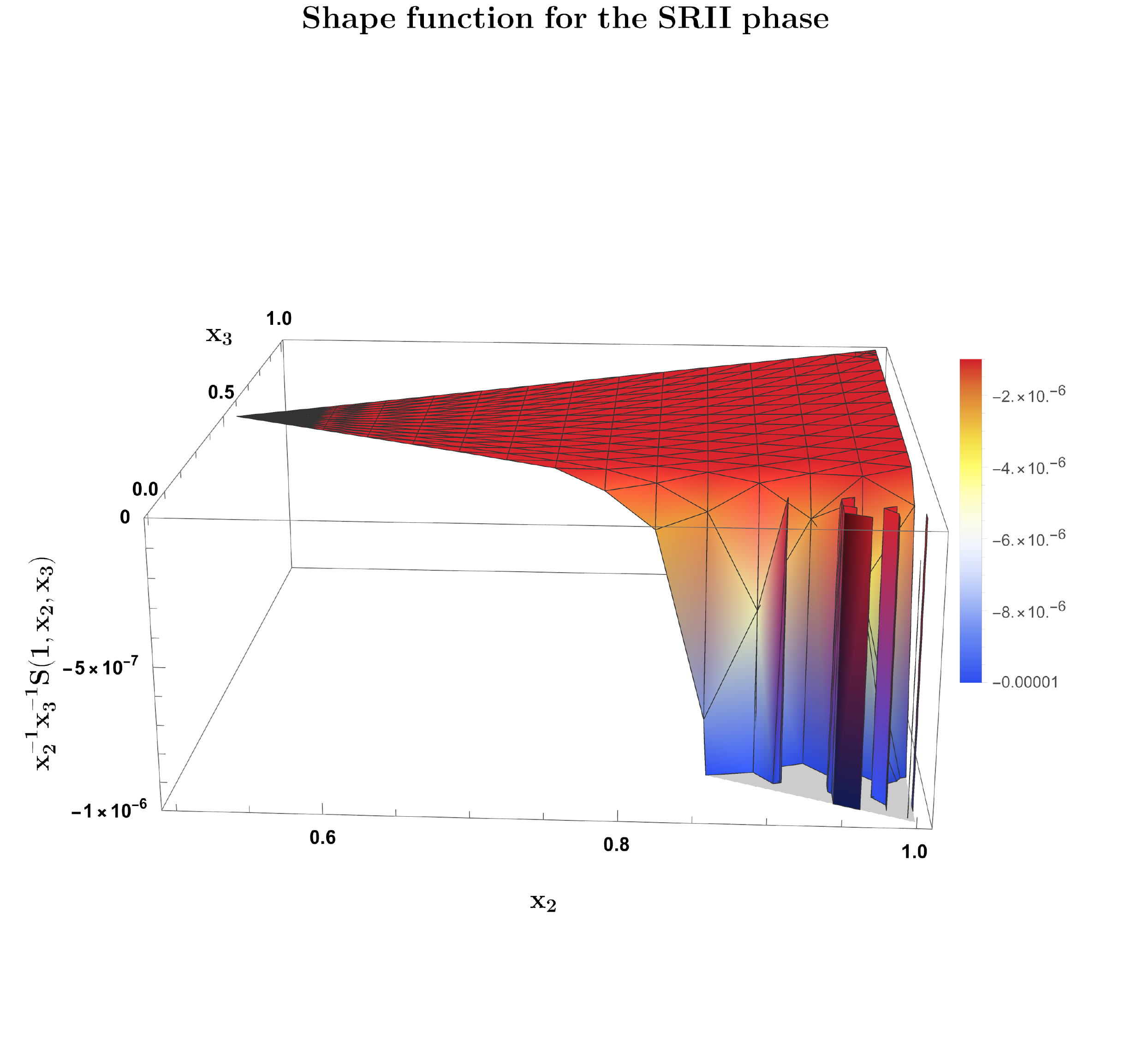}
        \label{shape3}
    }
    	\caption[Optional caption for list of figures]{\textcolor{black}{Plot of the function $x_{2}^{-1}x_{3}^{-1}S(1,x_{2},x_{3})$ for the SRII region. The shape is set to vanish outside the region $1-x_{2} \leq x_{3} \leq x_{2}$.}}  
        \label{SRIIshape}
    \end{figure*}

\textcolor{black}{We now analyze the various shapes obtained for the three regions. Starting with the SRI phase, the bispectrum contains interaction between different Fourier modes resulting from the expression in Eqn.(\ref{shapeftn}). However, within the figure Fig.(\ref{SRIshape}), the more informative feature that is visible is where the shape peaks around the squeezed limit, $x_{3} \rightarrow 0$ and $x_{2} \rightarrow 1$. This limit corresponds to the case where one of the modes becomes much longer, $k_{3} \rightarrow 0$ here, than the others and freezes out much before forming a background for the evolution of the remaining two modes. The non-linearities then develop when outside the horizon and produce the above shape. Under such a scenario, the consistency condition from Maldacena's theorem holds good, and therefore our previously obtained result for the non-Gaussianity $f_{\rm NL}$, in Fig.(\ref{fNLSRI:b}), for the SRI region is supported here by the shape function behaviour.} 

\textcolor{black}{Next we proceed with the shape for the USR region. The figure Fig.(\ref{USRshape}) shows an interesting behaviour of the shape. In the squeezed limit, $x_{3} \rightarrow 0$ and $x_{2} \rightarrow 1$, the shape drops quite sharply, as indicated by the violet colour. This feature is precisely the opposite of what is needed to establish the consistency condition, and hence, the correlations are highly suppressed in the squeezed limit. There are also sharp and discontinuous regions present in the shape, specifically when approaching the squeezed limit, which signifies sharp and rapid changes in the non-Gaussianity $f_{\rm NL}$, as already seen previously in fig.(\ref{fNLUSR:b}). The shape sees a continuous increase in the values when both $x_{2} \rightarrow 1$, and $x_{3} \rightarrow 1$, which is also the equilateral limit. Having a USR phase corresponds to a non-attractor feature within the theory, which leads to deviation from the consistency condition. Also, since our theory consists of higher derivative operators, this ultimately results in increased interactions between the modes that cross the horizon simultaneously, and beyond that, their interactions start to become negligible. This feature is an essential reason for the observed shape of the bispectrum in the equilateral limit.} 

\textcolor{black}{The figure Fig.(\ref{SRIIshape}), depicts the shape for the SRII phase. In a similar spirit as for the USR case, the shape here is almost flat until it sharply changes near the squeezed limit while also maximizing near the equilateral limit. Both the USR and SRII phases feature an underlying quantum vacuum structure increasingly shifted away from the Bunch-Davies vacuum state, and no longer is the consistency condition guaranteed to hold in such a case. It is, therefore, a consequence of the mentioned properties of the vacuum and increased interactions shown through the bispectrum shape in the equilateral limit that leads to large non-Gaussianities in both the USR and SRII, as also seen in fig.(\ref{fNLUSR:b},\ref{fNLSRII:b}).}

\subsection{Cumulative results obtained from all regions: SRI + USR + SRII }
\label{s6f}

The representative diagram in Fig.(\ref{totPBH}) describes the behavior of the non-Gaussianity amplitude $f_{\text{NL}}$ through all the three regions, SRI, USR, and SRII, and represents the combination of the specific behavior of $f_{\text{NL}}$, as seen in the Figs.(\ref{fNLSRI:b},\ref{fNLUSR:b},\ref{fNLSRII:b}) into a single plot.

In the single field slow-roll model, for getting large mass PBH, the values $k_s = 10^{6}$ Mpc$^{-1}$ and $k_e = 10^{7}$ Mpc$^{-1}$ is fixed; that restriction comes through the one-loop calculations. Now, the mass for PBH satisfies the relation $M_{\text{PBH}} \propto 1/k_{s}^{2}$. The problem arises when the value of $k_{s}$ is small, then we can generate large solar mass PBH, but, reducing the number of e-foldings in the case of single-field slow-roll models. We can only achieve 20-25 e-folds when also considering the quantum loop effects. As a result, only PBH with small $M_{\text{PBH}}$ can get produced by pushing the values to $k_s = 10^{21}$ Mpc$^{-1}$ and $k_e = 10^{22}$ Mpc$^{-1}$, where such large values are required to have inflation. Hence, we get $M_{\text{PBH}} \sim  {\cal O}(10^{2})$ gm in this framework by pushing the wave number towards larger values due to the proportionality relation before. This problem exists within the single field slow-roll model due to the need for applying renormalization and resummation techniques which would then require the duration of the SRII region to be very small and total number of e-foldings, $\Delta {\cal N}_{\rm Total}\sim 25$. Which means that generating large $M_{\text{PBH}}$ would require total e-foldings to be $25$; if in the present situation to have sufficient inflation $60$ e-foldings is the goal, we could only get small $M_{\text{PBH}}$ in the single field models of inflation. For more details see the refs. \cite{Choudhury:2023vuj,Choudhury:2023jlt,Choudhury:2023rks,Choudhury:2023hvf} where the details regarding the single field inflation has been discussed with proper justifications.

In the present Galileon inflationary paradigm, there is the facility to accommodate both the scenarios, which can able to generate both large and small mass PBHs. Additionally, in the Galileon theory, SRII duration can be changed; the duration for USR is always the same in almost all types of frameworks, which is $\Delta {\cal N}_{\rm USR}\sim 2$. If we increase the duration of the SRII phase to achieve the $60$ e-foldings, then we can generate large $M_{\text{PBH}}$ with a SRI to USR sharp transition scale $k_s = 10^{6}$ Mpc$^{-1}$ and USR to SRII sharp transition scale $k_e = 10^{7}$ Mpc$^{-1}$. If we shift $k_s = 10^{21}$ Mpc$^{-1}$ and $k_e = 10^{22}$ Mpc$^{-1}$, then SRII duration will be very small, and we can generate only small $M_{\text{PBH}}$ within the framework of Galileon inflation.

The non-Gaussianities are calculated using the third-order perturbed action, which is also used to calculate the loop effects \cite{Choudhury:2023vuj,Choudhury:2023jlt,Choudhury:2023rks,Choudhury:2023hvf}. The properties of the non-renormalization theorem, in the present Galileon theory, do not require a renormalized version of the power spectrum, so resummation is also unnecessary. Hence, earlier those methods which were giving huge constraints and did not allow for a prolonged SRII phase are not present in the case of the Galileon theory, and we can generate large $M_{\text{PBH}}$. The requirement for the generation of large non-Gaussianities from the theory, which can also be possible to detect and facilitate the generation of large $M_{\text{PBH}}$, requires the addition of an important new feature in the theory. The deviation from Gaussianity is significant when a sharp transition is present from one region to another, and the respective position of both the transitions, SRI to USR and USR to SRII, is essential because, in the final results, the amplitude would appear to be proportional to the inverse power of those transition wave numbers. Hence, an increase in the value of $k_{s}$ and $k_{e}$ in the denominator will suppress the total contribution. At wave numbers $k_s = 10^{6}$ Mpc$^{-1}$ and $k_e = 10^{7}$ Mpc$^{-1}$, the enhancement is significant. So, in conclusion, if large $M_{\text{PBH}}$ is required, then only large non-Gaussianities are produced, and for tiny $M_{\text{PBH}}$, the non-Gaussianities would be negligible.

Now, the proportionality relation for  $M_{\text{PBH}}$ follows $M_{\text{PBH}} \propto 1/k_{s}^{2}$ and a similar relation for $\fnl$ is presently modified to become proportional to inverse powers of both $k_{s}$ and $k_{e}$. If we assume that huge quantum fluctuations can produce an amplitude of the order ${\cal O}(10^{-4})$, rather than the actual ${\cal O}(10^{-2})$ amplitude required for PBH creation, then the finding for PBH production is inconclusive. It is not enough to have substantial non-Gaussianities even if the power spectrum amplitude is small. The precise location of the transition wave numbers is also important in calculating the quantities, $M_{\text{PBH}}$ and $f_{\rm NL}$.

Now, we mention clearly the highlighting outcomes and its physical consequences from the analysis performed in this paper point-wise:
\begin{enumerate}
    \item \underline{\textbf{Implication of Non-renormalization theorem}}: The Galileon theory has this theorem as its most attractive feature. In the case of single field slow-roll models, due to the presence of quantum loop effects and the need to complete inflation with the required amount of e-foldings, i.e., $60$ e-folds, we will not be able to achieve large $M_{\text{PBH}}$ and maintain $60$ e-folds of expansion at the same time. The need for renormalization and resummation procedures puts extra constraints. In Galileon theory, however, such constraints are not present and we are able to have a sufficiently long SRII period to complete inflation. This later helps to produce large non-Gaussianities which facilitates the formation of large $M_{\text{PBH}}$ due to the help of a new sharp transition feature which comes when going from SRI to USR and USR to SRII regions.
     \item \underline{\textbf{Mass of PBH}}: \textcolor{black}{In this work, we are primarily interested in the case of large mass PBHs, which are more interesting from the cosmological perspective as they can provide for a substantial fraction of the present-day dark matter. In the case of working with canonical and non-canonical single field models of inflation in the EFT framework, the necessary need to perform renormalization and resummation procedure puts strong constraints on the position of the USR, shifting its position to larger wavenumbers which ultimately makes it possible only to generate small $M_{\rm PBH}$, as evident from the relation $M_{\rm PBH} \propto 1/k_{s}^{2}$ \cite{Choudhury:2023vuj, Choudhury:2023jlt, Choudhury:2023rks}. However, quantum loop corrections and related constraints remain absent in the case of Galileon as per the non-renormalization theorem, which allows us to shift the transition scale at the position of interest \cite{Choudhury:2023hvf}. Moving it towards the larger values generates small $M_{\rm PBH}$ while towards the smaller value, we can observe larger $M_{\rm PBH}$. Regardless of the USR position, we will observe large non-Gaussianity in return. Fortunately, since we demand the case of large $M_{\rm PBH}$ in the present analysis, we also observe the corresponding scenario of large non-Gaussianity in the USR phase.  }
    \item \underline{\textbf{Controlling Non-Gaussianities}}: In the present framework, to produce PBH with the help of a sharp transition window satisfying $k_{e}/k_{s} \sim {\cal O}(10)$, the amount of non-Gaussianity suffers a great increase in the USR region which has to be controlled using the coefficients ${\cal G}_{i}$, $\forall i=1,2,3,4.$ If the transition wave numbers are of smaller order then the enhancement could be managed in a controlled fashion and it will be able to give us large $M_{\text{PBH}}$. However, if the same wave numbers are shifted to much larger orders, then the controlling would require extreme fine-tuning of the coefficients which can also lead to a violation of the power spectrum constraints and we can only expect small $M_{\text{PBH}}$ with small non-Gaussianities. Hence, to have controlled non-Gaussianities in the present theory, we have smaller values of the transition wave numbers. From our results in Fig.(\ref{totPBH}), the amount of non-Gaussianity from SRI region is of the order ${\cal O}(10^{-2})$ and from the USR and SRII region it is of the order ${\cal O}(1)$. Hence, the cumulative average of the amount of non-Gaussianity produced all the three regions is of the order $|\fnl| \sim {\cal O}(1)$. It is clear from these results that the bound on the amplitude $\fnl$ from Maldacena's \textit{no-go theorem}, which is $|\fnl| \sim {\cal O}(10^{-2})$, is strictly satisfied in SRI region, while the same bound is strongly violated in both the USR and SRII regions.
\end{enumerate}

\begin{table}

\scriptsize

\begin{tabular}{|c|c|c|c|c|}

 \hline\hline

 \multicolumn{5}{|c|}{\footnotesize A comparison between the One-loop power spectrum and Tree-level non-Gaussian effects in Galileon theory for all three phases} \\

 \hline\hline

 \bf{Phase} & \bf{One-loop corrected} & \bf{Non-Gaussian amplitude} & \bf{Allowed Wave number} & \bf{Highlights and findings}\\
 & \bf{Power spectrum amplitude} ($\Delta^{2}_{\zeta}(k)$) & $(f_{\text{NL}})$ & ($k$ Mpc$^{-1}$) &\\

 \hline\hline

& & & & \\

& & & $k_{*} \leq k \leq k_{s}$ & No large non-Gaussianities observed. \\
\bf{SRI} & ${\cal O}(10^{-9})$  & $(0.01,0.05)$  & where $k_{*} = 0.01$Mpc$^{-1}$,  & Non-Gaussianity of ${\cal O}(10^{-2})$ observed.\\
& & &   and $k_{s} = 10^{6}$Mpc$^{-1}.$  & Consistency condition (\ding{52})\\
& & &   & Total e-folds achieved: $\Delta{\cal N}_{\rm SRI} = 18$. \\

& & & & \\

\hline\hline

& & & & \\

&  &  &  & Sharp transition observed at \\
& & & $k_{s} \leq k \leq k_{e}$  & the beginning and end of the phase. \\
\bf{USR}  & ${\cal O}(10^{-2})$ & $(-5,5)$ &   where $k_{s} = 10^{6}$Mpc$^{-1}$,  & Large non-Gaussianities $\sim$ ${\cal O}(1)$.\\
& & &   and $k_{e} = 10^{7}$Mpc$^{-1}.$  & Favourable for large $M_{\text{PBH}}$. \\
& & &   & Consistency condition (\ding{56}) \\
& & &   & Total e-folds required to maintain \\
& & &   & perturbativity approximation: $\Delta{\cal N}_{\rm USR} = 2$. \\

& & & & \\

\hline\hline

& & & & \\

&  &  &  & Sharp transition observed while \\
& & & $k_{e} \leq k \leq k_{end}$ & exiting the USR phase in beginning. \\
\bf{SRII} & ${\cal O}(10^{-5})$ & $(-1,2)$ & where $k_{e} = 10^{7}$Mpc$^{-1}$,  & Large non-Gaussianities $\sim$ ${\cal O}(1)$.\\
& & &   and $k_{\text{end}} = 10^{27}$Mpc$^{-1}.$  & Consistency condition (\ding{56}) \\
& & &   & Rapid oscillatory behaviour near $k_{\text{end}}$. \\
& & &   & Total e-folds required to complete \\
& & &   & inflation: $\Delta{\cal N}_{\rm SRII} = 40$.
\\

& & & & \\

\hline\hline




\end{tabular}

\caption{Comparison between one-loop corrected power spectrum and tree-level bispectrum amplitudes. We mention each of their values for all the three phases along with the respective wave numbers involved. The important highlights from our analysis of the results for the bispectrum is also present which discusses the amount of non-Gaussianities produced, the behavior of the non-Gaussian amplitude, the total number of e-folds required and validity of the consistency condition in each phase represented using \ding{52} and \ding{56}.}

\label{tab:1}

\end{table}
The representative flow chart Fig.(\ref{totPBH2}) provides a general view of our overall findings. It shows that we are able to predict the generation of large non-Gaussianities which also facilitates the formation of large mass primordial black holes, $M_{\text{PBH}} \sim {\cal O}(10^{29}-10^{30})\text{kg}$. The non-renormalization theorem protects the galileon theory from unwanted quantum effects which, along with an additional phase with $\Delta{\cal N}_{\text{USR}} \sim 2$ helps to support our previous statement. The figure also shows that we are able to control the amount of non-Gaussianity produced in all the three phases through which we are led to witness the violation of the consistency condition in the USR and SRII regions. 

The table \ref{tab:1} provides a comparison between the amplitude of the one-loop corrected power spectrum as calculated in \cite{Choudhury:2023hvf} and the amount of non-Gaussianity which we obtain for each of the three phases, SRI, USR and SRII, in our present work. It also highlights the important features regarding the validity of the consistency condition and the behaviour of the amplitude $\fnl$ when transitioning between the phases, which are obtained in our findings and visualized through our plots in Figs.(\ref{fNLSRI:b},\ref{fNLUSR:b},\ref{fNLSRII:b},\ref{totPBH}). 

\section{\textcolor{black}{Comparison with other works} }
\label{s6e}

\textcolor{black}{In our present work, we show the violation of the well-established consistency condition between the amount of non-Gaussianity and the spectral tilt in single-field inflation models of inflation under the squeezed limit. Within our underlying framework of Galileon theory, including a USR phase that provides for a non-attractor evolution of the background, rightfully exhibits this violation, which only attractor models of inflation respect. The violation is drastic as the amplitude rises quickly when a sharp transition occurs from a slow roll to a briefly lived USR phase. We have gone through a rigorous analysis which involves the comoving curvature perturbation dynamics in the superhorizon regime to truly determine how the overall strength of the non-Gaussianity gets produced in various phases. The final result is the behavior of non-Gaussianity with the wavenumber using the existing relations, Eqs.(\ref{c1Bf},\ref{c2Bf},\ref{c3Bf}), in the squeezed limit.}

\textcolor{black}{The present analysis rests on a detailed computation for the bispectrum. In the study by Palma \textit{et al.} \cite{Mooij:2015yka}, they provide a general discussion on the underlying origins of non-Gaussianity for single-field inflation models with a non-attractor evolution of the background. Their arguments for this dramatic violation of the consistency condition are driven purely by considering the symmetries involved, using which they have presented a modified version of the consistency condition for it to hold during the USR phase. The proposed new result adopts a similar approach for its derivation by which the original consistency result was re-derived there mainly following symmetry arguments. Our analysis does not introduce any newer version of the consistency condition. Their main result in \cite{Mooij:2015yka}, in the limit of $\epsilon\rightarrow 0$ and $\eta\rightarrow -6$,  reduces to give $f_{\rm NL} = 5/2$ and agrees with the previously derived relation between the bispectrum and the power spectrum in the squeezed limit when a violation of the consistency condition occurs \cite{Namjoo:2012aa, Martin:2012pe}. The value of $f_{\rm NL} = 5/2$ signals large non-Gaussianity in the USR, and the magnitude also agrees with the predictions through our calculations, but we also show from our analysis that a maximum of $f_{\rm NL} \sim -6$ is possible to achieve. Both the results, in the present work and by Palma \textit{et al.}, point towards the reason for this violation involving the superhorizon evolution of the curvature perturbation modes. }

\textcolor{black}{In another work by Cai \textit{et al.} \cite{Cai:2018dkf}, the authors discuss the effects of the two kinds of transitions, smooth and sharp from a USR to another SR region, on the modification of the non-gaussianity generated in the USR phase. We have explicitly focused on the sharp nature of transition, which has a significantly different effect on the evolution of perturbations than the other smooth case. Also pointed out through the construct in our work is that the non-attractor phase of background evolution is relatively small, $\Delta{\cal N}_{\rm USR}\sim {\cal O}(2)$, compared to the total e-foldings required for successful inflation achieved after the addition of another SRII phase. In our present setup, there is no relaxation phase, and the transition from the initial SRI to the non-attractor USR, and from the USR to the SRII phase happens instantly, which is evident from the change in the $\eta$ parameter going from $\eta \rightarrow 0$ in SRI to $\eta \rightarrow -6$ in the USR and $\eta \rightarrow -6$ to $\eta \rightarrow 0$ from the USR to SRII. For such a step function transition feature, had we considered a case of starting with an initial non-attractor phase that transitions into another SR phase, then according to the analysis in \cite{Cai:2018dkf}, this would modify the non-gaussianity present in the non-attractor phase which does not lead to recover its value $f_{\rm NL}=5/2$ as the mode functions do not freeze. Our setup has a case where the sharp transition is incorporated to connect two slow-roll phases, namely SRI and SRII. We have found that after we observe the transition from the initial SRI, the consistency condition gets violated and remains as such due to the presence of such sharp transitions. This result differs from the authors' conclusion in \cite{Cai:2018dkf}. Their analysis centers around canonical non-attractor and specific models of the $P(X,\phi)$ class and the effects of transition features on non-gaussianity in such models. Our underlying model is the Galileon theory, a beyond $P(X,\phi)$ class of model. In this Covariantized Galileon Theory (CGT) driven USR phase we do observe large non-gaussianity, $f_{\rm NL}\sim -6$, which is a result of this sharp transition and this feature also suppresses the amount of non-gaussianity to $f_{\rm NL} \sim 2$ in the SRII phase, which still remains in the category of large non-gaussianity thus violating the consistency condition. 
In the driven USR phase model, we conclude that the generated non-gaussianity will always be more significant than the slow-roll case and remain so after we enter another SRII phase. This scenario is also directly related to the parameterization used for the effective sound speed $c_{s}$, which is adjusted to peak abruptly near the transition scales in our model. This feature generates a strong result as it also tags the large non-gaussianity with PBH production in the USR and leads to the theory being consistent and supportive of the observations coming from the latest NANOGrav 15 signal \cite{NANOGrav:2023gor}. There are similar works in \cite{Choudhury:2023kam,Choudhury:2023hfm,Choudhury:2023fwk,Ferrante:2022mui,franciolini:2023pbf,Gorji:2023sil} which highlight the advantages of this large non-gaussianity feature, $f_{\rm NL}>1$, from the perspective of observations with the NANOGrav 15 signal and in addressing the severe PBH overproduction problem. Our results here show that our model is well-suited to support the conclusions from the observational side. }

\section{Summary and conclusions}
\label{s7}

In this paper, we studied the production of large primordial non-Gaussianities in the framework of Galileon inflation. Generation of large non-Gaussianities has always been a challenge \cite{Maldacena:2002vr}: Single-field models of inflation, including the $P(X,\phi)$ theories and the multi-field framework, are either not able to produce a significant amount of non-Gaussianities or are too cumbersome to give any meaningful results. We have demonstrated that said problem is relaxed with the use of the Galileon theory, where the effective sound speed satisfies the unitarity and causality constraint of $0.024 < c_{s} \leq 1$ and the framework does not contain any unwanted ghost terms. We have shown that there is a significant improvement when an additional mechanism, requiring a sharp transition from the SRI to the USR phase, is added, enabling an increase in the overall non-Gaussianity. This also helps in addressing the parallel question of the generation of large masses of primordial black holes, $M_{\text{PBH}}$. In the present scenario, quantum loop effects are controllable, facilitating the formation of PBH with $M_{\text{PBH}} \sim {\cal O}(10^{31} \text{kg})$ \cite{Choudhury:2023hvf}; the addition of the sharp transition mechanism into the USR phase provides us with a possibility to address both the problems concerning the required amount of $M_{\text{PBH}}$ and the non-Gaussianity amplitude $f_{\text{NL}}$. Galileon theory also possesses a unique feature in its non-renormalization theorem, which protects the theory and any of its correlators from quantum correction. This allowed to have a sufficient period of expansion while being able to produce large $M_{\text{PBH}}$, which would not have been possible in otherwise models for single field slow-roll inflation due to a need for the aforesaid procedures. With this theory at hand, we move on towards obtaining the general solutions for the scalar perturbation modes by exploiting the continuity conditions for the modes and their conjugate momenta at the transition moments between the SRI to USR and USR to SRII phases. We found that the initial Bunch-Davies vacuum state gets shifted to a non-Bunch Davies type and that gives rise to the enhancement of the non-Gaussianities also during the SRII phase. Next, we used the action which is third-order in the curvature perturbations $\zeta$, to calculate the tree-level bispectrum in the squeezed limit to obtain the expression for the non-Gaussianity amplitude $\fnl$. Our cumulative findings are displayed by representative diagram Fig.(\ref{totPBH}, \ref{totPBH2}) for the squeezed bispectrum which confirms that we are able to achieve large non-Gaussianities from the USR period, but with a controlled enhancement to satisfy $-5 \leq |f_{\text{NL}}| \leq 5$ and $k_{e}/k_{s} \sim {\cal O}(10)$ in order to maintain the perturbativity approximation. The positions of the transition wave numbers, i.e., $k_{e}, k_{s}$, are also equally important as they control the mass of the produced PBH. It worth to recall that in the case of a standard single field inflation, production of large $M_{\text{PBH}}$ would require lower values of the transition wave numbers, i.e., $k_{s} = 10^{6}$ Mpc$^{-1}$ and $k_{e} = 10^{7}$ Mpc$^{-1}$, but that would not suffice the requirement for achieving 60 e-folds. Suppose the same wave numbers increase to $k_{s} = 10^{21}$ Mpc$^{-1}$ and $k_{e} = 10^{22}$ Mpc$^{-1}$, then inflation is possible, although at the expense of having produced small $M_{\text{PBH}}$. In the Galileon theory framework, we have demonstrated the possibility to accommodate both features satisfactorily. Fig.(\ref{totPBH}, \ref{totPBH2}) shows that we can achieve the exact number of e-folds as well as obtain the large $M_{\text{PBH}}$ through the specific positions of the transition wave numbers and the increased amount of non-Gaussianities produced. As a result of the \textit{no-go theorem} by Maldacena, the consistency condition is strictly valid in the SRI phase but gets violated in the USR and the SRII phases. The table \ref{tab:1} provides a quick comparison between the non-Gaussian amplitude observed from the individual phases, and the one-loop corrected power spectrum value, calculated in a previous work \cite{Choudhury:2023hvf}. In both cases, the same third-order action is used, so both quantities are equally important. The table also lists the critical highlights from the analysis of the results from each phase. Finally, since controlling the enhancement also requires tuning the coefficients ${\cal G}_{i}$, $\forall i=1,2,3,4$, if the transition wave numbers are shifted closer to the pivot scale, then the SRI period will be extremely small and the SRII period will increase dramatically. Still, the non-Gaussianities are controllable. However, if the same transition wave numbers are taken to much larger wave number values, then controlling the enhancement in the new SRII region would require extreme fine-tuning of the coefficients, which is not acceptable.

{\bf Acknowledgements:} SC would like to thank the work friendly environment of The Thanu Padmanabhan Centre For Cosmology and Science Popularization (CCSP), Shree Guru Gobind Singh Tricentenary (SGT) University, Gurugram, Delhi-NCR for providing tremendous support in
research and offer the Assistant Professor (Senior Grade) position. SC would like to specially thanks Soumitra SenGupta for inviting at IACS, Kolkata during the work. SC thanks Shamik Banerjee for inviting as a plenary speaker to give a talk on "Quantum loop effects on Primordial Black Hole Formation from Effective Field Theory of single field inflation" in the conference Current topics in String theory at NISER, Bhubaneswar, India. SC also thanks Misao Sasaki, Vicharit Yingcharoenrat from IPMU, Tokyo University, Japan and Rong-Gen Cai, Zong-Kuan Guo, Li Li, and Shi PSC, from the Institute of Theoretical Physics, Chinese Academy of Sciences, Beijing, China for inviting to give online seminars in their Cosmology journal clubs on the same topic. SC additionally thanks all
the members of our newly formed virtual international non-profit consortium Quantum Aspects of the SpaceTime \& Matter (QASTM) for elaborative discussions. The work of MS is supported by Science and Engineering Research Board (SERB), DST, Government of India under the Grant Agreement number CRG/2022/004120 (Core Research Grant). MS is also partially supported by the Ministry of Education and Science of the Republic of Kazakhstan, Grant
No. 0118RK00935 and CAS President's International Fellowship Initiative (PIFI). Last but not
least, we would like to acknowledge our debt to the people belonging to the various parts
of the world for their generous and steady support for research in natural sciences.

\newpage

\section{Appendix}

\subsection{Detailed computation of the bispectrum in the region I: SRI}\label{App:A}

In this subsection, we present the detailed analysis for the evaluation of the integrals required to get the tree-level contribution to the three-point correlation function. We begin by mentioning the general definition for the three-point function which will be used later, and also in the evaluation of the correlation functions for the USR and SRII regions ahead, for the calculations of contributions from specific operators.

We begin with the first operator i.e., $\displaystyle{a(\tau_{1}){\cal G}_1}\zeta^{'3}$ and use this into Eq.(\ref{a3}) where after performing the wick contractions, we write the expression as follows:
\bea \label{c1a} \langle\hat{\zeta}_{\bf k_{1}}\hat{\zeta}_{\bf k_{2}}\hat{\zeta}_{\bf k_{3}}\rangle_{\zeta^{'3}} &=& \text{2 $\times$ Im}\Bigg[\zeta_{\bold{k}_{1}}(\tau)\zeta_{\bold{k}_{2}}(\tau)\zeta_{\bold{k}_{3}}(\tau)\int\frac{d^{3}\bold{q}_{1}}{(2\pi)^3}\frac{d^{3}\bold{q}_{2}}{(2\pi)^3}\frac{d^{3}\bold{q}_{3}}{(2\pi)^3}\int^{\tau_{s}}_{-\infty}d\tau_{1}d^{3}x\displaystyle{\frac{a(\tau_{1}){\cal G}_1(\tau_1)}{H^3}\partial_{\tau}\zeta^{*}_{\bold{q}_{1}}(\tau_{1})\partial_{\tau}\zeta^{*}_{\bold{q}_{2}}(\tau_{1})\partial_{\tau}\zeta^{*}_{\bold{q}_{3}}(\tau_{1})}\nonumber\\
&& \quad\quad\quad\times \displaystyle{e^{-i(\bold{q}_{1}+\bold{q}_{2}+\bold{q}_{3}).\bold{x}}}(2\pi)^{9}\delta^{3}(\bold{k}_{1}+\bold{q}_{1})\delta^{3}(\bold{k}_{2}+\bold{q}_{2})\delta^{3}(\bold{k}_{3}+\bold{q}_{3}) + \text{1} \longleftrightarrow \text{2} + \text{1} \longleftrightarrow \text{3} \Bigg].\eea
where last two terms indicate the $2$ possible ways to contract with the modes outside the time integral. The delta function arises as a result of wick contraction between two modes of different momenta. Now, similar expressions can be written for the other interaction operators which we write here as they will also be important in many calculations:
\bea \label{c1b} \langle\hat{\zeta}_{\bold{k}_{1}}\hat{\zeta}_{\bold{k}_{2}}\hat{\zeta}_{\bold{k}_{3}}\rangle_{\zeta^{'2}\partial^{2}\zeta} &=& \text{2$\times$2$\times$Im}\Bigg[\zeta_{\bold{k}_{1}}(\tau)\zeta_{\bold{k}_{2}}(\tau)\zeta_{\bold{k}_{3}}(\tau)\int\frac{d^{3}\bold{q}_{1}}{(2\pi)^3}\frac{d^{3}\bold{q}_{2}}{(2\pi)^3}\frac{d^{3}\bold{q}_{3}}{(2\pi)^3}\int^{\tau_{s}}_{-\infty}d\tau_{1}d^{3}x\displaystyle{\frac{{\cal G}_{2}(\tau_1)}{H^3}}\partial_{\tau}\zeta^{*}_{\bold{q}_{1}}(\tau_{1})\partial_{\tau}\zeta^{*}_{\bold{q}_{2}}(\tau_{1})\zeta^{*}_{\bold{q}_{3}}(\tau_{1})\nonumber\\
&& \quad\quad\quad\times (\bold{q}^{2}_{3})\displaystyle{e^{-i(\bold{q}_{1}+\bold{q}_{2}+\bold{q}_{3}).\bold{x}}}(2\pi)^{9}\delta^{3}(\bold{k}_{1}+\bold{q}_{1})\delta^{3}(\bold{k}_{2}+\bold{q}_{2})\delta^{3}(\bold{k}_{3}+\bold{q}_{3}) + \text{1} \longleftrightarrow \text{2} + \text{1} \longleftrightarrow \text{3} \bigg].\quad\quad
\eea
\bea \label{c1c} \langle\hat{\zeta}_{\bold{k}_{1}}\hat{\zeta}_{\bold{k}_{2}}\hat{\zeta}_{\bold{k}_{3}}\rangle_{\zeta^{'}(\partial_{i}\zeta)^{2}} &=& \text{2$\times$2$\times$Im}\Bigg[\zeta_{\bold{k}_{1}}(\tau)\zeta_{\bold{k}_{2}}(\tau)\zeta_{\bold{k}_{3}}(\tau)\int\frac{d^{3}\bold{q}_{1}}{(2\pi)^3}\frac{d^{3}\bold{q}_{2}}{(2\pi)^3}\frac{d^{3}\bold{q}_{3}}{(2\pi)^3}\int^{\tau_{s}}_{-\infty}d\tau_{1}d^{3}x\displaystyle{\frac{a(\tau_1){\cal G}_{3}(\tau_1)}{H^3}\zeta^{*}_{\bold{q}_{1}}(\tau_{1})\zeta^{*}_{\bold{q}_{2}}(\tau_{1})\partial_{\tau}\zeta^{*}_{\bold{q}_{3}}(\tau_{1})}\nonumber\\
&& \quad\quad\times (\bold{q}_{1}.\bold{q}_{2})\displaystyle{e^{-i(\bold{q}_{1}+\bold{q}_{2}+\bold{q}_{3}).\bold{x}}}(2\pi)^{9}\delta^{3}(\bold{k}_{1}+\bold{q}_{1})\delta^{3}(\bold{k}_{2}+\bold{q}_{2})\delta^{3}(\bold{k}_{3}+\bold{q}_{3}) + \text{1} \longleftrightarrow \text{2} + \text{1} \longleftrightarrow \text{3} \bigg].\quad\quad
\eea
\bea \label{c1d} \langle\hat{\zeta}_{\bold{k}_{1}}\hat{\zeta}_{\bold{k}_{2}}\hat{\zeta}_{\bold{k}_{3}}\rangle_{(\partial_{i}\zeta)^{2}(\partial^{2}\zeta)} &=& \text{2$\times$Im}\Bigg[\zeta_{\bold{k}_{1}}(\tau)\zeta_{\bold{k}_{2}}(\tau)\zeta_{\bold{k}_{3}}(\tau)\int\frac{d^{3}\bold{q}_{1}}{(2\pi)^3}\frac{d^{3}\bold{q}_{2}}{(2\pi)^3}\frac{d^{3}\bold{q}_{3}}{(2\pi)^3}\int^{\tau_{s}}_{-\infty}d\tau_{1}d^{3}x\displaystyle{\frac{{\cal G}_{4}(\tau_1)}{H^3}\zeta^{*}_{\bold{q}_{1}}(\tau_{1})\zeta^{*}_{\bold{q}_{2}}(\tau_{1})\zeta^{*}_{\bold{q}_{3}}(\tau_{1})}\nonumber\\
&& \times((\bold{q}_{1}.\bold{q}_{2})\bold{q}^{2}_{3})\displaystyle{e^{-i(\bold{q}_{1}+\bold{q}_{2}+\bold{q}_{3}).\bold{x}}}(2\pi)^{9}\delta^{3}(\bold{k}_{1}+\bold{q}_{1})\delta^{3}(\bold{k}_{2}+\bold{q}_{2})\delta^{3}(\bold{k}_{3}+\bold{q}_{3}) + \text{1} \longleftrightarrow \text{2} + \text{1} \longleftrightarrow \text{3} \bigg].\quad\quad
\eea
where the extra factor of $2$ multiplied ahead of the expressions represents the identical terms coming from permutations while contracting with either two $\partial_{\tau}\zeta_{\bold{k}}^{*}$ or $\zeta^{*}_{\bold{k}}$ terms. Let us also mention the time derivative of the mode in SRI region:
\bea \Pi_{\bold{k}} = \partial_{\tau}\zeta_{\bold{k}} = \left(\frac{iH^{2}}{2\sqrt{\cal{A}}}\right)\frac{1}{(c_{s}k)^{3/2}}(k^{2}c_{s}^{2}\tau)e^{-ikc_{s}\tau}.
\eea
Now using these information, we can explicitly write the contributions to three-point correlation function coming from each interaction operator. For \underline{\textbf{Operator 1}: $a(\tau_{1})\zeta^{'3}$}, the formula from (\ref{c1a}) gives us:
\bea \langle\hat{\zeta}_{\bold{k}_{1}}\hat{\zeta}_{\bold{k}_{2}}\hat{\zeta}_{\bold{k}_{3}}\rangle_{\zeta^{'3}} &=& \text{2 $\times$ Im}\Bigg[\frac{H^{12}}{(4{\cal A})^{3}}\zeta_{\bold{k}_{1}}(\tau)\zeta_{\bold{k}_{2}}(\tau)\zeta_{\bold{k}_{3}}(\tau)\int^{\tau_{s}}_{-\infty}d\tau_{1}\displaystyle{\frac{a(\tau_{1}){\cal G}_1(\tau_1)}{H^3 c_{s}^{3}}k_{1}^{2}k_{2}^{2}k_{3}^2\tau_{1}^{3}\exp{(ic_{s}(k_{1}+k_{2}+k_{3})\tau_{1})}}\nonumber\\
&& \quad\quad\quad \times (2\pi)^{3}\delta^{3}(\bold{k}_{1}+\bold{k}_{2}+\bold{k}_{3}) + \text{1} \longleftrightarrow \text{2} + \text{1} \longleftrightarrow \text{3} \Bigg]_{\tau \rightarrow 0}.
\eea
here the integral over the momentum-delta functions is performed and the following property is used:
\be \int\d^{3}x e^{-i(\bold{k}_{1}+\bold{k}_{2}+\bold{k}_{3}).\bold{x}} = (2\pi)^{3}\delta^{3}(\bold{k}_{1}+\bold{k}_{2}+\bold{k}_{3}).\ee 
The evaluation of the correlation is performed in the limit $\tau \rightarrow 0$ while using the relation $\tau_{s} = -1/(k_{s}c_{s})$, $\tau=-1/(a(\tau)H)$ and the shorthand $K=k_{1}+k_{2}+k_{3}$, where $k_{i} = |\bold{k}_{i}|$ $\forall i=1,2,3$. We also use the following shorthand by absorbing the momentum conserving delta function in this manner:
\begin{align*}
    \langle\hat{\zeta}_{\bold{k}_{1}}\hat{\zeta}_{\bold{k}_{2}}\hat{\zeta}_{\bold{k}_{3}}\rangle = (2\pi)^{3}\delta{(\bold{K})}\langle\hat{\zeta}_{\bold{k}_{1}}\hat{\zeta}_{\bold{k}_{2}}\hat{\zeta}_{\bold{k}_{3}}\rangle^{'}
\end{align*}
where $\bold{K} = \bold{k}_{1}+\bold{k}_{2}+\bold{k}_{3}$ and $\Tilde{K} = |\bold{K}| = \sqrt{k_{1}^{2} + k_{2}^{2} + k_{3}^{2}}$.

When the aforesaid substitutions are taken care of, it gives us the following integral which evaluates to: 
\bea \label{c1Ia} \displaystyle{\int_{-\infty}^{\tau_{s}}d\tau_{1}k_{1}^{2}k_{2}^{2}k_{3}^2\tau_{1}^{2}c_{s}^{-3}\exp{(ic_{s}K\tau_{1})} = \frac{-ik_{1}^2 k_{2}^2 k_{3}^2}{c_{s}^{6} K^3} \left(\frac{K^2}{k_{s}^2}-\frac{2i K}{k_{s}}-2\right) \exp{\left(-i\frac{K}{k_{s}}\right)}}.\eea
After substitution of this result and taking only the imaginary part coming from this integral, we get the final result as:
\bea \label{c1r1} \langle\hat{\zeta}_{\bold{k}_{1}}\hat{\zeta}_{\bold{k}_{2}}\hat{\zeta}_{\bold{k}_{3}}\rangle_{\zeta^{'3}}^{'} =  \frac{H^{12}}{(4{\cal A})^{3}}\frac{{\cal G}_{1}}{H^4}\frac{6}{(k_{1}^{3}k_{2}^{3}k_{3}^{3})}\frac{k_{1}^2 k_{2}^2 k_{3}^2}{c_{s}^{6} K^3}\left\{\left(\frac{K^{2}}{k_{s}^{2}}-2\right)\cos\left(\frac{K}{k_{s}}\right)-2\frac{K}{k_{s}}\sin\left(\frac{K}{k_{s}}\right)\right\}.
\eea
This gives us the contribution from the first interaction to the three-point correlation function in the SRI region. Similar analysis including the substitutions can be carried out to evaluate the contributions from the other operators. 

For \underline{\textbf{Operator 2}: $\zeta^{'2}(\partial^{2}\zeta)$}, the formula from (\ref{c1b}) gives us the following result after taking care of all the permutations in the momentum variables:
\bea \label{c1r2} \langle\hat{\zeta}_{\bold{k}_{1}}\hat{\zeta}_{\bold{k}_{2}}\hat{\zeta}_{\bold{k}_{3}}\rangle_{\zeta^{'2}(\partial^{2}\zeta)}^{'} = \frac{H^{12}}{(4{\cal A})^{3}}\frac{{\cal G}_{2}}{H^3}\frac{4}{(k_{1}^{3}k_{2}^{3}k_{3}^{3})}\displaystyle{\frac{k_{1}^2 k_{2}^2 k_{3}^2}{c_{s}^{8} K^3}\left\{\left(12-6\frac{K^{2}}{k_{s}^2}\right)\cos\left(\frac{K}{k_{s}}\right) - \left(\frac{K^{3}}{k_{s}^3}-12\frac{K}{k_{s}}\right)\sin\left(\frac{K}{k_{s}}\right)\right\}}.\eea
For \underline{\textbf{Operator 3}: $a(\tau_{1})\zeta^{'}(\partial_{i}\zeta)^{2}$}, the formula from (\ref{c1c}) gives us the following result after taking care of all the permutations in the momentum variables:
\bea \label{c1r3} \langle\hat{\zeta}_{\bold{k}_{1}}\hat{\zeta}_{\bold{k}_{2}}\hat{\zeta}_{\bold{k}_{3}}\rangle_{\zeta^{'}(\partial_{i}\zeta)^{2}}^{'} &=& \frac{H^{12}}{(4{\cal A})^{3}}\frac{{\cal G}_{3}}{H^4}\frac{4}{k_{1}^{3}k_{2}^{3}k_{3}^{3}}
\frac{k_{1}k_{2}k_{3}}{c_{s}^{8}K^{3}k_{s}^{2}}\Bigg\{\bigg(16 k_2 k_3 k_1 K+ 2 k_1^3\left(k_2+k_3\right) +2k_3^3\left(k_1+k_2\right)+ 2k_2^3\left(k_1+k_3\right)\nonumber\\
   && + 2\left(k_1^2 k_2^2+k_3^2 k_2^2+k_1^2 k_3^2\right)\bigg) k_s \sin
   \left(\frac{K}{k_s}\right)-3 k_1 k_2 k_3 K^{2} \cos
   \left(\frac{K}{k_s}\right)+\bigg(k_1^3+ 5\left(k_2+k_3\right) k_1^2 \nonumber\\
   && +18 k_2 k_3 k_1+k_2^3+k_3^3+5
   \left(k_1+k_2\right) k_3^2+5 k_2^2 \left(k_1+k_3\right)\bigg) k_s^2 \cos
   \left(\frac{K}{k_s}\right)\Bigg\}. \eea

For \underline{\textbf{Operator 4}: $\partial^{2}\zeta(\partial_{i}\zeta)^{2}$}, the formula from (\ref{c1d}) gives us the following result after taking care of all the permutations in the momentum variables:
\bea && \label{c1r4}\langle\hat{\zeta}_{\bold{k}_{1}}\hat{\zeta}_{\bold{k}_{2}}\hat{\zeta}_{\bold{k}_{3}}\rangle_{\partial^{2}\zeta(\partial_{i}\zeta)^{2}}^{'} = \frac{H^{12}}{(4{\cal A})^{3}}\frac{{\cal G}_{4}}{H^3}\frac{6}{k_{1}^{3}k_{2}^{3}k_{3}^{3}}
\frac{k_{1}k_{2}k_{3}}{c_{s}^{10}K^{3}}\left\{k_{1}k_{2}k_{3}\frac{K^{3}}{k_{s}^{3}}\left(\sin{\frac{K}{k_s}}\right)+\frac{K^{2}}{k_{s}^{2}}\bigg(6k_{1}k_{2}k_{3} + k_{1}^{2}(k_{2}+k_{3})+k_{2}^{2}(k_{3}+k_{1})\right.\nonumber\\
&&\left. \quad\quad\quad\quad\quad\quad\quad\quad\quad +k_{3}^{2}(k_{2}+k_{1})\bigg)\left(\cos{\frac{K}{k_s}}\right) -\frac{1}{k_{s}}\bigg(k_{1}^{4}+ k_{2}^{4}+k_{3}^{4}+6k_{1}^{3}(k_{2}+k_{3})+6k_{2}^{3}(k_{1}+k_{3}) + 6k_{3}^{3}(k_{1}+k_{2})\right. \nonumber\\
&& \left. \quad\quad\quad\quad\quad\quad\quad\quad\quad +28k_{1}k_{2}k_{3}K+ 10k_{1}^{2}k_{2}^{2}+8k_{2}^{2}k_{3}^{2}+10k_{1}^{2}k_{3}^{2} \bigg)\left(\sin{\frac{K}{k_{s}}}\right)-2\bigg(k_{1}^{3}+k_{2}^{3}+k_{3}^{3}+ 12k_{1}k_{2}k_{3} + 4k_{1}^{2}(k_{2}+k_{3}) \right.\nonumber\\
&& \left. \quad\quad\quad\quad\quad\quad\quad\quad\quad + 4k_{2}^{2}(k_{3}+k_{1}) +4k_{3}^{2}(k_{1}+k_{2})\bigg)\left(\cos{\frac{K}{k_{s}}}\right) \right\}.
\eea
As it can be seen, we have collected all the contributions coming from the interaction operators towards the tree-level, three-point correlation function or the bispectrum in the SRI region.
To get the total bispectrum contribution for the SRI region, we must add the contributions coming from the  Eqs.(\ref{c1r1},\ref{c1r2},\ref{c1r3},\ref{c1r4}). 

\subsection{Detailed computation of the bispectrum in the region II: USR}\label{App:B}

In this subsection, we present the explicit expressions for the set of momentum-dependent functions in terms of which we have mentioned our results for the tree-level calculation of the bispectrum in the USR region. 

Before explaining the calculations it will be convenient to mention the time-derivative of the mode function explicitly in the USR region:
\bea \Pi^{*}_{\bold{k}} = \partial_{\tau}\zeta^{*}_{\bold{k}} = \left(\frac{\tau_{s}}{\tau}\right)^{9}\left\{\alpha_{\bold{k}}^{(2)*}\left(k^{2}c_{s}^{2}\tau-\frac{3}{\tau}(1-ikc_{s}\tau)\right)\exp{(ikc_{s}\tau)}+\beta_{\bold{k}}^{(2)*}\left(\frac{3}{\tau}(1+ikc_{s}\tau)-k^{2}c_{s}^{2}\tau\right)\exp{(-ikc_{s}\tau)}\right\}.
\eea where we have written the complex conjugate of the time-derivative of the curvature perturbation and $k = |\bold{k}|$. The factor $(-iH^{2})/(2\sqrt{{\cal A}})$ is also present in the front but omitted here for the sake of clarity.

The time-derivatives in some operators drastically increases the number of integrals to evaluate and would lead to different kinds of integrals depending upon the factors involved. However, there will be a limited kind of these integrals and the rest will differ in a particular way. All of their results are written in terms of the functions described in this section. We have categorized the different set of functions based on their corresponding kind of integral. Also, we have considered scaling all he expressions with the wave number $k_{e}$ and consequently adjusted for the expressions for having the necessary correct dimensions. This has made the physical impact of the functions much clearer to interpret based on the powers of the wave numbers.

In the upcoming results, we have defined a few variables as follows: $\displaystyle{x_{e,i} = {\cal K}_{i}/{k_{e}}}$, $\displaystyle{x_{s.i} = {\cal K}_{i}/{k_{s}}}$, and ${\cal K}_i$ where $i = 1,2,3,4$, is basically any of the following linear combination of the magnitude of the three momenta $k_1,k_2,k_3$,
\begin{align} \label{c2k}
    {\cal K}_1 = k_1+k_2+k_3, \quad{\cal K}_2 = k_1-k_2+k_3, \quad{\cal K}_3 = k_1+k_2-k_3, \quad{\cal K}_4 = -k_1+k_2+k_3.
\end{align} 
The total contribution from all the integrals in the USR region are explicitly taken together in the results of section \ref{s5c2} and we only mention specific number of contributions and explain how each is related to the rest.

For \underline{\textbf{Operator 1}: $a(\tau_{1})\zeta^{'3}$}, we use the formula in Eq.(\ref{c1a}) to work out the contribution to the bispectrum. We have the following kinds of integrals for this operator:
\begin{itemize}
    \item \underline{\textbf{Integral of the first kind}}:
    For integral of the first kind, the corresponding set of functions are as follows:
    \begin{subequations}
    \begin{align}
    {\cal P}_{1,1}({\cal K}_{i}, k_{1},k_{2},k_{3}) &= \frac{-11{\cal K}_{i}^3}{k_e^3}+\frac{132 {\cal K}_{i} \left(k_2 k_3+k_1 k_2+k_1 k_3\right)}{k_e^3}-\frac{1320 k_1 k_2 k_3}{k_e^3}.\\
    {\cal Q}_{1,1}({\cal K}_{i}) &= \frac{i{\cal K}_{i}^8}{k_e^8}-\frac{{\cal K}_{i}^7}{k_e^7} -\frac{2 i{\cal K}_{i}^6}{k_e^6} + \frac{6 {\cal K}_{i}^5}{k_e^5} + \frac{24 i {\cal K}_{i}^4}{k_e^4} - \frac{120 {\cal K}_{i}^3}{k_e^3} - \frac{720 i {\cal K}_{i}^2}{k_e^{2}} + \frac{5040 {\cal K}_{i}}{k_e} + 40320 i.\\
    {\cal R}_{1,1}({\cal K}_{i}) &= k_e^3 \bigg\{-\frac{{\cal K}_{i}^9 \exp{(ix_{s,i})} \left(\text{Ei}\left(-ix_{e,i}\right)-\text{Ei}\left(-ix_{s,i}\right)\right)}{k_s^{12}}-\frac{i
    {\cal K}_{i}^8}{k_s^{11}}+\frac{{\cal K}_{i}^7}{k_s^{10}}+\frac{2 i{\cal K}_{i}^6}{k_s^9}-\frac{6 {\cal K}_{i}^5}{k_s^8}-\frac{24 i
    {\cal K}_{i}^4}{k_s^7}\notag\\
    & \quad\quad\quad\quad +\frac{120 {\cal K}_{i}^3}{k_s^6} 
    +\frac{720 i {\cal K}_{i}^2}{k_s^5}-\frac{5040
    {\cal K}_{i}}{k_s^4}-\frac{40320 i}{k_s^3}\bigg\}.
    \end{align} 
    \label{c211f}
    \end{subequations}
    The result of this integral is expressed using the above functions in the following form:
    \bea \label{c211aI} (\bold{I_{1}})_{j}({\cal K}_i,k_1,k_2,k_3) &=& -\frac{k_e^{12} \exp{(-i x_{e,i})}}{17740800 c_{s}^6 k_s^9}\bigg\{{\cal P}_{1,1}({\cal K}_i,k_1,k_2,k_3){\cal Q}_{1,1}({\cal K}_i) + 39916800 \nonumber\\
   && -\frac{362880 \left(132 \left(k_2 k_3+k_1 k_2+k_1 k_3\right)-11 {\cal K}_i^{2}\right)}{k_e^2} + i\frac{39916800
   {\cal K}_i}{k_e}  \nonumber\\
   && + \frac{k_s^{12}}{k_e^{12}}\exp{(i(x_{e,i}-x_{s,i}))}\bigg({\cal P}_{1,1}({\cal K}_i){\cal R}_{1,1}({\cal K}_i)-39916800 -\frac{39916800 i {\cal K}_{i}}{k_s} \nonumber\\
   && + \frac{362880 \left(132 \left(k_2 k_3+k_1 k_2+k_1 k_3\right)-11
   {\cal K}_{i}^{2}\right)}{k_s^2} \bigg) \bigg\}
    \eea
    \begin{enumerate}
        \item Consider \textbf{Integral 1a}: \bea \label{c211a} \displaystyle{\int_{\tau_{s}}^{\tau_{e}}d\tau_{1}\frac{-27\tau_{s}^{9}}{c_s^{9}\tau_{1}^{13}}\exp{(ic_{s}{\cal K}_{1}\tau_{1})}}(1-ik_{1}c_{s}\tau_{1})(1-ik_{2}c_{s}\tau_{1})(1-ik_{3}c_{s}\tau_{1}) = (\bold{I_{1}})_{1}({\cal K}_1,k_1,k_2,k_3).
    \eea 
The result for this integral is written using the general formula as described above. 
Other integrals of the similar kind, for different ${\cal K}_i$, also give results in the form of the above equation. The signature of the momentum variables in the integral corresponds to those present in the factor ${\cal K}_{i}$. For this, consider another example.
    \item For a  
    similar \textbf{Integral 1b}: \begin{align*}\displaystyle{\int_{\tau_{s}}^{\tau_{e}}d\tau_{1}\frac{27\tau_{s}^{9}}{c_{s}^{9}\tau_{1}^{13}}\exp{(ic_{s}{\cal K}_{2}\tau_{1})}}(1-ik_{1}c_{s}\tau_{1})(1+ik_{2}c_{s}\tau_{1})(1-ik_{3}c_{s}\tau_{1}) = -(\bold{I_{1}})_{2}({\cal K}_2,k_1,-k_2,k_3). 
    \end{align*}  
    \item For \textbf{Integral 1c}:
    \bea \label{c211b} \displaystyle{\int_{\tau_{s}}^{\tau_{e}}d\tau_{1}\frac{-27\tau_{s}^{9}}{c_{s}^{9}\tau_{1}^{13}}\exp{(-ic_{s}{\cal K}_{3}\tau_{1})}}(1+ik_{1}c_{s}\tau_{1})(1+ik_{2}c_{s}\tau_{1})(1-ik_{3}c_{s}\tau_{1}) = (\bold{I_{1}})_{3}^{*}({\cal K}_{3},k_1,k_2,-k_3).
    \eea
The notation of keeping maximum number of positive momentum variables inside the factor ${\cal K}_{i}$ is used. This will give us integrals with negative exponential factors. The result for the above integral is written in terms of the complex conjugate of the general result obtained through the formula in Eq.(\ref{c211aI}). The momentum signature in the argument of RHS is similarly related to those in factor ${\cal K}_{i}$.
    \end{enumerate}
There will be a total of $4$ integrals of first kind, each containing a positive and negative exponential.
    \item \underline{\textbf{Integral of the second kind}}:
    For integral of the second kind, the corresponding set of functions are as follows:
    \begin{subequations}
    \begin{align} 
    {\cal P}_{1,2}({\cal K}_{i}, k_{a},k_{b}) &= \frac{{\cal K}_{i}^2}{k_e^2}+\frac{10 {\cal K}_{i} \left(k_a + k_b\right)}{k_e^2}+\frac{90 k_a k_b}{k_e^2}.\\
    {\cal Q}_{1,2}({\cal K}_{i}) &= -\frac{i{\cal K}_{i}^7}{k_e^7} +\frac{{\cal K}_{i}^6}{k_e^6} + \frac{2i {\cal K}_{i}^5}{k_e^5} - \frac{6{\cal K}_{i}^4}{k_e^4} - \frac{24i{\cal K}_{i}^3}{k_e^3} + \frac{120 {\cal K}_{i}^2}{k_e^{2}} +  \frac{720 i{\cal K}_{i}}{k_e} -5040.\\
    {\cal R}_{1,2}({\cal K}_{i}) &= k_e^2 \bigg\{ \frac{k^8 (\exp{ix_{s,i}}) \left(\text{Ei}\left(-ix_{e,i}\right)-\text{Ei}\left(-ix_{s,i}\right)\right)}{k_s^{10}}+\frac{i
   {\cal K}_{i}^7}{k_s^9}-\frac{k^6}{k_s^8}-\frac{2 i {\cal K}_{i}^5}{k_s^7}+\frac{6 {\cal K}_{i}^4}{k_s^6}+\frac{24 i {\cal K}_{i}^3}{k_s^5}-\frac{120
   {\cal K}_{i}^2}{k_s^4}\notag\\
   &-\frac{720 i {\cal K}_{i}}{k_s^3}+\frac{5040}{k_s^2}\bigg\}.
    \end{align}
    \label{c22f}
    \end{subequations}
     where $a,b$ belongs to $1,2,3$. 
    The result of this integral is expressed using the above functions in the following form:
    \bea \label{c212aI} (\bold{I_{2}})_{j}({\cal K}_i,k_a,k_b) &=& \frac{k_e^{10} \exp{(-i x_{e,i})}}{403200 c_{s}^6 k_s^9}\bigg\{{\cal P}_{1,2}({\cal K}_i,k_a,k_b){\cal Q}_{1,2}({\cal K}_i) -\frac{40320 i \left(-10\left(k_a+k_b\right) + {\cal K}_i\right)}{k_e} \nonumber\\
    && + 362880 + \frac{k_s^{10}}{k_e^{10}}\exp{(i(x_{e,i}-x_{s,i}))}\bigg({\cal P}_{1,2}({\cal K}_i){\cal R}_{1,2}({\cal K}_i)-362880 \nonumber\\
    && +\frac{40320 i\left(-10(k_a+k_b) + {\cal K}_{i}\right)}{k_s}\bigg) \bigg\}.
    \eea
    \begin{enumerate} 
    \item Consider \textbf{Integral 2a}: \bea \label{c212a} \displaystyle{\int_{\tau_{s}}^{\tau_{e}}d\tau_{1}\frac{9\tau_{s}^{9}}{\tau_{1}^{12}c_{s}^{9}}\exp{(ic_{s}{\cal K}_{1}\tau_{1})}}(1-ik_{1}c_{s}\tau_{1})(k_{3}^{2}c_{s}^{2}\tau_{1})(1-ik_{2}c_{s}\tau_{1}) = k_{3}^{2}(\bold{I_{2}})_{1}({\cal K}_1,-k_1,-k_2).
        \eea
where ${\cal K}_i = {\cal K}_1$, $k_a = -k_1$ and $k_b = -k_3$ comes from the fact that, inside the integral, both $k_1,k_3$ have negative signature. The result follows from the general formula in the equation above. We mention another example of this.
    \item For a similar \textbf{Integral 2b}:
    \begin{align*} \displaystyle{\int_{\tau_{s}}^{\tau_{e}}d\tau_{1}\frac{-9\tau_{s}^{9}}{\tau_{1}^{12}c_{s}^{9}}\exp{(ic_{s}{\cal K}_{2}\tau_{1})}}(k_{1}^{2}c_{s}^{2}\tau_{1})(1+ik_{2}c_{s}\tau_{1})(1-ik_{3}c_{s}\tau_{1}) = -k_{1}^{2}(\bold{I_{2}})_{5}({\cal K}_2,k_2,-k_3).
    \end{align*} 
     where $k_a = k_2$, $k_b=-k_3$ depend on their signature inside the integral.
    \item For \textbf{Integral 2c}
    \bea  \displaystyle{\int_{\tau_{s}}^{\tau_{e}}d\tau_{1}\frac{9\tau_{s}^{9}}{\tau_{1}^{12}c_{s}^{9}}\exp{(-ic_{s}{\cal K}_{3}\tau_{1})}}(k_{1}^{2}c_{s}^{2}\tau_{1})(1+ik_{2}c_{s}\tau_{1})(1-ik_{3}c_{s}\tau_{1}) = k_{1}^{2}(\bold{I_{2}})^{*}_{9}({\cal K}_3,k_2,-k_3). \eea
The negative exponential integrals is related to the formula in Eq.(\ref{c212aI}) by a complex conjugate as shown for the integral above.
\end{enumerate}
There will be a total of $12$ integrals of the second kind, each containing a positive and negative exponential. 
    \item \underline{\textbf{Integral of the third kind}}:
    For integral of the third kind, the corresponding set of functions are as follows:
    \begin{subequations}
    \begin{align} 
    {\cal P}_{1,3}({\cal K}_{i}, k_{a}) &= \frac{{\cal K}_{i}}{k_e}+\frac{8 k_a}{k_e}.\\
    {\cal Q}_{1,3}({\cal K}_{i}) &= \frac{i{\cal K}^6}{k_e^6} - \frac{{\cal K}_{i}^5}{k_e^5} - \frac{2i{\cal K}_{i}^4}{k_e^4} + \frac{6{\cal K}_{i}^3}{k_e^3} + \frac{24i{\cal K}_{i}^2}{k_e^2}  - \frac{120{\cal K}_{i}}{k_e} - 720 i.\\
    {\cal R}_{1,3}({\cal K}_{i}) &= k_e \bigg\{ \frac{{\cal K}_{i}^7 \exp{(ix_{s,i})} \left(\text{Ei}\left(-ix_{e,i}\right)-\text{Ei}\left(-ix_{s,i}\right)\right)}{k_s^{8}}+\frac{i
   {\cal K}_{i}^6 }{k_s^7}-\frac{{\cal K}_{i}^5}{k_s^6}-\frac{2 i {\cal K}_{i}^4}{k_s^5}+\frac{6{\cal K}_{i}^3}{k_s^4}+\frac{24 i{\cal K}_{i}^2}{k_s^3}\notag\\
   &-\frac{120{\cal K}_{i}}{k_{s}^{2}} -\frac{720 i}{k_s} \bigg\}.
    \end{align}
    \label{c23f}
    \end{subequations}
     where $a$ belongs to $1,2,3$. The result of this integral is expressed using the above functions in the following form:
    \bea \label{c213aI} (\bold{I_{3}})_{j}({\cal K}_i,k_a) &=& -\frac{k_e^{8} \exp{(-i x_{e,i})}}{13440 c_{s}^6 k_s^9}\bigg\{{\cal P}_{1,3}({\cal K}_i,k_a){\cal Q}_{1,3}({\cal K}_i) + 5040 \nonumber\\
   && - \frac{k_s^{8}}{k_e^{8}}\exp{(i(x_{e,i}-x_{s,i}))}\bigg({\cal P}_{1,3}({\cal K}_1){\cal R}_{1,3}({\cal K}_i)+5040\bigg)\bigg\}.
    \eea 
\begin{enumerate} 
    \item Consider \textbf{Integral 3a}: \bea \label{c213a} \displaystyle{\int_{\tau_{s}}^{\tau_{e}}d\tau_{1}\frac{-3\tau_{s}^{9}}{\tau_{1}^{11}c_{s}^{9}}\exp{(ic_{s}{\cal K}_{1}\tau_{1})}}(k_{2}^{2}c_{s}^{2}\tau_{1})(k_{3}^{2}c_{s}^{2}\tau_{1})(1-ik_{1}c_{s}\tau_{1}) = k_{2}^{2}k_{3}^{2}(\bold{I_{3}})_{1}({\cal K}_1,-k_1).
    \eea
    In the result above, $k_a = -k_3$ is based upon the signature of the momentum variable inside the integral. The result follows from the general formula in Eq.(\ref{c213aI}). Consider another example of this.
    \item For a similar \textbf{Integral 3b}:
     \begin{align*} \displaystyle{\int_{\tau_{s}}^{\tau_{e}}d\tau_{1}\frac{3\tau_{s}^{9}}{\tau_{1}^{11}c_{s}^{9}}\exp{(ic_{s}{\cal K}_{2}\tau_{1})}}(k_{1}^{2}c_{s}^{2}\tau_{1})(k_{3}^{2}c_{s}^{2}\tau_{1})(1+ik_{2}c_{s}\tau_{1}) =  -k_1^{2}k_{3}^{2}(\bold{I_{3}})_{6}({\cal K}_2,k_2).
    \end{align*}
    \item For \textbf{Integral 3c}:
     \bea \label{c213c} \displaystyle{\int_{\tau_{s}}^{\tau_{e}}d\tau_{1}\frac{-3\tau_{s}^{9}}{\tau_{1}^{11}c_{s}^{9}}\exp{(-ic_{s}{\cal K}_{3}\tau_{1})}}(k_{1}^{2}c_{s}^{2}\tau_{1})(k_{2}^{2}c_{s}^{2}\tau_{1})(1+ik_{3}c_{s}\tau_{1}) = k_1^{2}k_{2}^{2}(\bold{I_{3}})^{*}_{9}({\cal K}_3,k_3).
    \eea
    The negative exponential integrals is related to the formula in Eq.(\ref{c213aI}) by a complex conjugate as shown for the integral above.
\end{enumerate} 
There will be a total of $12$ such integrals of the third kind, each with a positive and negative exponential factor.
    \item \underline{\textbf{Integral of the fourth kind}}:
     For the integral of the fourth kind, the corresponding set of functions are as follows:
    \begin{subequations}
    \begin{align} 
    {\cal Q}_{1,4}({\cal K}_{i}) &= \frac{i{\cal K}^5}{k_e^5} - \frac{{\cal K}_{i}^4}{k_e^4} - \frac{2i{\cal K}_{i}^3}{k_e^3} + \frac{6{\cal K}_{i}^2}{k_e^2} + i\frac{24{\cal K}_{i}}{k_e} - 120.\\
    {\cal R}_{1,4}({\cal K}_{i}) &= \bigg\{ \frac{{\cal K}_{i}^6 \exp{(ix_{s,i})} \left(\text{Ei}\left(-ix_{e,i}\right)-\text{Ei}\left(-ix_{s,i}\right)\right)}{k_s^{6}}-\frac{i
   {\cal K}_{i}^5}{k_s^5}+\frac{{\cal K}_{i}^4}{k_s^4}+\frac{2 i {\cal K}_{i}^3}{k_s^3}-\frac{6{\cal K}_{i}^2}{k_s^2}-\frac{24 i{\cal K}_{i}}{k_s} +120\bigg\}.
    \end{align}
    \label{c24f}
    \end{subequations}
    The result of this integral is expressed using the above functions in the following form: 
    \bea \label{c214aI} (\bold{I_{4}})_{j}({\cal K}_i) &=& -\frac{k_e^{6} \exp{(-i x_{e,i})}}{720c_{s}^6 k_s^9}\bigg\{{\cal Q}_{1,4}({\cal K}_i) + \frac{k_s^{6}}{k_e^{6}}\exp{(i(x_{e,i}-x_{s,i}))}{\cal R}_{1,4}({\cal K}_i)\bigg\}.
    \eea   
    \begin{enumerate}
    \item Consider \textbf{Integral 4a}:
    \bea \label{c214a} \displaystyle{\int_{\tau_{s}}^{\tau_{e}}d\tau_{1}\frac{\tau_{s}^{9}}{c_{s}^{3}\tau_{1}^{7}}k_{1}^{2}k_{2}^{2}k_{3}^2\exp{(ic_{s}{\cal K}_{1}\tau_{1})}} = k_{1}^{2}k_{2}^{2}k_{3}^{2}(\bold{I_{4}})_{1}({\cal K}_1).
    \eea
     The argument depends only in the factor ${\cal K}_{i}$ for each such integral.  The result follows from the general formula in Eq.(\ref{c214aI}). 
    \item For \textbf{Integral 4b}:
    \bea \label{c214b} \displaystyle{\int_{\tau_{s}}^{\tau_{e}}d\tau_{1}\frac{\tau_{s}^{9}}{c_{s}^{3}\tau_{1}^{7}}k_{1}^{2}k_{2}^{2}k_{3}^{2}\exp{(-ic_{s}{\cal K}_{3}\tau_{1})}} = k_{1}^{2}k_{2}^{2}k_{3}^{2}(\bold{I_{4}})^{*}_{3}({\cal K}_3).
    \eea
     The negative exponential integrals is related to the formula in Eq.(\ref{c214aI}) by a complex conjugate as shown for the integral above.
    \end{enumerate}
    There will be a total of $4$ integrals of the fourth kind, each with a positive and negative exponential factor.
\end{itemize}

For \underline{\textbf{Operator 2}: $\displaystyle{\zeta^{'2}\partial^{2}\zeta}$}, we use the formula in Eq.(\ref{c1b}) to work out the contribution to the bispectrum. We have the following types of integrals for this operator:
\begin{itemize}
    \item \underline{\textbf{Integral of the first kind}}:
    For integral of the first kind, the corresponding set of functions are as follows:
    \begin{subequations}
    \begin{align}
    {\cal P}_{2,1}({\cal K}_{i}, k_{1},k_{2},k_{3}) &= \frac{-9{\cal K}_{i}^3}{k_e^3}+\frac{90 {\cal K}_{i} \left(k_2 k_3+k_1 k_2+k_1 k_3\right)}{k_e^3}-\frac{720 k_1 k_2 k_3}{k_e^3}.\\
    {\cal Q}_{2,1}({\cal K}_{i}) &= \frac{ -i{\cal K}_{i}^6}{k_e^6}+\frac{{\cal K}_{i}^5}{k_e^5} +\frac{2 i{\cal K}_{i}^4}{k_e^4} - \frac{6{\cal K}_{i}^3}{k_e^3} - \frac{24 i{\cal K}_{i}^2}{k_e^2} + \frac{120{\cal K}_{i}}{k_e} + 720i.\\
    {\cal R}_{2,1}({\cal K}_{i}) &= k_e^3 \bigg\{\frac{{\cal K}_{i}^7 \exp{(ix_{s,i})} \left(\text{Ei}\left(-ix_{e,i}\right)-\text{Ei}\left(-ix_{s,i}\right)\right)}{k_s^{10}}+\frac{i
    {\cal K}_{i}^6}{k_s^{9}}-\frac{{\cal K}_{i}^5}{k_s^{8}}-\frac{2i{\cal K}_{i}^4}{k_s^7}+\frac{6{\cal K}_{i}^3}{k_s^6}+\frac{24i
    {\cal K}_{i}^2}{k_s^5}-\frac{120{\cal K}_{i}}{k_s^4}\notag\\
    &-\frac{720 i}{k_s^3}\bigg\}.
    \end{align} 
    \label{c221f}
    \end{subequations}
    The result of this integral is expressed using the above functions in the following form:
    \bea \label{c221aI} (\bold{I_{1}})_{j}({\cal K}_i,k_1,k_2,k_3) &=& \frac{k_e^{10} \exp{(-i x_{e,i})}}{403200 c_{s}^8 k_s^9}\bigg\{{\cal P}_{2,1}({\cal K}_i,k_1,k_2,k_3){\cal Q}_{2,1}({\cal K}_i) + i 362880\frac{{\cal K}_{i}}{k_e} \nonumber\\
    && -\frac{5040 \left(90\left(k_2 k_3+k_1 k_2+k_1 k_3\right)-9 {\cal K}_i^{2}\right)}{k_e^2} + 362880 \nonumber\\
    && + \frac{k_s^{10}}{k_e^{10}}\exp{(i(x_{e,i}-x_{s,i}))}\bigg({\cal P}_{2,1}({\cal K}_i){\cal R}_{2,1}({\cal K}_i)-362880 -i362880\frac{{\cal K}_i}{k_s} \nonumber\\
    && + \frac{5040\left(90\left(k_2 k_3+k_1 k_2+k_1 k_3\right)-9 {\cal K}_i^{2}\right)}{k_e^2} \bigg) \bigg\}.
    \eea  
    \begin{enumerate}
        \item Consider \textbf{Integral 1a}:
        \bea \label{c221a} \displaystyle{\int_{\tau_{s}}^{\tau_{e}}d\tau_{1}\frac{9\tau_{s}^{9}}{\tau_{1}^{11}c_{s}^{9}}\exp{(ic_{s}{\cal K}_{1}\tau_{1})}}(1-ik_{1}c_{s}\tau_{1})(1-ik_{2}c_{s}\tau_{1})(1-ik_{3}c_{s}\tau_{1}) = (\bold{I_{1}})_{1}({\cal K}_1,k_1,k_2,k_3).
    \eea 
    The signature of the momentum arguments depend on their respective signature in ${\cal K}_{i}$ for each such integral. The result is expressed using the formula in Eq.(\ref{c221aI}).
    \item For \textbf{Integral 1b}:
    \bea \label{c221b} \displaystyle{\int_{\tau_{s}}^{\tau_{e}}d\tau_{1}\frac{9\tau_{s}^{9}}{\tau_{1}^{11}c_{s}^{9}}\exp{(-ic_{s}{\cal K}_{3}\tau_{1})}}(1+ik_{1}c_{s}\tau_{1})(1+ik_{2}c_{s}\tau_{1})(1-ik_{3}c_{s}\tau_{1}) = (\bold{I_{1}})^{*}_{3}({\cal K}_3,k_1,k_2,-k_3).
    \eea 
    Signature of momenta in the result corresponds similarly to the respective ones inside the factor ${\cal K}_{i}$. The result of this integral is related to the formula in Eq.(\ref{c221aI}) by a complex conjugate as shown for the
integral above.
    \end{enumerate}
There will a total of $4$ integrals of this kind, each with a positive and negative exponential factor.
\item  \underline{\textbf{Integral of the second kind}}:
    For integral of the second kind, the corresponding set of functions are as follows:
     \begin{subequations}
    \begin{align} 
    {\cal P}_{2,2}({\cal K}_{i}, k_{a},k_{b}) &= \frac{{\cal K}_{i}^2}{k_e^2}+\frac{8 {\cal K}_{i} \left(k_a + k_b\right)}{k_e^2}+\frac{56 k_a k_b}{k_e^2}.\\
    {\cal Q}_{2,2}({\cal K}_{i}) &= \frac{i{\cal K}_{i}^5}{k_e^5} -\frac{{\cal K}_{i}^4}{k_e^4} - \frac{2i{\cal K}_{i}^3}{k_e^3} + \frac{6{\cal K}_{i}^2}{k_e^2} + \frac{24i{\cal K}_{i}}{k_e} - 120.\\
    {\cal R}_{2,2}({\cal K}_{i}) &= k_e^2 \bigg\{ -\frac{k^6 (\exp{ix_{s,i}}) \left(\text{Ei}\left(-ix_{e,i}\right)-\text{Ei}\left(-ix_{s,i}\right)\right)}{k_s^{8}}-\frac{i
   {\cal K}_{i}^5}{k_s^7}+\frac{{\cal K}_{i}^4}{k_s^6}+\frac{2 i{\cal K}_{i}^3}{k_s^5}-\frac{6 {\cal K}_{i}^2}{k_s^4}-\frac{24 i{\cal K}_{i}}{k_s^3}+\frac{120}{k_s^2}\bigg\}.
    \end{align}
    \label{c222f}
    \end{subequations}
    where $a, b$ belongs to $1, 2, 3$.
    The result of this integral is expressed using the above functions in the following form:
    \bea \label{c222aI} (\bold{I_{2}})_{j}({\cal K}_i,k_a,k_b) &=& -\frac{k_e^{8} \exp{(-i x_{e,i})}}{13440 c_{s}^8 k_s^9}\bigg\{{\cal P}_{2,2}({\cal K}_i,k_a,k_b){\cal Q}_{2,2}({\cal K}_i) -\frac{720i \left(8\left(k_a+k_b\right)+ {\cal K}_i\right)}{k_e} \nonumber\\
    && + 5040 + \frac{k_s^{8}}{k_e^{8}}\exp{(i(x_{e,i}-x_{s,i}))}\bigg({\cal P}_{2,2}({\cal K}_i){\cal R}_{2,2}({\cal K}_i)-5040 \nonumber\\
    && +\frac{720 i\left(8(k_a+k_b) + {\cal K}_{i}\right)}{k_s}\bigg) \bigg\}.
    \eea 
    \begin{enumerate}
    \item Consider \textbf{Integral 2a}: \bea \label{c222a} \displaystyle{\int_{\tau_{s}}^{\tau_{e}}d\tau_{1}\frac{-3\tau_{s}^{9}}{\tau_{1}^{10}c_{s}^{9}}\exp{(ic_{s}{\cal K}_{1}\tau_{1})}}(k_{2}^2c_{s}^2\tau_{1})(1-ik_{1}c_{s}\tau_{1})(1-ik_{3}c_{s}\tau_{1}) = k_{2}^{2}(\bold{I_{2}})_{1}({\cal K}_1,-k_1,-k_3).
    \eea 
    The signature of momentum arguments inside the result depends on the corresponding values inside the integral for each such integral.
    \item For \textbf{Integral 2b}:
    \bea \label{c222b} \displaystyle{\int_{\tau_{s}}^{\tau_{e}}d\tau_{1}\frac{-3\tau_{s}^{9}}{\tau_{1}^{10}}\exp{(-ic_{s}{\cal K}_{3}\tau_{1})}}(k_{2}^2c_{s}^2\tau_{1})(1+ik_{1}c_{s}\tau_{1})(1-ik_{3}c_{s}\tau_{1}) = k_{2}^{2}(\bold{I_{2}})^{*}_{5}({\cal K}_3,k_1, -k_3).
    \eea 
    The result of such integrals is related to the formula in Eq.(\ref{c222aI}) by a complex conjugate as shown for the
    integral above. 
    \end{enumerate}
    There will be a total of $8$ integrals of this kind, each with a positive and negative exponential factor.
\item  \underline{\textbf{Integral of the third kind}}:
    For integral of the third kind, the corresponding set of functions are as follows:
    \begin{subequations}
    \begin{align} 
    {\cal P}_{2,3}({\cal K}_{i}, k_{a}) &= \frac{{\cal K}_{i}}{k_e}+\frac{6 k_a}{k_e}\\
    {\cal Q}_{2,3}({\cal K}_{i}) &= -\frac{i{\cal K}_{i}^4}{k_e^4} + \frac{{\cal K}_{i}^3}{k_e^3} + \frac{2i {\cal K}_{i}^2}{k_e^2} - \frac{6{\cal K}_{i}}{k_e} - 24i\\
    {\cal R}_{2,3}({\cal K}_{i}) &= k_e \bigg\{ -\frac{{\cal K}_{i}^5 \exp{(ix_{s,i})} \left(\text{Ei}\left(-ix_{e,i}\right)-\text{Ei}\left(-ix_{s,i}\right)\right)}{k_s^{6}}-\frac{i {\cal K}_{i}^4}{k_s^5}+\frac{{\cal K}_{i}^3}{k_s^4}+\frac{2 i{\cal K}_{i}^2}{k_s^3}-\frac{6 {\cal K}_{i}}{k_s^2}-\frac{24 i}{k_s}\bigg\}.
    \end{align}
    \label{c223f}
    \end{subequations}
    where $a$ belongs to $1, 2, 3$. The result of this integral is expressed using the above functions in the following form:
    \bea \label{c223aI} (\bold{I_{3}})_{j}({\cal K}_i,k_a) &=& \frac{k_e^{6} \exp{(-i x_{e,i})}}{720 c_{s}^8 k_s^9}\bigg\{{\cal P}_{2,3}({\cal K}_i,k_a){\cal Q}_{2,3}({\cal K}_i) + 120 \nonumber\\
    &&-\frac{k_s^{6}}{k_e^{6}}\exp{(i(x_{e,i}-x_{s,i}))}\bigg({\cal Q}_{2,3}({\cal K}_1){\cal R}_{2,3}({\cal K}_i)+120 \bigg)\bigg\}. \eea
    \begin{enumerate}
    \item Consider \textbf{Integral 3a}: \bea \label{c223a}\displaystyle{\int_{\tau_{s}}^{\tau_{e}}d\tau_{1}\frac{\tau_{s}^{9}}{\tau_{1}^{9}c_{s}^{9}}\exp{(ic_{s}{\cal K}_{1}\tau_{1})}}(k_{1}^2c_{s}^2\tau_{1})(k_{2}^2c_{s}^2\tau_{1})(1-ik_{3}c_{s}\tau_{1}) = k_{1}^{2}k_{2}^{2}(\bold{I_{3}})_{1}({\cal K}_1,-k_3).
    \eea 
    The signature of the momentum variables inside the result depends on the corresponding values inside for each such integral.
    \item For \textbf{Integral 3b}:
    \bea \label{c223b} \displaystyle{\int_{\tau_{s}}^{\tau_{e}}d\tau_{1}\frac{\tau_{s}^{9}}{\tau_{1}^{9}c_{s}^{9}}\exp{(-ic_{s}{\cal K}_{3}\tau_{1})}}(k_{1}^2c_{s}^2\tau_{1})(k_{2}^2c_{s}^2\tau_{1})(1-ik_{3}c_{s}\tau_{1}) =  k_{1}^{2}k_{2}^{2}(\bold{I_{3}})^{*}_{3}({\cal K}_1, -k_3).
    \eea
    The result of such integrals is related to the formula in Eq.(\ref{c223aI}) by a complex conjugate as shown for the integral above. 
    \end{enumerate}
    There will be a total of $4$ integrals of the second kind, each with a positive and negative exponential factor.
\end{itemize}
For \underline{\textbf{Operator 3}: $\displaystyle{a(\tau_{1})\zeta^{'}(\partial_{i}\zeta)^{2}}$}, we use the formula in Eq.(\ref{c1c}) to work out the contribution to the bispectrum. We have the following types of integrals for this operator:
\begin{itemize}
    \item   \underline{\textbf{Integral of the first kind}}:
    \begin{enumerate}
        \item Consider \textbf{Integral 1a}:
        \bea \label{c231a} \displaystyle{\int_{\tau_{s}}^{\tau_{e}}d\tau_{1}\frac{-3\tau_{s}^{9}}{\tau_{1}^{11}c_{s}^{9}}\exp{(ic_{s}{\cal K}_{1}\tau_{1})}}(1-ik_{1}c_{s}\tau_{1})(1-ik_{2}c_{s}\tau_{1})(1-ik_{3}c_{s}\tau_{1}) = (\bold{I_{2}})_{1}({\cal K}_1,k_1,k_2,k_3).
    \eea The form of this integral is exactly equivalent to the one mentioned in Eq.(\ref{c221a}). Hence, using the same functions defined before in Eq.(\ref{c221f}), and the same general solution as defined in Eq.(\ref{c221aI}), we can write the result of the integrals of the form in Eq.(\ref{c231a}). The signature of the momentum arguments depend on their respective signature in ${\cal K}_{i}$. 
      \item For \textbf{Integral 1b}:
        \bea \displaystyle{\int_{\tau_{s}}^{\tau_{e}}d\tau_{1}\frac{3\tau_{s}^{9}}{\tau_{1}^{11}c_{s}^{9}}\exp{(-ic_{s}{\cal K}_{4}\tau_{1})}}(1-ik_{1}c_{s}\tau_{1})(1+ik_{2}c_{s}\tau_{1})(1+ik_{3}c_{s}\tau_{1}) = -(\bold{I_{1}})^{*}_{4}({\cal K}_4,-k_1,k_2,k_3).
    \eea
    This integral is also exactly equivalent to the one mentioned in Eq.(\ref{c221b}). Hence, using the same functions as defined before in Eq.(\ref{c221f}), but this time taking a complex conjugate of the general formula in Eq.(\ref{c221aI}), we can write down the results of such integrals.
    \end{enumerate}
There will be a total of $4$ integrals of this first kind, each with a positive and negative exponential factor. 
    \item  \underline{\textbf{Integral of the second kind}}:
    \begin{enumerate}
        \item Consider \textbf{Integral 2a}: \bea \label{c232a}\displaystyle{\int_{\tau_{s}}^{\tau_{e}}d\tau_{1}\frac{\tau_{s}^{9}}{\tau_{1}^{10}c_{s}^{9}}\exp{(ic_{s}{\cal K}_{1}\tau_{1})}}(1-ik_{1}c_{s}\tau_{1})(1-ik_{2}c_{s}\tau_{1})(k_{3}^{2}c_{s}^{2}\tau_{1}) =  k_{3}^{2}(\bold{I_{1}})_{1}({\cal K}_1,-k_1,-k_2).
        \eea
    The form of this integral is exactly equivalent to that mentioned in the Eq.(\ref{c222a}). Hence, using the same functions as defined in Eq.(\ref{c222f}), and the same general solution as defined in Eq.(\ref{c222aI}), we can write the result of the integrals of the form in Eq.(\ref{c232a}). The signature of the momentum arguments depend on their respective signature inside the integral.
       \item For \textbf{Integral 2b}:
       \bea \displaystyle{\int_{\tau_{s}}^{\tau_{e}}d\tau_{1}\frac{-\tau_{s}^{9}}{\tau_{1}^{10}c_{s}^{9}}\exp{(ic_{s}{\cal K}_{3}\tau_{1})}}(1+ik_{1}c_{s}\tau_{1})(1+ik_{2}c_{s}\tau_{1})(k_{3}^{2}c_{s}^{2}\tau_{1})= -k_{3}^{2}(\bold{I_{1}})^{*}_{1}({\cal K}_1,k_1,k_2).
       \eea
    \end{enumerate}
    There will be a total of $4$ integrals of this first kind, each with a positive and negative exponential factor. 
\end{itemize}
For \underline{\textbf{Operator 4}: $\displaystyle{(\partial^{2}\zeta)(\partial_{i}\zeta)^{2}}$}, we use the formula in Eq.(\ref{c1d}) to work out the contributions to the bispectrum. Here we have only one kind of integral. 
\begin{itemize}
     \item Consider \underline{\textbf{Integral of the first kind}}:
     For integral of this kind, the corresponding set of functions are as follows:
    \begin{subequations}
    \begin{align}
    {\cal P}_{4,1}({\cal K}_{i}, k_{1},k_{2},k_{3}) &= \frac{-7{\cal K}_{i}^3}{k_e^3}+\frac{56 {\cal K}_{i} \left(k_2 k_3+k_1 k_2+k_1 k_3\right)}{k_e^3}-\frac{336 k_1 k_2 k_3}{k_e^3}.\\
    {\cal Q}_{2,1}({\cal K}_{i}) &= \frac{ i{\cal K}_{i}^4}{k_e^4}-\frac{{\cal K}_{i}^3}{k_e^3} -\frac{2 i{\cal K}_{i}^2}{k_e^2} + \frac{6{\cal K}_{i}}{k_e} + 24i. \\
    {\cal R}_{2,1}({\cal K}_{i}) &= k_e^3 \bigg\{\frac{{\cal K}_{i}^5 \exp{(ix_{s,i})} \left(\text{Ei}\left(-ix_{e,i}\right)-\text{Ei}\left(-ix_{s,i}\right)\right)}{k_s^{8}}+\frac{i
    {\cal K}_{i}^4}{k_s^{7}}-\frac{{\cal K}_{i}^3}{k_s^{6}}-\frac{2i{\cal K}_{i}^2}{k_s^5}+\frac{6{\cal K}_{i}}{k_s^4}+\frac{24i}{k_s^3}\bigg\}.
\end{align} 
\label{c241f}
\end{subequations}
    The result of this integral is expressed using the above functions in the following form:
    \bea \label{c224aI} (\bold{I_{1}})_{j}({\cal K}_i,k_1,k_2,k_3) &=& \frac{k_e^{8} \exp{(-i x_{e,i})}}{40320 c_{s}^{10} k_s^9}\bigg\{{\cal P}_{4,1}({\cal K}_i,k_1,k_2,k_3){\cal Q}_{4,1}({\cal K}_i) + i 5040\frac{{\cal K}_{i}}{k_e} \nonumber\\
    && -\frac{ 120\left(56\left(k_2 k_3+k_1 k_2+k_1 k_3\right)-7 {\cal K}_i^{2}\right)}{k_e^2} + 5040  \nonumber\\
    && -\frac{k_s^{8}}{k_e^{8}}\exp{(i(x_{e,i}-x_{s,i}))}\bigg({\cal Q}_{4,1}({\cal K}_i){\cal R}_{4,1}({\cal K}_i)+5040 +i\frac{5040
    {\cal K}_i}{k_s} \nonumber\\
    && - \frac{120\left(56\left(k_2 k_3+k_1 k_2+k_1 k_3\right)-7 {\cal K}_i^{2}\right)}{k_s^2} \bigg) \bigg\}. \eea
    \begin{enumerate}
    \item \textbf{Integral 1a}: \bea \label{c224a} \displaystyle{\int_{\tau_{s}}^{\tau_{e}}d\tau_{1}\frac{\tau_{s}^{9}}{\tau_{1}^{9}c_{s}^{9}}\exp{(ic_{s}{\cal K}_{1}\tau_{1})}}(1-ik_{1}c_{s}\tau_{1})(1-ik_{2}c_{s}\tau_{1})(1-ik_{3}c_{s}\tau_{1}) = (\bold{I_{1}})_{1}({\cal K}_1,k_1,k_2,k_3). \eea 
    The signature of the momentum arguments in the results of such integrals depend on their respective signature in ${\cal K}_{i}$.
    \item For \textbf{Integral 1b}:
    \bea \label{c224b}\displaystyle{\int_{\tau_{s}}^{\tau_{e}}d\tau_{1}\frac{\tau_{s}^{9}}{\tau_{1}^{9}c_{s}^{9}}\exp{(-ic_{s}{\cal K}_{3}\tau_{1})}}(1+ik_{1}c_{s}\tau_{1})(1+ik_{2}c_{s}\tau_{1})(1-ik_{3}c_{s}\tau_{1}) = (\bold{I_{1}})^{*}_{3}({\cal K}_3,k_1,k_2,-k_3).
    \eea
    The result of this integral is related to the formula in Eq.(\ref{c224aI}) by a complex conjugate as shown for the integral above.
    \end{enumerate}
 There will be a total of $4$ integrals of the fourth kind, each with a positive and negative exponential factor.
\end{itemize}

 
\subsection{Detailed computation of the bispectrum in the region III: SRII}\label{App:C}

In this subsection, we present the explicit computations for the contributions from individual operators to the tree-level three point correlation function in the SRII region.

We mention the time-derivative of the mode function in the SRII phase as it would be very convenient for further calculations. This is given by:
\bea \Pi_{\bold{k}}^{*} = \partial_{\tau}\zeta_{\bold{k}}^{*} = \left(\frac{-iH^{2}}{2\sqrt{\cal A}}\right)\left(\frac{\tau_{s}}{\tau_{e}}\right)^{3}\frac{1}{(c_{s}k)^{3/2}}\times\left(k^{2}c_{s}^{2}\tau\right)\left(\alpha_{\bold{k}}^{(3)*}\exp{(ikc_{s}\tau)}-\beta_{\bold{k}}^{(3)*}\exp{(-ikc_{s}\tau)}\right).
\eea where $k = |\bold{k}|$. In the subsequent integrals, the factor ${\cal K}_{i}$ is the same factor from the Eq.(\ref{c2k}).

We start with \underline{\textbf{Operator 1}: $\displaystyle{a(\tau_{1})\zeta^{'3}}$}, and use the formula in Eq.(\ref{c1a}) to work out the contribution to the bispectrum.
\begin{enumerate}
\item Consider \textbf{Integral 1a}:
\bea \label{c31a} 
\displaystyle{\int_{\tau_{e}}^{\tau_{end}\rightarrow 0}d\tau_{1}k_{1}^{2}k_{2}^{2}k_{3}^{2}c_{s}^{-3}\tau_{1}^{2}\exp{(ic_{s}{\cal K}_{1}\tau_{1})}} = k_1^2 k_2^2 k_3^2(\bold{J}_{1})_{1}({\cal K}_{1}).
\eea 
where the result of the above integral is written in terms of the following general formula: 
\bea \label{c31aI} 
(\bold{J}_{1})_{i}({\cal K}_{i}) = \frac{1}{c_{s}^6 {\cal K}_{i}^3}\left\{\left(2 i-\exp{\left(-\frac{i {\cal K}_{i}}{k_e}\right)} \left(2 i-\frac{{\cal K}_{i}}{k_e} \left(2+\frac{i{\cal K}_{i}}{k_e}\right)\right)\right)\right\}
\eea
Different integrals similar to the above will be related through a change in the factor ${\cal K}_{i}$ in the above general formula.
\item For \textbf{Integral 1b}:
\bea \label{c31b} 
\displaystyle{\int_{\tau_{e}}^{\tau_{end}\rightarrow 0}d\tau_{1}k_{1}^{2}k_{2}^{2}k_{3}^{2}c_{s}^{-3}\tau_{1}^{2}\exp{(-ic_{s}{\cal K}_{4}\tau_{1})}} = k_{1}^{2}k_{2}^{2}k_{3}^{2}(\bold{J}_{1})^{*}_{4}({\cal K}_{4}).
\eea
The result of which can be written in terms of the complex conjugate of the general formula in Eq.(\ref{c31aI}).
\end{enumerate}
There will be $4$ integrals of first kind, each with positive and negative exponential factors along with different ${\cal K}_{i}$'s.

For \underline{\textbf{Operator 2}: $\displaystyle{\zeta^{'2}\partial^{2}\zeta}$}, we use the formula in Eq.(\ref{c1b}) to work out the contribution to the bispectrum.
\begin{enumerate}
    \item Consider \textbf{Integral 2a}:
    \bea \label{c32a} 
    \displaystyle{\int_{\tau_{e}}^{\tau_{end}\rightarrow 0}d\tau_{1}k_{1}^{2}k_{2}^{2}(1-ik_{3}c_{s}\tau_{1})c_{s}^{-5}\tau_{1}^{2}\exp{(ic_{s}{\cal K}_{1}\tau_{1})}} = k_1^{2}k_2^{2}(\bold{J}_{2})_{1}({\cal K}_{1},-k_3).\eea
The result of the above integral is written in terms of the following general formula: 
\bea \label{c32aI}
(\bold{J}_{2})_{i}({\cal K}_{i},k_a) &=& \frac{1}{c^8 {\cal K}_{i}^4}\Bigg\{\left({\cal K}_{i}\left(2 i-\exp{\left(-\frac{i {\cal K}_{i}}{k_e}\right)} \left(2 i-\frac{{\cal K}_{i}}{k_e}\left(2+\frac{i{\cal K}_{i}}{k_e}\right)\right)\right)\right)\nonumber\\
&&\quad\quad\quad\quad\quad\quad\quad\quad\quad\quad +i k_a \left(-6+\exp{\left(-\frac{i {\cal K}_{i}}{k_e}\right)}\left(6+\frac{{\cal K}_{i}}{k_e}\left(6 i-\frac{{\cal K}_{i}}{k_e}\left(3+\frac{i {\cal K}_{i}}{k_e}\right)\right)\right)\right)\Bigg\}
\eea 
Different integrals similar to the above will be related through a change in the factor ${\cal K}_{i}$ and the signature of the momentum variables in the above general formula which depends on their corresponding values inside the integral. In the above case, $k_{a} = -k_{3}$.
    \item For \textbf{Integral 2b}:
    \bea \label{c32b}
    \displaystyle{\int_{\tau_{e}}^{\tau_{end}\rightarrow 0}d\tau_{1}k_{1}^{2}k_{2}^{2}(1+ik_{3}c_{s}\tau_{1})c_{s}^{-5}\tau_{1}^{2}\exp{(-ic_{s}{\cal K}_{2}\tau_{1})}} = k_1^{2}k_2^{2}(\bold{J}_{2})^{*}_{2}({\cal K}_{2}).\eea
The result of this integral is written in terms of the complex conjugate of the general formula in Eq.(\ref{c32aI}).
\end{enumerate}
There will be $4$ integrals of second kind, each with positive and negative exponential factors along with different ${\cal K}_{i}$'s.

For \underline{\textbf{Operator 3}: $\displaystyle{a(\tau_{1})\zeta^{'}(\partial_{i}\zeta)^{2}}$}, we use the formula in Eq.(\ref{c1c}) to work out the contribution to the bispectrum. 
\begin{enumerate}
    \item Consider \textbf{Integral 3a}:
    \bea \label{c33a} 
    \displaystyle{\int_{\tau_{e}}^{\tau_{end}\rightarrow 0}d\tau_{1}(1-ik_{1}c_{s}\tau_{1})(1-ik_{2}c_{s}\tau_{1})k_{3}^{2}c_{s}^{-7}\exp{(ic_{s}{\cal K}_{1}\tau_{1})}} = k_{3}^{2}(\bold{J}_{3})_{1}({\cal K}_{1},k_1,k_2,k_3).
    \eea
    The result of the above integral is written in terms of the following general formula: 
\bea \label{c33aI} 
     (\bold{J}_{3})_{i}({\cal K}_{i},k_1,k_2,k_3) &=& -\frac{1}{c^{8}{\cal K}_{i}{}^3}\Bigg\{i \left(2 k_1^2+3 \left(2 k_2+k_3\right) k_1+\left(k_2+k_3\right) \left(2
    k_2+k_3\right)\right)-\exp{\left(-\frac{i{\cal K}_{i}}{k_e}\right)} \nonumber\\
    &&\bigg(-i k_1 k_2\frac{ {\cal K}_{i}{}^2}{k_e^2} +i \left(2 k_1^2+3 \left(2 k_2+k_3\right)
    k_1+\left(k_2+k_3\right) \left(2 k_2+k_3\right)\right)\nonumber\\
    && -{\cal K}_{i}\frac{\left(k_1^2+\left(4 k_2+k_3\right) k_1+k_2
    \left(k_2+k_3\right)\right)}{k_e}\bigg)\Bigg\}.\eea
Different integrals similar to the above will be related through a change in the factor ${\cal K}_{i}$ and the signature of the momentum variables in the above general formula which also depends on their corresponding values within ${\cal K}_{i}$.
     \item For \textbf{Integral 3b}:
     \bea \label{c33b} 
    \displaystyle{\int_{\tau_{e}}^{\tau_{end}\rightarrow 0}d\tau_{1}(1+ik_{1}c_{s}\tau_{1})(1+ik_{2}c_{s}\tau_{1})k_{3}^{2}c_{s}^{-7}\exp{(-ic_{s}{\cal K}_{3}\tau_{1})}} = k_{3}^{2}(\bold{J}_{3})^{*}_{3}({\cal K}_{3},k_1,k_2,-k_3).
     \eea
     The result of this integral can be written in terms of the complex conjugate of the general formula in Eq.(\ref{c33aI}).
     \end{enumerate}
There will be $4$ integrals of third kind, each with positive and negative exponential factors along with different ${\cal K}_{i}$'s.

For \underline{\textbf{Operator 4}: $\partial^{2}\zeta(\partial_{i}\zeta)^{2}$}, we use the formula in Eq.(\ref{c1d}) to work out the contribution to the bispectrum. 
\begin{enumerate}
    \item Consider \textbf{Integral 4a}:
    \bea \label{c34a} 
    \displaystyle{\int_{\tau_{e}}^{\tau_{end}\rightarrow 0}d\tau_{1}(1-ik_{1}c_{s}\tau_{1})(1-ik_{2}c_{s}\tau_{1})(1-ik_{3}c_{s}\tau_{1})c_{s}^{-9}\exp{(ic_{s}{\cal K}_{2}\tau_{1})}} = (\bold{J}_{4})_{1}({\cal K}_{2},k_1,k_2,k_3).
    \eea
    The result is expressed in term of the following general formula:
     \bea \label{c34aI} (\bold{J}_{4})_{j}({\cal K}_{i},k_1,k_2,k_3) &=& \frac{i}{c^{10}{\cal K}_{i}^4}\Bigg\{\bigg(-2 \left(k_1^3+4 \left(k_2+k_3\right) k_1^2+4 \left(k_2^2+3 k_3 k_2+k_3^2\right) k_1+\left(k_2+k_3\right) \left(k_2^2+3 k_3 k_2+k_3^2\right)\right)\nonumber\\
    && +\exp{\left(-\frac{i {\cal K}_{i}}{k_e}\right)} \bigg(-\frac{{\cal K}_{i}^2\left(\left(k_2+k_3\right) k_1^2+\left(k_2^2+6 k_3 k_2+k_3^2\right) k_1+k_2 k_3 \left(k_2+k_3\right)\right)}{k_e^2}-i k_1 k_2 k_3 \frac{{\cal K}_{i}{}^3}{k_e^3}\nonumber\\
    && +\frac{i {\cal K}_{i} \left(k_1^3+5 \left(k_2+k_3\right) k_1^2+\left(5 k_2^2+18 k_3 k_2+5 k_3^2\right) k_1+\left(k_2+k_3\right)\left(k_2^2+4 k_3 k_2+k_3^2\right)\right)}{k_e}\nonumber\\
    && +2 \left(k_1^3+4 \left(k_2+k_3\right) k_1^2+4 \left(k_2^2+3 k_3 k_2+k_3^2\right) k_1+\left(k_2+k_3\right) \left(k_2^2+3 k_3 k_2+k_3^2\right)\right)\bigg)\bigg)\Bigg\}.
    \eea
Different integrals similar to the above will be related through a change in the factor ${\cal K}_{i}$ and the signature of the momentum variables in the above general formula.
    \item For \textbf{Integral 4b}:
     \bea \label{c34b} 
    \displaystyle{\int_{\tau_{e}}^{\tau_{end}\rightarrow 0}d\tau_{1}(1+ik_{1}c_{s}\tau_{1})(1+ik_{2}c_{s}\tau_{1})(1-ik_{3}c_{s}\tau_{1})c_{s}^{-9}\exp{(-ic_{s}{\cal K}_{3}\tau_{1})}} = k_{3}^{2}(\bold{J}_{4})^{*}_{3}({\cal K}_{3},k_1,k_2,-k_3).
     \eea
     The result of this integral can be written in terms of the complex conjugate of the general formula in Eq.(\ref{c34aI}).
\end{enumerate}
There will be $4$ integrals of fourth kind, each with positive and negative exponential factors along with different ${\cal K}_{i}$'s.

\bibliography{references2}

\providecommand{\href}[2]{#2}\begingroup\raggedright\begin{thebibliography}{100}

\bibitem{Planck:2019kim}
{\bfseries Planck} Collaboration, Y.~Akrami {\em et~al.}, ``{Planck 2018
  results. IX. Constraints on primordial non-Gaussianity},''
  \href{http://dx.doi.org/10.1051/0004-6361/201935891}{{\em Astron. Astrophys.}
  {\bfseries 641} (2020) A9}, \href{http://arxiv.org/abs/1905.05697}{{\ttfamily
  arXiv:1905.05697 [astro-ph.CO]}}.

\bibitem{Achucarro:2022qrl}
A.~Ach\'ucarro {\em et~al.}, ``{Inflation: Theory and Observations},''
  \href{http://arxiv.org/abs/2203.08128}{{\ttfamily arXiv:2203.08128
  [astro-ph.CO]}}.

\bibitem{CMB-S4:2023zem}
{\bfseries CMB-S4} Collaboration, D.~Zegeye {\em et~al.}, ``{CMB-S4:
  Forecasting Constraints on $f_\mathrm{NL}$ Through $\mu$-distortion
  Anisotropy},'' \href{http://arxiv.org/abs/2303.00916}{{\ttfamily
  arXiv:2303.00916 [astro-ph.CO]}}.

\bibitem{Maldacena:2002vr}
J.~M. Maldacena, ``{Non-Gaussian features of primordial fluctuations in single
  field inflationary models},''
  \href{http://dx.doi.org/10.1088/1126-6708/2003/05/013}{{\em JHEP} {\bfseries
  05} (2003) 013}, \href{http://arxiv.org/abs/astro-ph/0210603}{{\ttfamily
  arXiv:astro-ph/0210603}}.

\bibitem{Alishahiha:2004eh}
M.~Alishahiha, E.~Silverstein, and D.~Tong, ``{DBI in the sky},''
  \href{http://dx.doi.org/10.1103/PhysRevD.70.123505}{{\em Phys. Rev. D}
  {\bfseries 70} (2004) 123505},
  \href{http://arxiv.org/abs/hep-th/0404084}{{\ttfamily arXiv:hep-th/0404084}}.

\bibitem{Mazumdar:2001mm}
A.~Mazumdar, S.~Panda, and A.~Perez-Lorenzana, ``{Assisted inflation via
  tachyon condensation},''
  \href{http://dx.doi.org/10.1016/S0550-3213(01)00410-2}{{\em Nucl. Phys. B}
  {\bfseries 614} (2001) 101--116},
  \href{http://arxiv.org/abs/hep-ph/0107058}{{\ttfamily arXiv:hep-ph/0107058}}.

\bibitem{Choudhury:2002xu}
D.~Choudhury, D.~Ghoshal, D.~P. Jatkar, and S.~Panda, ``{On the cosmological
  relevance of the tachyon},''
  \href{http://dx.doi.org/10.1016/S0370-2693(02)02512-1}{{\em Phys. Lett. B}
  {\bfseries 544} (2002) 231--238},
  \href{http://arxiv.org/abs/hep-th/0204204}{{\ttfamily arXiv:hep-th/0204204}}.

\bibitem{Panda:2005sg}
S.~Panda, M.~Sami, and S.~Tsujikawa, ``{Inflation and dark energy arising from
  geometrical tachyons},''
  \href{http://dx.doi.org/10.1103/PhysRevD.73.023515}{{\em Phys. Rev. D}
  {\bfseries 73} (2006) 023515},
  \href{http://arxiv.org/abs/hep-th/0510112}{{\ttfamily arXiv:hep-th/0510112}}.

\bibitem{Chingangbam:2004ng}
P.~Chingangbam, S.~Panda, and A.~Deshamukhya, ``{Non-minimally coupled
  tachyonic inflation in warped string background},''
  \href{http://dx.doi.org/10.1088/1126-6708/2005/02/052}{{\em JHEP} {\bfseries
  02} (2005) 052}, \href{http://arxiv.org/abs/hep-th/0411210}{{\ttfamily
  arXiv:hep-th/0411210}}.

\bibitem{Armendariz-Picon:1999hyi}
C.~Armendariz-Picon, T.~Damour, and V.~F. Mukhanov, ``{k - inflation},''
  \href{http://dx.doi.org/10.1016/S0370-2693(99)00603-6}{{\em Phys. Lett. B}
  {\bfseries 458} (1999) 209--218},
  \href{http://arxiv.org/abs/hep-th/9904075}{{\ttfamily arXiv:hep-th/9904075}}.

\bibitem{Garriga:1999vw}
J.~Garriga and V.~F. Mukhanov, ``{Perturbations in k-inflation},''
  \href{http://dx.doi.org/10.1016/S0370-2693(99)00602-4}{{\em Phys. Lett. B}
  {\bfseries 458} (1999) 219--225},
  \href{http://arxiv.org/abs/hep-th/9904176}{{\ttfamily arXiv:hep-th/9904176}}.

\bibitem{Burrage:2010cu}
C.~Burrage, C.~de~Rham, D.~Seery, and A.~J. Tolley, ``{Galileon inflation},''
  \href{http://dx.doi.org/10.1088/1475-7516/2011/01/014}{{\em JCAP} {\bfseries
  01} (2011) 014}, \href{http://arxiv.org/abs/1009.2497}{{\ttfamily
  arXiv:1009.2497 [hep-th]}}.

\bibitem{Choudhury:2012yh}
S.~Choudhury and S.~Pal, ``{DBI Galileon inflation in background SUGRA},''
  \href{http://dx.doi.org/10.1016/j.nuclphysb.2013.05.010}{{\em Nucl. Phys. B}
  {\bfseries 874} (2013) 85--114},
  \href{http://arxiv.org/abs/1208.4433}{{\ttfamily arXiv:1208.4433 [hep-th]}}.

\bibitem{Choudhury:2012whm}
S.~Choudhury and S.~Pal, ``{Primordial non-Gaussian features from DBI Galileon
  inflation},'' \href{http://dx.doi.org/10.1140/epjc/s10052-015-3452-3}{{\em
  Eur. Phys. J. C} {\bfseries 75} no.~6, (2015) 241},
  \href{http://arxiv.org/abs/1210.4478}{{\ttfamily arXiv:1210.4478 [hep-th]}}.

\bibitem{Chen:2010xka}
X.~Chen, ``{Primordial Non-Gaussianities from Inflation Models},''
  \href{http://dx.doi.org/10.1155/2010/638979}{{\em Adv. Astron.} {\bfseries
  2010} (2010) 638979}, \href{http://arxiv.org/abs/1002.1416}{{\ttfamily
  arXiv:1002.1416 [astro-ph.CO]}}.

\bibitem{Chen:2006nt}
X.~Chen, M.-x. Huang, S.~Kachru, and G.~Shiu, ``{Observational signatures and
  non-Gaussianities of general single field inflation},''
  \href{http://dx.doi.org/10.1088/1475-7516/2007/01/002}{{\em JCAP} {\bfseries
  01} (2007) 002}, \href{http://arxiv.org/abs/hep-th/0605045}{{\ttfamily
  arXiv:hep-th/0605045}}.

\bibitem{Chen:2009zp}
X.~Chen and Y.~Wang, ``{Quasi-Single Field Inflation and Non-Gaussianities},''
  \href{http://dx.doi.org/10.1088/1475-7516/2010/04/027}{{\em JCAP} {\bfseries
  04} (2010) 027}, \href{http://arxiv.org/abs/0911.3380}{{\ttfamily
  arXiv:0911.3380 [hep-th]}}.

\bibitem{Chen:2009we}
X.~Chen and Y.~Wang, ``{Large non-Gaussianities with Intermediate Shapes from
  Quasi-Single Field Inflation},''
  \href{http://dx.doi.org/10.1103/PhysRevD.81.063511}{{\em Phys. Rev. D}
  {\bfseries 81} (2010) 063511},
  \href{http://arxiv.org/abs/0909.0496}{{\ttfamily arXiv:0909.0496
  [astro-ph.CO]}}.

\bibitem{Chen:2008wn}
X.~Chen, R.~Easther, and E.~A. Lim, ``{Generation and Characterization of Large
  Non-Gaussianities in Single Field Inflation},''
  \href{http://dx.doi.org/10.1088/1475-7516/2008/04/010}{{\em JCAP} {\bfseries
  04} (2008) 010}, \href{http://arxiv.org/abs/0801.3295}{{\ttfamily
  arXiv:0801.3295 [astro-ph]}}.

\bibitem{Chen:2006xjb}
X.~Chen, R.~Easther, and E.~A. Lim, ``{Large Non-Gaussianities in Single Field
  Inflation},'' \href{http://dx.doi.org/10.1088/1475-7516/2007/06/023}{{\em
  JCAP} {\bfseries 06} (2007) 023},
  \href{http://arxiv.org/abs/astro-ph/0611645}{{\ttfamily
  arXiv:astro-ph/0611645}}.

\bibitem{Chen:2013aj}
X.~Chen, H.~Firouzjahi, M.~H. Namjoo, and M.~Sasaki, ``{A Single Field
  Inflation Model with Large Local Non-Gaussianity},''
  \href{http://dx.doi.org/10.1209/0295-5075/102/59001}{{\em EPL} {\bfseries
  102} no.~5, (2013) 59001}, \href{http://arxiv.org/abs/1301.5699}{{\ttfamily
  arXiv:1301.5699 [hep-th]}}.

\bibitem{Chen:2012ge}
X.~Chen and Y.~Wang, ``{Quasi-Single Field Inflation with Large Mass},''
  \href{http://dx.doi.org/10.1088/1475-7516/2012/09/021}{{\em JCAP} {\bfseries
  09} (2012) 021}, \href{http://arxiv.org/abs/1205.0160}{{\ttfamily
  arXiv:1205.0160 [hep-th]}}.

\bibitem{Chen:2009bc}
X.~Chen, B.~Hu, M.-x. Huang, G.~Shiu, and Y.~Wang, ``{Large Primordial
  Trispectra in General Single Field Inflation},''
  \href{http://dx.doi.org/10.1088/1475-7516/2009/08/008}{{\em JCAP} {\bfseries
  08} (2009) 008}, \href{http://arxiv.org/abs/0905.3494}{{\ttfamily
  arXiv:0905.3494 [astro-ph.CO]}}.

\bibitem{Creminelli:2010ba}
P.~Creminelli, A.~Nicolis, and E.~Trincherini, ``{Galilean Genesis: An
  Alternative to inflation},''
  \href{http://dx.doi.org/10.1088/1475-7516/2010/11/021}{{\em JCAP} {\bfseries
  11} (2010) 021}, \href{http://arxiv.org/abs/1007.0027}{{\ttfamily
  arXiv:1007.0027 [hep-th]}}.

\bibitem{Kobayashi:2010cm}
T.~Kobayashi, M.~Yamaguchi, and J.~Yokoyama, ``{G-inflation: Inflation driven
  by the Galileon field},''
  \href{http://dx.doi.org/10.1103/PhysRevLett.105.231302}{{\em Phys. Rev.
  Lett.} {\bfseries 105} (2010) 231302},
  \href{http://arxiv.org/abs/1008.0603}{{\ttfamily arXiv:1008.0603 [hep-th]}}.

\bibitem{Mizuno:2010ag}
S.~Mizuno and K.~Koyama, ``{Primordial non-Gaussianity from the DBI
  Galileons},'' \href{http://dx.doi.org/10.1103/PhysRevD.82.103518}{{\em Phys.
  Rev. D} {\bfseries 82} (2010) 103518},
  \href{http://arxiv.org/abs/1009.0677}{{\ttfamily arXiv:1009.0677 [hep-th]}}.

\bibitem{Burrage:2011hd}
C.~Burrage, R.~H. Ribeiro, and D.~Seery, ``{Large slow-roll corrections to the
  bispectrum of noncanonical inflation},''
  \href{http://dx.doi.org/10.1088/1475-7516/2011/07/032}{{\em JCAP} {\bfseries
  07} (2011) 032}, \href{http://arxiv.org/abs/1103.4126}{{\ttfamily
  arXiv:1103.4126 [astro-ph.CO]}}.

\bibitem{Kobayashi:2011pc}
T.~Kobayashi, M.~Yamaguchi, and J.~Yokoyama, ``{Primordial non-Gaussianity from
  G-inflation},'' \href{http://dx.doi.org/10.1103/PhysRevD.83.103524}{{\em
  Phys. Rev. D} {\bfseries 83} (2011) 103524},
  \href{http://arxiv.org/abs/1103.1740}{{\ttfamily arXiv:1103.1740 [hep-th]}}.

\bibitem{DeFelice:2011zh}
A.~De~Felice and S.~Tsujikawa, ``{Primordial non-Gaussianities in general
  modified gravitational models of inflation},''
  \href{http://dx.doi.org/10.1088/1475-7516/2011/04/029}{{\em JCAP} {\bfseries
  04} (2011) 029}, \href{http://arxiv.org/abs/1103.1172}{{\ttfamily
  arXiv:1103.1172 [astro-ph.CO]}}.

\bibitem{Renaux-Petel:2011lur}
S.~Renaux-Petel, ``{Orthogonal non-Gaussianities from Dirac-Born-Infeld
  Galileon inflation},''
  \href{http://dx.doi.org/10.1088/0264-9381/28/24/249601}{{\em Class. Quant.
  Grav.} {\bfseries 28} (2011) 182001},
  \href{http://arxiv.org/abs/1105.6366}{{\ttfamily arXiv:1105.6366
  [astro-ph.CO]}}. [Erratum: Class.Quant.Grav. 28, 249601 (2011)].

\bibitem{DeFelice:2011uc}
A.~De~Felice and S.~Tsujikawa, ``{Inflationary non-Gaussianities in the most
  general second-order scalar-tensor theories},''
  \href{http://dx.doi.org/10.1103/PhysRevD.84.083504}{{\em Phys. Rev. D}
  {\bfseries 84} (2011) 083504},
  \href{http://arxiv.org/abs/1107.3917}{{\ttfamily arXiv:1107.3917 [gr-qc]}}.

\bibitem{Gao:2011qe}
X.~Gao and D.~A. Steer, ``{Inflation and primordial non-Gaussianities of
  'generalized Galileons'},''
  \href{http://dx.doi.org/10.1088/1475-7516/2011/12/019}{{\em JCAP} {\bfseries
  12} (2011) 019}, \href{http://arxiv.org/abs/1107.2642}{{\ttfamily
  arXiv:1107.2642 [astro-ph.CO]}}.

\bibitem{deRham:2012az}
C.~de~Rham, ``{Galileons in the Sky},''
  \href{http://dx.doi.org/10.1016/j.crhy.2012.04.006}{{\em Comptes Rendus
  Physique} {\bfseries 13} (2012) 666--681},
  \href{http://arxiv.org/abs/1204.5492}{{\ttfamily arXiv:1204.5492
  [astro-ph.CO]}}.

\bibitem{Ohashi:2012wf}
J.~Ohashi and S.~Tsujikawa, ``{Potential-driven Galileon inflation},''
  \href{http://dx.doi.org/10.1088/1475-7516/2012/10/035}{{\em JCAP} {\bfseries
  10} (2012) 035}, \href{http://arxiv.org/abs/1207.4879}{{\ttfamily
  arXiv:1207.4879 [gr-qc]}}.

\bibitem{DeFelice:2013ar}
A.~De~Felice and S.~Tsujikawa, ``{Shapes of primordial non-Gaussianities in the
  Horndeski's most general scalar-tensor theories},''
  \href{http://dx.doi.org/10.1088/1475-7516/2013/03/030}{{\em JCAP} {\bfseries
  03} (2013) 030}, \href{http://arxiv.org/abs/1301.5721}{{\ttfamily
  arXiv:1301.5721 [hep-th]}}.

\bibitem{Arroja:2013dya}
F.~Arroja, N.~Bartolo, E.~Dimastrogiovanni, and M.~Fasiello, ``{On the
  Trispectrum of Galileon Inflation},''
  \href{http://dx.doi.org/10.1088/1475-7516/2013/11/005}{{\em JCAP} {\bfseries
  11} (2013) 005}, \href{http://arxiv.org/abs/1307.5371}{{\ttfamily
  arXiv:1307.5371 [astro-ph.CO]}}.

\bibitem{Choudhury:2013qza}
S.~Choudhury and A.~Dasgupta, ``{Galileogenesis: A new cosmophenomenological
  zip code for reheating through R-parity violating coupling},''
  \href{http://dx.doi.org/10.1016/j.nuclphysb.2014.02.024}{{\em Nucl. Phys. B}
  {\bfseries 882} (2014) 195--204},
  \href{http://arxiv.org/abs/1309.1934}{{\ttfamily arXiv:1309.1934 [hep-ph]}}.

\bibitem{Pirtskhalava:2015zwa}
D.~Pirtskhalava, L.~Santoni, E.~Trincherini, and F.~Vernizzi, ``{Large
  Non-Gaussianity in Slow-Roll Inflation},''
  \href{http://dx.doi.org/10.1007/JHEP04(2016)117}{{\em JHEP} {\bfseries 04}
  (2016) 117}, \href{http://arxiv.org/abs/1506.06750}{{\ttfamily
  arXiv:1506.06750 [hep-th]}}.

\bibitem{Baumann:2009ds}
D.~Baumann,
  \href{http://dx.doi.org/10.1142/9789814327183_0010}{``{Inflation},''} in {\em
  {Theoretical Advanced Study Institute in Elementary Particle Physics}:
  {Physics of the Large and the Small}}, pp.~523--686.
\newblock 2011.
\newblock \href{http://arxiv.org/abs/0907.5424}{{\ttfamily arXiv:0907.5424
  [hep-th]}}.

\bibitem{Senatore:2016aui}
L.~Senatore, \href{http://dx.doi.org/10.1142/9789813149441_0008}{``{Lectures on
  Inflation},''} in {\em {Theoretical Advanced Study Institute in Elementary
  Particle Physics}: {New Frontiers in Fields and Strings}}, pp.~447--543.
\newblock 2017.
\newblock \href{http://arxiv.org/abs/1609.00716}{{\ttfamily arXiv:1609.00716
  [hep-th]}}.

\bibitem{Baumann:2018muz}
D.~Baumann, ``{Primordial Cosmology},''
  \href{http://dx.doi.org/10.22323/1.305.0009}{{\em PoS} {\bfseries TASI2017}
  (2018) 009}, \href{http://arxiv.org/abs/1807.03098}{{\ttfamily
  arXiv:1807.03098 [hep-th]}}.

\bibitem{Das:2023cum}
B.~Das and H.~V. Ragavendra, ``{Indirect imprints of primordial non-Gaussianity
  on cosmic microwave background},''
  \href{http://arxiv.org/abs/2304.05941}{{\ttfamily arXiv:2304.05941
  [astro-ph.CO]}}.

\bibitem{Choudhury:2011sq}
S.~Choudhury and S.~Pal, ``{Brane inflation in background supergravity},''
  \href{http://dx.doi.org/10.1103/PhysRevD.85.043529}{{\em Phys. Rev. D}
  {\bfseries 85} (2012) 043529},
  \href{http://arxiv.org/abs/1102.4206}{{\ttfamily arXiv:1102.4206 [hep-th]}}.

\bibitem{Choudhury:2012ib}
S.~Choudhury and S.~Pal, ``{Brane inflation: A field theory approach in
  background supergravity},''
  \href{http://dx.doi.org/10.1088/1742-6596/405/1/012009}{{\em J. Phys. Conf.
  Ser.} {\bfseries 405} (2012) 012009},
  \href{http://arxiv.org/abs/1209.5883}{{\ttfamily arXiv:1209.5883 [hep-th]}}.

\bibitem{Choudhury:2013jya}
S.~Choudhury, A.~Mazumdar, and S.~Pal, ``{Low \& High scale MSSM inflation,
  gravitational waves and constraints from Planck},''
  \href{http://dx.doi.org/10.1088/1475-7516/2013/07/041}{{\em JCAP} {\bfseries
  07} (2013) 041}, \href{http://arxiv.org/abs/1305.6398}{{\ttfamily
  arXiv:1305.6398 [hep-ph]}}.

\bibitem{Choudhury:2013zna}
S.~Choudhury, T.~Chakraborty, and S.~Pal, ``{Higgs inflation from new K\"ahler
  potential},'' \href{http://dx.doi.org/10.1016/j.nuclphysb.2014.01.002}{{\em
  Nucl. Phys. B} {\bfseries 880} (2014) 155--174},
  \href{http://arxiv.org/abs/1305.0981}{{\ttfamily arXiv:1305.0981 [hep-th]}}.

\bibitem{Esposito:2019jkb}
A.~Esposito, L.~Hui, and R.~Scoccimarro, ``{Nonperturbative test of consistency
  relations and their violation},''
  \href{http://dx.doi.org/10.1103/PhysRevD.100.043536}{{\em Phys. Rev. D}
  {\bfseries 100} no.~4, (2019) 043536},
  \href{http://arxiv.org/abs/1905.11423}{{\ttfamily arXiv:1905.11423
  [astro-ph.CO]}}.

\bibitem{Goldstein:2022hgr}
S.~Goldstein, A.~Esposito, O.~H.~E. Philcox, L.~Hui, J.~C. Hill,
  R.~Scoccimarro, and M.~H. Abitbol, ``{Squeezing fNL out of the matter
  bispectrum with consistency relations},''
  \href{http://dx.doi.org/10.1103/PhysRevD.106.123525}{{\em Phys. Rev. D}
  {\bfseries 106} no.~12, (2022) 123525},
  \href{http://arxiv.org/abs/2209.06228}{{\ttfamily arXiv:2209.06228
  [astro-ph.CO]}}.

\bibitem{Planck:2015sxf}
{\bfseries Planck} Collaboration, P.~A.~R. Ade {\em et~al.}, ``{Planck 2015
  results. XX. Constraints on inflation},''
  \href{http://dx.doi.org/10.1051/0004-6361/201525898}{{\em Astron. Astrophys.}
  {\bfseries 594} (2016) A20},
  \href{http://arxiv.org/abs/1502.02114}{{\ttfamily arXiv:1502.02114
  [astro-ph.CO]}}.

\bibitem{Cheung:2007st}
C.~Cheung, P.~Creminelli, A.~L. Fitzpatrick, J.~Kaplan, and L.~Senatore, ``{The
  Effective Field Theory of Inflation},''
  \href{http://dx.doi.org/10.1088/1126-6708/2008/03/014}{{\em JHEP} {\bfseries
  03} (2008) 014}, \href{http://arxiv.org/abs/0709.0293}{{\ttfamily
  arXiv:0709.0293 [hep-th]}}.

\bibitem{Weinberg:2008hq}
S.~Weinberg, ``{Effective Field Theory for Inflation},''
  \href{http://dx.doi.org/10.1103/PhysRevD.77.123541}{{\em Phys. Rev. D}
  {\bfseries 77} (2008) 123541},
  \href{http://arxiv.org/abs/0804.4291}{{\ttfamily arXiv:0804.4291 [hep-th]}}.

\bibitem{Choudhury:2017glj}
S.~Choudhury, ``{CMB from EFT},''
  \href{http://dx.doi.org/10.3390/universe5060155}{{\em Universe} {\bfseries 5}
  no.~6, (2019) 155}, \href{http://arxiv.org/abs/1712.04766}{{\ttfamily
  arXiv:1712.04766 [hep-th]}}.

\bibitem{Easther:2005zr}
R.~Easther and L.~McAllister, ``{Random matrices and the spectrum of
  N-flation},'' \href{http://dx.doi.org/10.1088/1475-7516/2006/05/018}{{\em
  JCAP} {\bfseries 05} (2006) 018},
  \href{http://arxiv.org/abs/hep-th/0512102}{{\ttfamily arXiv:hep-th/0512102}}.

\bibitem{Marsh:2013qca}
M.~C.~D. Marsh, L.~McAllister, E.~Pajer, and T.~Wrase, ``{Charting an
  Inflationary Landscape with Random Matrix Theory},''
  \href{http://dx.doi.org/10.1088/1475-7516/2013/11/040}{{\em JCAP} {\bfseries
  11} (2013) 040}, \href{http://arxiv.org/abs/1307.3559}{{\ttfamily
  arXiv:1307.3559 [hep-th]}}.

\bibitem{Amin:2015ftc}
M.~A. Amin and D.~Baumann, ``{From Wires to Cosmology},''
  \href{http://dx.doi.org/10.1088/1475-7516/2016/02/045}{{\em JCAP} {\bfseries
  02} (2016) 045}, \href{http://arxiv.org/abs/1512.02637}{{\ttfamily
  arXiv:1512.02637 [astro-ph.CO]}}.

\bibitem{Pedro:2016sli}
F.~G. Pedro and A.~Westphal, ``{Inflation with a graceful exit in a random
  landscape},'' \href{http://dx.doi.org/10.1007/JHEP03(2017)163}{{\em JHEP}
  {\bfseries 03} (2017) 163}, \href{http://arxiv.org/abs/1611.07059}{{\ttfamily
  arXiv:1611.07059 [hep-th]}}.

\bibitem{Choudhury:2018rjl}
S.~Choudhury, A.~Mukherjee, P.~Chauhan, and S.~Bhattacherjee, ``{Quantum
  Out-of-Equilibrium Cosmology},''
  \href{http://dx.doi.org/10.1140/epjc/s10052-019-6751-2}{{\em Eur. Phys. J. C}
  {\bfseries 79} no.~4, (2019) 320},
  \href{http://arxiv.org/abs/1809.02732}{{\ttfamily arXiv:1809.02732
  [hep-th]}}.

\bibitem{Choudhury:2018bcf}
S.~Choudhury and A.~Mukherjee, ``{Quantum randomness in the Sky},''
  \href{http://dx.doi.org/10.1140/epjc/s10052-019-7072-1}{{\em Eur. Phys. J. C}
  {\bfseries 79} no.~7, (2019) 554},
  \href{http://arxiv.org/abs/1812.04107}{{\ttfamily arXiv:1812.04107
  [physics.gen-ph]}}.

\bibitem{Paban:2018ole}
S.~Paban and R.~Rosati, ``{Inflation in Multi-field Modified DBM Potentials},''
  \href{http://dx.doi.org/10.1088/1475-7516/2018/09/042}{{\em JCAP} {\bfseries
  09} (2018) 042}, \href{http://arxiv.org/abs/1807.07654}{{\ttfamily
  arXiv:1807.07654 [astro-ph.CO]}}.

\bibitem{Sugiyama:2012tj}
N.~S. Sugiyama, E.~Komatsu, and T.~Futamase, ``{$\delta$N formalism},''
  \href{http://dx.doi.org/10.1103/PhysRevD.87.023530}{{\em Phys. Rev. D}
  {\bfseries 87} no.~2, (2013) 023530},
  \href{http://arxiv.org/abs/1208.1073}{{\ttfamily arXiv:1208.1073 [gr-qc]}}.

\bibitem{Dias:2012qy}
M.~Dias, R.~H. Ribeiro, and D.~Seery, ``{The \ensuremath{\delta}N formula is
  the dynamical renormalization group},''
  \href{http://dx.doi.org/10.1088/1475-7516/2013/10/062}{{\em JCAP} {\bfseries
  10} (2013) 062}, \href{http://arxiv.org/abs/1210.7800}{{\ttfamily
  arXiv:1210.7800 [astro-ph.CO]}}.

\bibitem{Naruko:2012fe}
A.~Naruko, Y.-i. Takamizu, and M.~Sasaki, ``{Beyond \textbackslash{}delta N
  formalism},'' \href{http://dx.doi.org/10.1093/ptep/ptt008}{{\em PTEP}
  {\bfseries 2013} (2013) 043E01},
  \href{http://arxiv.org/abs/1210.6525}{{\ttfamily arXiv:1210.6525
  [astro-ph.CO]}}.

\bibitem{Chen:2013eea}
X.~Chen, H.~Firouzjahi, E.~Komatsu, M.~H. Namjoo, and M.~Sasaki, ``{In-in and
  $\delta N$ calculations of the bispectrum from non-attractor single-field
  inflation},'' \href{http://dx.doi.org/10.1088/1475-7516/2013/12/039}{{\em
  JCAP} {\bfseries 12} (2013) 039},
  \href{http://arxiv.org/abs/1308.5341}{{\ttfamily arXiv:1308.5341
  [astro-ph.CO]}}.

\bibitem{Choudhury:2014uxa}
S.~Choudhury, ``{Constraining N = 1 supergravity inflation with non-minimal
  Kaehler operators using $\delta$N formalism},''
  \href{http://dx.doi.org/10.1007/JHEP04(2014)105}{{\em JHEP} {\bfseries 04}
  (2014) 105}, \href{http://arxiv.org/abs/1402.1251}{{\ttfamily arXiv:1402.1251
  [hep-th]}}.

\bibitem{Choudhury:2015hvr}
S.~Choudhury and S.~Panda, ``{COSMOS-e\textquoteright{}-GTachyon from string
  theory},'' \href{http://dx.doi.org/10.1140/epjc/s10052-016-4072-2}{{\em Eur.
  Phys. J. C} {\bfseries 76} no.~5, (2016) 278},
  \href{http://arxiv.org/abs/1511.05734}{{\ttfamily arXiv:1511.05734
  [hep-th]}}.

\bibitem{Hawking:1974rv}
S.~W. Hawking, ``{Black hole explosions},''
  \href{http://dx.doi.org/10.1038/248030a0}{{\em Nature} {\bfseries 248} (1974)
  30--31}.

\bibitem{Carr:1974nx}
B.~J. Carr and S.~W. Hawking, ``{Black holes in the early Universe},''
  \href{http://dx.doi.org/10.1093/mnras/168.2.399}{{\em Mon. Not. Roy. Astron.
  Soc.} {\bfseries 168} (1974) 399--415}.

\bibitem{Carr:1975qj}
B.~J. Carr, ``{The Primordial black hole mass spectrum},''
  \href{http://dx.doi.org/10.1086/153853}{{\em Astrophys. J.} {\bfseries 201}
  (1975) 1--19}.

\bibitem{Chapline:1975ojl}
G.~F. Chapline, ``{Cosmological effects of primordial black holes},''
  \href{http://dx.doi.org/10.1038/253251a0}{{\em Nature} {\bfseries 253}
  no.~5489, (1975) 251--252}.

\bibitem{Carr:1993aq}
B.~J. Carr and J.~E. Lidsey, ``{Primordial black holes and generalized
  constraints on chaotic inflation},''
  \href{http://dx.doi.org/10.1103/PhysRevD.48.543}{{\em Phys. Rev. D}
  {\bfseries 48} (1993) 543--553}.

\bibitem{Kawasaki:1997ju}
M.~Kawasaki, N.~Sugiyama, and T.~Yanagida, ``{Primordial black hole formation
  in a double inflation model in supergravity},''
  \href{http://dx.doi.org/10.1103/PhysRevD.57.6050}{{\em Phys. Rev. D}
  {\bfseries 57} (1998) 6050--6056},
  \href{http://arxiv.org/abs/hep-ph/9710259}{{\ttfamily arXiv:hep-ph/9710259}}.

\bibitem{Yokoyama:1998pt}
J.~Yokoyama, ``{Chaotic new inflation and formation of primordial black
  holes},'' \href{http://dx.doi.org/10.1103/PhysRevD.58.083510}{{\em Phys. Rev.
  D} {\bfseries 58} (1998) 083510},
  \href{http://arxiv.org/abs/astro-ph/9802357}{{\ttfamily
  arXiv:astro-ph/9802357}}.

\bibitem{Kawasaki:1998vx}
M.~Kawasaki and T.~Yanagida, ``{Primordial black hole formation in
  supergravity},'' \href{http://dx.doi.org/10.1103/PhysRevD.59.043512}{{\em
  Phys. Rev. D} {\bfseries 59} (1999) 043512},
  \href{http://arxiv.org/abs/hep-ph/9807544}{{\ttfamily arXiv:hep-ph/9807544}}.

\bibitem{Rubin:2001yw}
S.~G. Rubin, A.~S. Sakharov, and M.~Y. Khlopov, ``{The Formation of primary
  galactic nuclei during phase transitions in the early universe},''
  \href{http://dx.doi.org/10.1134/1.1385631}{{\em J. Exp. Theor. Phys.}
  {\bfseries 91} (2001) 921--929},
  \href{http://arxiv.org/abs/hep-ph/0106187}{{\ttfamily arXiv:hep-ph/0106187}}.

\bibitem{Khlopov:2002yi}
M.~Y. Khlopov, S.~G. Rubin, and A.~S. Sakharov, ``{Strong primordial
  inhomogeneities and galaxy formation},''
  \href{http://arxiv.org/abs/astro-ph/0202505}{{\ttfamily
  arXiv:astro-ph/0202505}}.

\bibitem{Khlopov:2004sc}
M.~Y. Khlopov, S.~G. Rubin, and A.~S. Sakharov, ``{Primordial structure of
  massive black hole clusters},''
  \href{http://dx.doi.org/10.1016/j.astropartphys.2004.12.002}{{\em Astropart.
  Phys.} {\bfseries 23} (2005) 265},
  \href{http://arxiv.org/abs/astro-ph/0401532}{{\ttfamily
  arXiv:astro-ph/0401532}}.

\bibitem{Saito:2008em}
R.~Saito, J.~Yokoyama, and R.~Nagata, ``{Single-field inflation, anomalous
  enhancement of superhorizon fluctuations, and non-Gaussianity in primordial
  black hole formation},''
  \href{http://dx.doi.org/10.1088/1475-7516/2008/06/024}{{\em JCAP} {\bfseries
  06} (2008) 024}, \href{http://arxiv.org/abs/0804.3470}{{\ttfamily
  arXiv:0804.3470 [astro-ph]}}.

\bibitem{Khlopov:2008qy}
M.~Y. Khlopov, ``{Primordial Black Holes},''
  \href{http://dx.doi.org/10.1088/1674-4527/10/6/001}{{\em Res. Astron.
  Astrophys.} {\bfseries 10} (2010) 495--528},
  \href{http://arxiv.org/abs/0801.0116}{{\ttfamily arXiv:0801.0116
  [astro-ph]}}.

\bibitem{Carr:2009jm}
B.~J. Carr, K.~Kohri, Y.~Sendouda, and J.~Yokoyama, ``{New cosmological
  constraints on primordial black holes},''
  \href{http://dx.doi.org/10.1103/PhysRevD.81.104019}{{\em Phys. Rev. D}
  {\bfseries 81} (2010) 104019},
  \href{http://arxiv.org/abs/0912.5297}{{\ttfamily arXiv:0912.5297
  [astro-ph.CO]}}.

\bibitem{Choudhury:2011jt}
S.~Choudhury and S.~Pal, ``{Fourth level MSSM inflation from new flat
  directions},'' \href{http://dx.doi.org/10.1088/1475-7516/2012/04/018}{{\em
  JCAP} {\bfseries 04} (2012) 018},
  \href{http://arxiv.org/abs/1111.3441}{{\ttfamily arXiv:1111.3441 [hep-ph]}}.

\bibitem{Choudhury:2013woa}
S.~Choudhury and A.~Mazumdar, ``{Primordial blackholes and gravitational waves
  for an inflection-point model of inflation},''
  \href{http://dx.doi.org/10.1016/j.physletb.2014.04.050}{{\em Phys. Lett. B}
  {\bfseries 733} (2014) 270--275},
  \href{http://arxiv.org/abs/1307.5119}{{\ttfamily arXiv:1307.5119
  [astro-ph.CO]}}.

\bibitem{Lyth:2011kj}
D.~H. Lyth, ``{Primordial black hole formation and hybrid inflation},''
  \href{http://arxiv.org/abs/1107.1681}{{\ttfamily arXiv:1107.1681
  [astro-ph.CO]}}.

\bibitem{Drees:2011yz}
M.~Drees and E.~Erfani, ``{Running Spectral Index and Formation of Primordial
  Black Hole in Single Field Inflation Models},''
  \href{http://dx.doi.org/10.1088/1475-7516/2012/01/035}{{\em JCAP} {\bfseries
  01} (2012) 035}, \href{http://arxiv.org/abs/1110.6052}{{\ttfamily
  arXiv:1110.6052 [astro-ph.CO]}}.

\bibitem{Drees:2011hb}
M.~Drees and E.~Erfani, ``{Running-Mass Inflation Model and Primordial Black
  Holes},'' \href{http://dx.doi.org/10.1088/1475-7516/2011/04/005}{{\em JCAP}
  {\bfseries 04} (2011) 005}, \href{http://arxiv.org/abs/1102.2340}{{\ttfamily
  arXiv:1102.2340 [hep-ph]}}.

\bibitem{Ezquiaga:2017fvi}
J.~M. Ezquiaga, J.~Garcia-Bellido, and E.~Ruiz~Morales, ``{Primordial Black
  Hole production in Critical Higgs Inflation},''
  \href{http://dx.doi.org/10.1016/j.physletb.2017.11.039}{{\em Phys. Lett. B}
  {\bfseries 776} (2018) 345--349},
  \href{http://arxiv.org/abs/1705.04861}{{\ttfamily arXiv:1705.04861
  [astro-ph.CO]}}.

\bibitem{Kannike:2017bxn}
K.~Kannike, L.~Marzola, M.~Raidal, and H.~Veerm\"ae, ``{Single Field Double
  Inflation and Primordial Black Holes},''
  \href{http://dx.doi.org/10.1088/1475-7516/2017/09/020}{{\em JCAP} {\bfseries
  09} (2017) 020}, \href{http://arxiv.org/abs/1705.06225}{{\ttfamily
  arXiv:1705.06225 [astro-ph.CO]}}.

\bibitem{Hertzberg:2017dkh}
M.~P. Hertzberg and M.~Yamada, ``{Primordial Black Holes from Polynomial
  Potentials in Single Field Inflation},''
  \href{http://dx.doi.org/10.1103/PhysRevD.97.083509}{{\em Phys. Rev. D}
  {\bfseries 97} no.~8, (2018) 083509},
  \href{http://arxiv.org/abs/1712.09750}{{\ttfamily arXiv:1712.09750
  [astro-ph.CO]}}.

\bibitem{Pi:2017gih}
S.~Pi, Y.-l. Zhang, Q.-G. Huang, and M.~Sasaki, ``{Scalaron from $R^2$-gravity
  as a heavy field},''
  \href{http://dx.doi.org/10.1088/1475-7516/2018/05/042}{{\em JCAP} {\bfseries
  05} (2018) 042}, \href{http://arxiv.org/abs/1712.09896}{{\ttfamily
  arXiv:1712.09896 [astro-ph.CO]}}.

\bibitem{Gao:2018pvq}
T.-J. Gao and Z.-K. Guo, ``{Primordial Black Hole Production in Inflationary
  Models of Supergravity with a Single Chiral Superfield},''
  \href{http://dx.doi.org/10.1103/PhysRevD.98.063526}{{\em Phys. Rev. D}
  {\bfseries 98} no.~6, (2018) 063526},
  \href{http://arxiv.org/abs/1806.09320}{{\ttfamily arXiv:1806.09320
  [hep-ph]}}.

\bibitem{Dalianis:2018frf}
I.~Dalianis, A.~Kehagias, and G.~Tringas, ``{Primordial black holes from
  \ensuremath{\alpha}-attractors},''
  \href{http://dx.doi.org/10.1088/1475-7516/2019/01/037}{{\em JCAP} {\bfseries
  01} (2019) 037}, \href{http://arxiv.org/abs/1805.09483}{{\ttfamily
  arXiv:1805.09483 [astro-ph.CO]}}.

\bibitem{Cicoli:2018asa}
M.~Cicoli, V.~A. Diaz, and F.~G. Pedro, ``{Primordial Black Holes from String
  Inflation},'' \href{http://dx.doi.org/10.1088/1475-7516/2018/06/034}{{\em
  JCAP} {\bfseries 06} (2018) 034},
  \href{http://arxiv.org/abs/1803.02837}{{\ttfamily arXiv:1803.02837
  [hep-th]}}.

\bibitem{Ozsoy:2018flq}
O.~\"Ozsoy, S.~Parameswaran, G.~Tasinato, and I.~Zavala, ``{Mechanisms for
  Primordial Black Hole Production in String Theory},''
  \href{http://dx.doi.org/10.1088/1475-7516/2018/07/005}{{\em JCAP} {\bfseries
  07} (2018) 005}, \href{http://arxiv.org/abs/1803.07626}{{\ttfamily
  arXiv:1803.07626 [hep-th]}}.

\bibitem{Byrnes:2018txb}
C.~T. Byrnes, P.~S. Cole, and S.~P. Patil, ``{Steepest growth of the power
  spectrum and primordial black holes},''
  \href{http://dx.doi.org/10.1088/1475-7516/2019/06/028}{{\em JCAP} {\bfseries
  06} (2019) 028}, \href{http://arxiv.org/abs/1811.11158}{{\ttfamily
  arXiv:1811.11158 [astro-ph.CO]}}.

\bibitem{Ballesteros:2018wlw}
G.~Ballesteros, J.~Beltran~Jimenez, and M.~Pieroni, ``{Black hole formation
  from a general quadratic action for inflationary primordial fluctuations},''
  \href{http://dx.doi.org/10.1088/1475-7516/2019/06/016}{{\em JCAP} {\bfseries
  06} (2019) 016}, \href{http://arxiv.org/abs/1811.03065}{{\ttfamily
  arXiv:1811.03065 [astro-ph.CO]}}.

\bibitem{Belotsky:2018wph}
K.~M. Belotsky, V.~I. Dokuchaev, Y.~N. Eroshenko, E.~A. Esipova, M.~Y. Khlopov,
  L.~A. Khromykh, A.~A. Kirillov, V.~V. Nikulin, S.~G. Rubin, and I.~V.
  Svadkovsky, ``{Clusters of primordial black holes},''
  \href{http://dx.doi.org/10.1140/epjc/s10052-019-6741-4}{{\em Eur. Phys. J. C}
  {\bfseries 79} no.~3, (2019) 246},
  \href{http://arxiv.org/abs/1807.06590}{{\ttfamily arXiv:1807.06590
  [astro-ph.CO]}}.

\bibitem{Martin:2019nuw}
J.~Martin, T.~Papanikolaou, and V.~Vennin, ``{Primordial black holes from the
  preheating instability in single-field inflation},''
  \href{http://dx.doi.org/10.1088/1475-7516/2020/01/024}{{\em JCAP} {\bfseries
  01} (2020) 024}, \href{http://arxiv.org/abs/1907.04236}{{\ttfamily
  arXiv:1907.04236 [astro-ph.CO]}}.

\bibitem{Ezquiaga:2019ftu}
J.~M. Ezquiaga, J.~Garc\'\i{}a-Bellido, and V.~Vennin, ``{The exponential tail
  of inflationary fluctuations: consequences for primordial black holes},''
  \href{http://dx.doi.org/10.1088/1475-7516/2020/03/029}{{\em JCAP} {\bfseries
  03} (2020) 029}, \href{http://arxiv.org/abs/1912.05399}{{\ttfamily
  arXiv:1912.05399 [astro-ph.CO]}}.

\bibitem{Motohashi:2019rhu}
H.~Motohashi, S.~Mukohyama, and M.~Oliosi, ``{Constant Roll and Primordial
  Black Holes},'' \href{http://dx.doi.org/10.1088/1475-7516/2020/03/002}{{\em
  JCAP} {\bfseries 03} (2020) 002},
  \href{http://arxiv.org/abs/1910.13235}{{\ttfamily arXiv:1910.13235 [gr-qc]}}.

\bibitem{Fu:2019ttf}
C.~Fu, P.~Wu, and H.~Yu, ``{Primordial Black Holes from Inflation with
  Nonminimal Derivative Coupling},''
  \href{http://dx.doi.org/10.1103/PhysRevD.100.063532}{{\em Phys. Rev. D}
  {\bfseries 100} no.~6, (2019) 063532},
  \href{http://arxiv.org/abs/1907.05042}{{\ttfamily arXiv:1907.05042
  [astro-ph.CO]}}.

\bibitem{Ashoorioon:2019xqc}
A.~Ashoorioon, A.~Rostami, and J.~T. Firouzjaee, ``{EFT compatible PBHs:
  effective spawning of the seeds for primordial black holes during
  inflation},'' \href{http://dx.doi.org/10.1007/JHEP07(2021)087}{{\em JHEP}
  {\bfseries 07} (2021) 087}, \href{http://arxiv.org/abs/1912.13326}{{\ttfamily
  arXiv:1912.13326 [astro-ph.CO]}}.

\bibitem{Auclair:2020csm}
P.~Auclair and V.~Vennin, ``{Primordial black holes from metric preheating:
  mass fraction in the excursion-set approach},''
  \href{http://dx.doi.org/10.1088/1475-7516/2021/02/038}{{\em JCAP} {\bfseries
  02} (2021) 038}, \href{http://arxiv.org/abs/2011.05633}{{\ttfamily
  arXiv:2011.05633 [astro-ph.CO]}}.

\bibitem{Vennin:2020kng}
V.~Vennin, {\em {Stochastic inflation and primordial black holes}}.
\newblock PhD thesis, U. Paris-Saclay, 6, 2020.
\newblock \href{http://arxiv.org/abs/2009.08715}{{\ttfamily arXiv:2009.08715
  [astro-ph.CO]}}.

\bibitem{Nanopoulos:2020nnh}
D.~V. Nanopoulos, V.~C. Spanos, and I.~D. Stamou, ``{Primordial Black Holes
  from No-Scale Supergravity},''
  \href{http://dx.doi.org/10.1103/PhysRevD.102.083536}{{\em Phys. Rev. D}
  {\bfseries 102} no.~8, (2020) 083536},
  \href{http://arxiv.org/abs/2008.01457}{{\ttfamily arXiv:2008.01457
  [astro-ph.CO]}}.

\bibitem{Inomata:2021uqj}
K.~Inomata, E.~McDonough, and W.~Hu, ``{Primordial black holes arise when the
  inflaton falls},'' \href{http://dx.doi.org/10.1103/PhysRevD.104.123553}{{\em
  Phys. Rev. D} {\bfseries 104} no.~12, (2021) 123553},
  \href{http://arxiv.org/abs/2104.03972}{{\ttfamily arXiv:2104.03972
  [astro-ph.CO]}}.

\bibitem{Stamou:2021qdk}
I.~D. Stamou, ``{Mechanisms of producing primordial black holes by breaking the
  $SU(2, 1)/SU(2)\times U(1)$ symmetry},''
  \href{http://dx.doi.org/10.1103/PhysRevD.103.083512}{{\em Phys. Rev. D}
  {\bfseries 103} no.~8, (2021) 083512},
  \href{http://arxiv.org/abs/2104.08654}{{\ttfamily arXiv:2104.08654
  [hep-ph]}}.

\bibitem{Ng:2021hll}
K.-W. Ng and Y.-P. Wu, ``{Constant-rate inflation: primordial black holes from
  conformal weight transitions},''
  \href{http://dx.doi.org/10.1007/JHEP11(2021)076}{{\em JHEP} {\bfseries 11}
  (2021) 076}, \href{http://arxiv.org/abs/2102.05620}{{\ttfamily
  arXiv:2102.05620 [astro-ph.CO]}}.

\bibitem{Wang:2021kbh}
Q.~Wang, Y.-C. Liu, B.-Y. Su, and N.~Li, ``{Primordial black holes from the
  perturbations in the inflaton potential in peak theory},''
  \href{http://dx.doi.org/10.1103/PhysRevD.104.083546}{{\em Phys. Rev. D}
  {\bfseries 104} no.~8, (2021) 083546},
  \href{http://arxiv.org/abs/2111.10028}{{\ttfamily arXiv:2111.10028
  [astro-ph.CO]}}.

\bibitem{Kawai:2021edk}
S.~Kawai and J.~Kim, ``{Primordial black holes from Gauss-Bonnet-corrected
  single field inflation},''
  \href{http://dx.doi.org/10.1103/PhysRevD.104.083545}{{\em Phys. Rev. D}
  {\bfseries 104} no.~8, (2021) 083545},
  \href{http://arxiv.org/abs/2108.01340}{{\ttfamily arXiv:2108.01340
  [astro-ph.CO]}}.

\bibitem{Solbi:2021rse}
M.~Solbi and K.~Karami, ``{Primordial black holes formation in the inflationary
  model with field-dependent kinetic term for quartic and natural
  potentials},'' \href{http://dx.doi.org/10.1140/epjc/s10052-021-09690-9}{{\em
  Eur. Phys. J. C} {\bfseries 81} no.~10, (2021) 884},
  \href{http://arxiv.org/abs/2106.02863}{{\ttfamily arXiv:2106.02863
  [astro-ph.CO]}}.

\bibitem{Ballesteros:2021fsp}
G.~Ballesteros, S.~C\'espedes, and L.~Santoni, ``{Large power spectrum and
  primordial black holes in the effective theory of inflation},''
  \href{http://dx.doi.org/10.1007/JHEP01(2022)074}{{\em JHEP} {\bfseries 01}
  (2022) 074}, \href{http://arxiv.org/abs/2109.00567}{{\ttfamily
  arXiv:2109.00567 [hep-th]}}.

\bibitem{Rigopoulos:2021nhv}
G.~Rigopoulos and A.~Wilkins, ``{Inflation is always semi-classical: diffusion
  domination overproduces Primordial Black Holes},''
  \href{http://dx.doi.org/10.1088/1475-7516/2021/12/027}{{\em JCAP} {\bfseries
  12} no.~12, (2021) 027}, \href{http://arxiv.org/abs/2107.05317}{{\ttfamily
  arXiv:2107.05317 [astro-ph.CO]}}.

\bibitem{Animali:2022otk}
C.~Animali and V.~Vennin, ``{Primordial black holes from stochastic
  tunnelling},'' \href{http://arxiv.org/abs/2210.03812}{{\ttfamily
  arXiv:2210.03812 [astro-ph.CO]}}.

\bibitem{Frolovsky:2022ewg}
D.~Frolovsky, S.~V. Ketov, and S.~Saburov, ``{Formation of primordial black
  holes after Starobinsky inflation},''
  \href{http://dx.doi.org/10.1142/S0217732322501358}{{\em Mod. Phys. Lett. A}
  {\bfseries 37} no.~21, (2022) 2250135},
  \href{http://arxiv.org/abs/2205.00603}{{\ttfamily arXiv:2205.00603
  [astro-ph.CO]}}.

\bibitem{Escriva:2022duf}
A.~Escriv\`a, F.~Kuhnel, and Y.~Tada, ``{Primordial Black Holes},''
  \href{http://arxiv.org/abs/2211.05767}{{\ttfamily arXiv:2211.05767
  [astro-ph.CO]}}.

\bibitem{Karam:2022nym}
A.~Karam, N.~Koivunen, E.~Tomberg, V.~Vaskonen, and H.~Veerm\"ae, ``{Anatomy of
  single-field inflationary models for primordial black holes},''
  \href{http://arxiv.org/abs/2205.13540}{{\ttfamily arXiv:2205.13540
  [astro-ph.CO]}}.

\bibitem{Ozsoy:2023ryl}
O.~\"Ozsoy and G.~Tasinato, ``{Inflation and Primordial Black Holes},''
  \href{http://arxiv.org/abs/2301.03600}{{\ttfamily arXiv:2301.03600
  [astro-ph.CO]}}.

\bibitem{Kristiano:2022maq}
J.~Kristiano and J.~Yokoyama, ``{Ruling Out Primordial Black Hole Formation
  From Single-Field Inflation},''
  \href{http://arxiv.org/abs/2211.03395}{{\ttfamily arXiv:2211.03395
  [hep-th]}}.

\bibitem{Kristiano:2023scm}
J.~Kristiano and J.~Yokoyama, ``{Response to criticism on ''Ruling Out
  Primordial Black Hole Formation From Single-Field Inflation'': A note on
  bispectrum and one-loop correction in single-field inflation with primordial
  black hole formation},'' \href{http://arxiv.org/abs/2303.00341}{{\ttfamily
  arXiv:2303.00341 [hep-th]}}.

\bibitem{Riotto:2023hoz}
A.~Riotto, ``{The Primordial Black Hole Formation from Single-Field Inflation
  is Not Ruled Out},'' \href{http://arxiv.org/abs/2301.00599}{{\ttfamily
  arXiv:2301.00599 [astro-ph.CO]}}.

\bibitem{Riotto:2023gpm}
A.~Riotto, ``{The Primordial Black Hole Formation from Single-Field Inflation
  is Still Not Ruled Out},'' \href{http://arxiv.org/abs/2303.01727}{{\ttfamily
  arXiv:2303.01727 [astro-ph.CO]}}.

\bibitem{Choudhury:2023vuj}
S.~Choudhury, M.~R. Gangopadhyay, and M.~Sami, ``{No-go for the formation of
  heavy mass Primordial Black Holes in Single Field Inflation},''
  \href{http://arxiv.org/abs/2301.10000}{{\ttfamily arXiv:2301.10000
  [astro-ph.CO]}}.

\bibitem{Choudhury:2023jlt}
S.~Choudhury, S.~Panda, and M.~Sami, ``{PBH formation in EFT of single field
  inflation with sharp transition},''
  \href{http://dx.doi.org/10.1016/j.physletb.2023.138123}{{\em Phys. Lett. B}
  {\bfseries 845} (2023) 138123},
  \href{http://arxiv.org/abs/2302.05655}{{\ttfamily arXiv:2302.05655
  [astro-ph.CO]}}.

\bibitem{Choudhury:2023rks}
S.~Choudhury, S.~Panda, and M.~Sami, ``{Quantum loop effects on the power
  spectrum and constraints on primordial black holes},''
  \href{http://dx.doi.org/10.1088/1475-7516/2023/11/066}{{\em JCAP} {\bfseries
  11} (2023) 066}, \href{http://arxiv.org/abs/2303.06066}{{\ttfamily
  arXiv:2303.06066 [astro-ph.CO]}}.

\bibitem{Choudhury:2023hvf}
S.~Choudhury, S.~Panda, and M.~Sami, ``{Galileon inflation evades the no-go for
  PBH formation in the single-field framework},''
  \href{http://dx.doi.org/10.1088/1475-7516/2023/08/078}{{\em JCAP} {\bfseries
  08} (2023) 078}, \href{http://arxiv.org/abs/2304.04065}{{\ttfamily
  arXiv:2304.04065 [astro-ph.CO]}}.

\bibitem{Kawaguchi:2023mgk}
R.~Kawaguchi, T.~Fujita, and M.~Sasaki, ``{Highly asymmetric probability
  distribution from a finite-width upward step during inflation},''
  \href{http://arxiv.org/abs/2305.18140}{{\ttfamily arXiv:2305.18140
  [astro-ph.CO]}}.

\bibitem{Fu:2022ypp}
C.~Fu and S.-J. Wang, ``{Primordial black holes and induced gravitational waves
  from double-pole inflation},''
  \href{http://dx.doi.org/10.1088/1475-7516/2023/06/012}{{\em JCAP} {\bfseries
  06} (2023) 012}, \href{http://arxiv.org/abs/2211.03523}{{\ttfamily
  arXiv:2211.03523 [astro-ph.CO]}}.

\bibitem{Saburov:2023buy}
S.~Saburov and S.~V. Ketov, ``{Improved model of primordial black hole
  formation after Starobinsky inflation},''
  \href{http://arxiv.org/abs/2306.06597}{{\ttfamily arXiv:2306.06597 [gr-qc]}}.

\bibitem{Ghoshal:2023wri}
A.~Ghoshal, A.~Moursy, and Q.~Shafi, ``{Cosmological probes of Grand
  Unification: Primordial Blackholes \& scalar-induced Gravitational Waves},''
  \href{http://arxiv.org/abs/2306.04002}{{\ttfamily arXiv:2306.04002
  [hep-ph]}}.

\bibitem{Karam:2023haj}
A.~Karam, N.~Koivunen, E.~Tomberg, A.~Racioppi, and H.~Veerm\"ae, ``{Primordial
  black holes and inflation from double-well potentials},''
  \href{http://arxiv.org/abs/2305.09630}{{\ttfamily arXiv:2305.09630
  [astro-ph.CO]}}.

\bibitem{Poisson:2023tja}
A.~Poisson, I.~Timiryasov, and S.~Zell, ``{Critical Points in Palatini Higgs
  Inflation with Small Non-Minimal Coupling},''
  \href{http://arxiv.org/abs/2306.03893}{{\ttfamily arXiv:2306.03893
  [hep-ph]}}.

\bibitem{Iacconi:2023slv}
L.~Iacconi and D.~J. Mulryne, ``{Multi-field inflation with large scalar
  fluctuations: non-Gaussianity and perturbativity},''
  \href{http://arxiv.org/abs/2304.14260}{{\ttfamily arXiv:2304.14260
  [astro-ph.CO]}}.

\bibitem{Mishra:2019pzq}
S.~S. Mishra and V.~Sahni, ``{Primordial Black Holes from a tiny bump/dip in
  the Inflaton potential},''
  \href{http://dx.doi.org/10.1088/1475-7516/2020/04/007}{{\em JCAP} {\bfseries
  04} (2020) 007}, \href{http://arxiv.org/abs/1911.00057}{{\ttfamily
  arXiv:1911.00057 [gr-qc]}}.

\bibitem{Mishra:2023lhe}
S.~S. Mishra, E.~J. Copeland, and A.~M. Green, ``{Primordial black holes and
  stochastic inflation beyond slow roll: I -- noise matrix elements},''
  \href{http://arxiv.org/abs/2303.17375}{{\ttfamily arXiv:2303.17375
  [astro-ph.CO]}}.

\bibitem{Gangopadhyay:2021kmf}
M.~R. Gangopadhyay, J.~C. Jain, D.~Sharma, and Yogesh, ``{Production of
  primordial black holes via single field inflation and observational
  constraints},'' \href{http://dx.doi.org/10.1140/epjc/s10052-022-10796-x}{{\em
  Eur. Phys. J. C} {\bfseries 82} no.~9, (2022) 849},
  \href{http://arxiv.org/abs/2108.13839}{{\ttfamily arXiv:2108.13839
  [astro-ph.CO]}}.

\bibitem{Bhattacharya:2023ysp}
G.~Bhattacharya, S.~Choudhury, K.~Dey, S.~Ghosh, A.~Karde, and N.~S. Mishra,
  ``{Evading no-go for PBH formation and production of SIGWs using Multiple
  Sharp Transitions in EFT of single field inflation},''
  \href{http://arxiv.org/abs/2309.00973}{{\ttfamily arXiv:2309.00973
  [astro-ph.CO]}}.

\bibitem{Choudhury:2023fjs}
S.~Choudhury, K.~Dey, and A.~Karde, ``{Untangling PBH overproduction in
  $w$-SIGWs generated by Pulsar Timing Arrays for MST-EFT of single field
  inflation},'' \href{http://arxiv.org/abs/2311.15065}{{\ttfamily
  arXiv:2311.15065 [astro-ph.CO]}}.

\bibitem{Choudhury:2023fwk}
S.~Choudhury, K.~Dey, A.~Karde, S.~Panda, and M.~Sami, ``{Primordial
  non-Gaussianity as a saviour for PBH overproduction in SIGWs generated by
  Pulsar Timing Arrays for Galileon inflation},''
  \href{http://arxiv.org/abs/2310.11034}{{\ttfamily arXiv:2310.11034
  [astro-ph.CO]}}.

\bibitem{Choudhury:2023hfm}
S.~Choudhury, A.~Karde, S.~Panda, and M.~Sami, ``{Scalar induced gravity waves
  from ultra slow-roll Galileon inflation},''
  \href{http://arxiv.org/abs/2308.09273}{{\ttfamily arXiv:2308.09273
  [astro-ph.CO]}}.

\bibitem{Choudhury:2023kam}
S.~Choudhury, ``{Single field inflation in the light of NANOGrav 15-year Data:
  Quintessential interpretation of blue tilted tensor spectrum through
  Non-Bunch Davies initial condition},''
  \href{http://arxiv.org/abs/2307.03249}{{\ttfamily arXiv:2307.03249
  [astro-ph.CO]}}.

\bibitem{Firouzjahi:2023aum}
H.~Firouzjahi, ``{One-loop Corrections in Power Spectrum in Single Field
  Inflation},'' \href{http://arxiv.org/abs/2303.12025}{{\ttfamily
  arXiv:2303.12025 [astro-ph.CO]}}.

\bibitem{Motohashi:2023syh}
H.~Motohashi and Y.~Tada, ``{Squeezed bispectrum and one-loop corrections in
  transient constant-roll inflation},''
  \href{http://arxiv.org/abs/2303.16035}{{\ttfamily arXiv:2303.16035
  [astro-ph.CO]}}.

\bibitem{Firouzjahi:2023ahg}
H.~Firouzjahi and A.~Riotto, ``{Primordial Black Holes and Loops in
  Single-Field Inflation},'' \href{http://arxiv.org/abs/2304.07801}{{\ttfamily
  arXiv:2304.07801 [astro-ph.CO]}}.

\bibitem{Franciolini:2023lgy}
G.~Franciolini, A.~Iovino, Junior., M.~Taoso, and A.~Urbano, ``{One loop to
  rule them all: Perturbativity in the presence of ultra slow-roll dynamics},''
  \href{http://arxiv.org/abs/2305.03491}{{\ttfamily arXiv:2305.03491
  [astro-ph.CO]}}.

\bibitem{Tasinato:2023ukp}
G.~Tasinato, ``{A large $|\eta|$ approach to single field inflation},''
  \href{http://arxiv.org/abs/2305.11568}{{\ttfamily arXiv:2305.11568
  [hep-th]}}.

\bibitem{Cheng:2023ikq}
S.-L. Cheng, D.-S. Lee, and K.-W. Ng, ``{Primordial perturbations from
  ultra-slow-roll single-field inflation with quantum loop effects},''
  \href{http://arxiv.org/abs/2305.16810}{{\ttfamily arXiv:2305.16810
  [astro-ph.CO]}}.

\bibitem{Horndeski:1974wa}
G.~W. Horndeski, ``{Second-order scalar-tensor field equations in a
  four-dimensional space},'' \href{http://dx.doi.org/10.1007/BF01807638}{{\em
  Int. J. Theor. Phys.} {\bfseries 10} (1974) 363--384}.

\bibitem{Kobayashi:2010wa}
T.~Kobayashi, ``{Cosmic expansion and growth histories in Galileon
  scalar-tensor models of dark energy},''
  \href{http://dx.doi.org/10.1103/PhysRevD.81.103533}{{\em Phys. Rev. D}
  {\bfseries 81} (2010) 103533},
  \href{http://arxiv.org/abs/1003.3281}{{\ttfamily arXiv:1003.3281
  [astro-ph.CO]}}.

\bibitem{Jain:2010ka}
B.~Jain and J.~Khoury, ``{Cosmological Tests of Gravity},''
  \href{http://dx.doi.org/10.1016/j.aop.2010.04.002}{{\em Annals Phys.}
  {\bfseries 325} (2010) 1479--1516},
  \href{http://arxiv.org/abs/1004.3294}{{\ttfamily arXiv:1004.3294
  [astro-ph.CO]}}.

\bibitem{Gannouji:2010au}
R.~Gannouji and M.~Sami, ``{Galileon gravity and its relevance to late time
  cosmic acceleration},''
  \href{http://dx.doi.org/10.1103/PhysRevD.82.024011}{{\em Phys. Rev. D}
  {\bfseries 82} (2010) 024011},
  \href{http://arxiv.org/abs/1004.2808}{{\ttfamily arXiv:1004.2808 [gr-qc]}}.

\bibitem{Ali:2010gr}
A.~Ali, R.~Gannouji, and M.~Sami, ``{Modified gravity a la Galileon: Late time
  cosmic acceleration and observational constraints},''
  \href{http://dx.doi.org/10.1103/PhysRevD.82.103015}{{\em Phys. Rev. D}
  {\bfseries 82} (2010) 103015},
  \href{http://arxiv.org/abs/1008.1588}{{\ttfamily arXiv:1008.1588
  [astro-ph.CO]}}.

\bibitem{deRham:2011by}
C.~de~Rham and L.~Heisenberg, ``{Cosmology of the Galileon from Massive
  Gravity},'' \href{http://dx.doi.org/10.1103/PhysRevD.84.043503}{{\em Phys.
  Rev. D} {\bfseries 84} (2011) 043503},
  \href{http://arxiv.org/abs/1106.3312}{{\ttfamily arXiv:1106.3312 [hep-th]}}.

\bibitem{Tsujikawa:2010sc}
S.~Tsujikawa, ``{Dark energy: investigation and modeling},''
  \href{http://arxiv.org/abs/1004.1493}{{\ttfamily arXiv:1004.1493
  [astro-ph.CO]}}.

\bibitem{Burrage:2010rs}
C.~Burrage and D.~Seery, ``{Revisiting fifth forces in the Galileon model},''
  \href{http://dx.doi.org/10.1088/1475-7516/2010/08/011}{{\em JCAP} {\bfseries
  08} (2010) 011}, \href{http://arxiv.org/abs/1005.1927}{{\ttfamily
  arXiv:1005.1927 [astro-ph.CO]}}.

\bibitem{DeFelice:2010jn}
A.~De~Felice and S.~Tsujikawa, ``{Generalized Brans-Dicke theories},''
  \href{http://dx.doi.org/10.1088/1475-7516/2010/07/024}{{\em JCAP} {\bfseries
  07} (2010) 024}, \href{http://arxiv.org/abs/1005.0868}{{\ttfamily
  arXiv:1005.0868 [astro-ph.CO]}}.

\bibitem{DeFelice:2010gb}
A.~De~Felice, S.~Mukohyama, and S.~Tsujikawa, ``{Density perturbations in
  general modified gravitational theories},''
  \href{http://dx.doi.org/10.1103/PhysRevD.82.023524}{{\em Phys. Rev. D}
  {\bfseries 82} (2010) 023524},
  \href{http://arxiv.org/abs/1006.0281}{{\ttfamily arXiv:1006.0281
  [astro-ph.CO]}}.

\bibitem{Babichev:2010jd}
E.~Babichev, C.~Deffayet, and R.~Ziour, ``{The Recovery of General Relativity
  in massive gravity via the Vainshtein mechanism},''
  \href{http://dx.doi.org/10.1103/PhysRevD.82.104008}{{\em Phys. Rev. D}
  {\bfseries 82} (2010) 104008},
  \href{http://arxiv.org/abs/1007.4506}{{\ttfamily arXiv:1007.4506 [gr-qc]}}.

\bibitem{DeFelice:2010pv}
A.~De~Felice and S.~Tsujikawa, ``{Cosmology of a covariant Galileon field},''
  \href{http://dx.doi.org/10.1103/PhysRevLett.105.111301}{{\em Phys. Rev.
  Lett.} {\bfseries 105} (2010) 111301},
  \href{http://arxiv.org/abs/1007.2700}{{\ttfamily arXiv:1007.2700
  [astro-ph.CO]}}.

\bibitem{DeFelice:2010nf}
A.~De~Felice and S.~Tsujikawa, ``{Generalized Galileon cosmology},''
  \href{http://dx.doi.org/10.1103/PhysRevD.84.124029}{{\em Phys. Rev. D}
  {\bfseries 84} (2011) 124029},
  \href{http://arxiv.org/abs/1008.4236}{{\ttfamily arXiv:1008.4236 [hep-th]}}.

\bibitem{Hinterbichler:2010xn}
K.~Hinterbichler, M.~Trodden, and D.~Wesley, ``{Multi-field galileons and
  higher co-dimension branes},''
  \href{http://dx.doi.org/10.1103/PhysRevD.82.124018}{{\em Phys. Rev. D}
  {\bfseries 82} (2010) 124018},
  \href{http://arxiv.org/abs/1008.1305}{{\ttfamily arXiv:1008.1305 [hep-th]}}.

\bibitem{Deffayet:2010qz}
C.~Deffayet, O.~Pujolas, I.~Sawicki, and A.~Vikman, ``{Imperfect Dark Energy
  from Kinetic Gravity Braiding},''
  \href{http://dx.doi.org/10.1088/1475-7516/2010/10/026}{{\em JCAP} {\bfseries
  10} (2010) 026}, \href{http://arxiv.org/abs/1008.0048}{{\ttfamily
  arXiv:1008.0048 [hep-th]}}.

\bibitem{Nesseris:2010pc}
S.~Nesseris, A.~De~Felice, and S.~Tsujikawa, ``{Observational constraints on
  Galileon cosmology},''
  \href{http://dx.doi.org/10.1103/PhysRevD.82.124054}{{\em Phys. Rev. D}
  {\bfseries 82} (2010) 124054},
  \href{http://arxiv.org/abs/1010.0407}{{\ttfamily arXiv:1010.0407
  [astro-ph.CO]}}.

\bibitem{Khoury:2010xi}
J.~Khoury, ``{Theories of Dark Energy with Screening Mechanisms},''
  \href{http://arxiv.org/abs/1011.5909}{{\ttfamily arXiv:1011.5909
  [astro-ph.CO]}}.

\bibitem{DeFelice:2010as}
A.~De~Felice, R.~Kase, and S.~Tsujikawa, ``{Matter perturbations in Galileon
  cosmology},'' \href{http://dx.doi.org/10.1103/PhysRevD.83.043515}{{\em Phys.
  Rev. D} {\bfseries 83} (2011) 043515},
  \href{http://arxiv.org/abs/1011.6132}{{\ttfamily arXiv:1011.6132
  [astro-ph.CO]}}.

\bibitem{Kimura:2010di}
R.~Kimura and K.~Yamamoto, ``{Large Scale Structures in Kinetic Gravity
  Braiding Model That Can Be Unbraided},''
  \href{http://dx.doi.org/10.1088/1475-7516/2011/04/025}{{\em JCAP} {\bfseries
  04} (2011) 025}, \href{http://arxiv.org/abs/1011.2006}{{\ttfamily
  arXiv:1011.2006 [astro-ph.CO]}}.

\bibitem{Zhou:2010di}
S.-Y. Zhou, ``{Goldstone's Theorem and Hamiltonian of Multi-galileon Modified
  Gravity},'' \href{http://dx.doi.org/10.1103/PhysRevD.83.064005}{{\em Phys.
  Rev. D} {\bfseries 83} (2011) 064005},
  \href{http://arxiv.org/abs/1011.0863}{{\ttfamily arXiv:1011.0863 [hep-th]}}.

\bibitem{Hirano:2010yf}
K.~Hirano and Z.~Komiya, ``{Observational tests of Galileon gravity with growth
  rate},'' \href{http://dx.doi.org/10.1007/s10714-016-2129-z}{{\em Gen. Rel.
  Grav.} {\bfseries 48} no.~10, (2016) 138},
  \href{http://arxiv.org/abs/1012.5451}{{\ttfamily arXiv:1012.5451
  [astro-ph.CO]}}.

\bibitem{Kamada:2010qe}
K.~Kamada, T.~Kobayashi, M.~Yamaguchi, and J.~Yokoyama, ``{Higgs
  G-inflation},'' \href{http://dx.doi.org/10.1103/PhysRevD.83.083515}{{\em
  Phys. Rev. D} {\bfseries 83} (2011) 083515},
  \href{http://arxiv.org/abs/1012.4238}{{\ttfamily arXiv:1012.4238
  [astro-ph.CO]}}.

\bibitem{VanAcoleyen:2011mj}
K.~Van~Acoleyen and J.~Van~Doorsselaere, ``{Galileons from Lovelock actions},''
  \href{http://dx.doi.org/10.1103/PhysRevD.83.084025}{{\em Phys. Rev. D}
  {\bfseries 83} (2011) 084025},
  \href{http://arxiv.org/abs/1102.0487}{{\ttfamily arXiv:1102.0487 [gr-qc]}}.

\bibitem{Hirano:2011wj}
K.~Hirano, Z.~Komiya, and H.~Shirai, ``{Constraining Galileon gravity from
  observational data with growth rate},''
  \href{http://dx.doi.org/10.1143/PTP.127.1041}{{\em Prog. Theor. Phys.}
  {\bfseries 127} (2012) 1041--1056},
  \href{http://arxiv.org/abs/1103.6133}{{\ttfamily arXiv:1103.6133
  [astro-ph.CO]}}.

\bibitem{Li:2011sd}
M.~Li, X.-D. Li, S.~Wang, and Y.~Wang, ``{Dark Energy},''
  \href{http://dx.doi.org/10.1088/0253-6102/56/3/24}{{\em Commun. Theor. Phys.}
  {\bfseries 56} (2011) 525--604},
  \href{http://arxiv.org/abs/1103.5870}{{\ttfamily arXiv:1103.5870
  [astro-ph.CO]}}.

\bibitem{Pujolas:2011he}
O.~Pujolas, I.~Sawicki, and A.~Vikman, ``{The Imperfect Fluid behind Kinetic
  Gravity Braiding},'' \href{http://dx.doi.org/10.1007/JHEP11(2011)156}{{\em
  JHEP} {\bfseries 11} (2011) 156},
  \href{http://arxiv.org/abs/1103.5360}{{\ttfamily arXiv:1103.5360 [hep-th]}}.

\bibitem{Khoury:2011da}
J.~Khoury, J.-L. Lehners, and B.~A. Ovrut, ``{Supersymmetric Galileons},''
  \href{http://dx.doi.org/10.1103/PhysRevD.84.043521}{{\em Phys. Rev. D}
  {\bfseries 84} (2011) 043521},
  \href{http://arxiv.org/abs/1103.0003}{{\ttfamily arXiv:1103.0003 [hep-th]}}.

\bibitem{Trodden:2011xh}
M.~Trodden and K.~Hinterbichler, ``{Generalizing Galileons},''
  \href{http://dx.doi.org/10.1088/0264-9381/28/20/204003}{{\em Class. Quant.
  Grav.} {\bfseries 28} (2011) 204003},
  \href{http://arxiv.org/abs/1104.2088}{{\ttfamily arXiv:1104.2088 [hep-th]}}.

\bibitem{Burrage:2011bt}
C.~Burrage, C.~de~Rham, and L.~Heisenberg, ``{de Sitter Galileon},''
  \href{http://dx.doi.org/10.1088/1475-7516/2011/05/025}{{\em JCAP} {\bfseries
  05} (2011) 025}, \href{http://arxiv.org/abs/1104.0155}{{\ttfamily
  arXiv:1104.0155 [hep-th]}}.

\bibitem{Liu:2011ns}
Z.-G. Liu, J.~Zhang, and Y.-S. Piao, ``{A Galileon Design of Slow Expansion},''
  \href{http://dx.doi.org/10.1103/PhysRevD.84.063508}{{\em Phys. Rev. D}
  {\bfseries 84} (2011) 063508},
  \href{http://arxiv.org/abs/1105.5713}{{\ttfamily arXiv:1105.5713
  [astro-ph.CO]}}.

\bibitem{Kobayashi:2011nu}
T.~Kobayashi, M.~Yamaguchi, and J.~Yokoyama, ``{Generalized G-inflation:
  Inflation with the most general second-order field equations},''
  \href{http://dx.doi.org/10.1143/PTP.126.511}{{\em Prog. Theor. Phys.}
  {\bfseries 126} (2011) 511--529},
  \href{http://arxiv.org/abs/1105.5723}{{\ttfamily arXiv:1105.5723 [hep-th]}}.

\bibitem{PerreaultLevasseur:2011wto}
L.~Perreault~Levasseur, R.~Brandenberger, and A.-C. Davis, ``{Defrosting in an
  Emergent Galileon Cosmology},''
  \href{http://dx.doi.org/10.1103/PhysRevD.84.103512}{{\em Phys. Rev. D}
  {\bfseries 84} (2011) 103512},
  \href{http://arxiv.org/abs/1105.5649}{{\ttfamily arXiv:1105.5649
  [astro-ph.CO]}}.

\bibitem{Clifton:2011jh}
T.~Clifton, P.~G. Ferreira, A.~Padilla, and C.~Skordis, ``{Modified Gravity and
  Cosmology},'' \href{http://dx.doi.org/10.1016/j.physrep.2012.01.001}{{\em
  Phys. Rept.} {\bfseries 513} (2012) 1--189},
  \href{http://arxiv.org/abs/1106.2476}{{\ttfamily arXiv:1106.2476
  [astro-ph.CO]}}.

\bibitem{Endlich:2011vg}
S.~Endlich and J.~Wang, ``{Classical Stability of the Galileon},''
  \href{http://dx.doi.org/10.1007/JHEP11(2011)065}{{\em JHEP} {\bfseries 11}
  (2011) 065}, \href{http://arxiv.org/abs/1106.1659}{{\ttfamily arXiv:1106.1659
  [hep-th]}}.

\bibitem{Brax:2011sv}
P.~Brax, C.~Burrage, and A.-C. Davis, ``{Laboratory Tests of the Galileon},''
  \href{http://dx.doi.org/10.1088/1475-7516/2011/09/020}{{\em JCAP} {\bfseries
  09} (2011) 020}, \href{http://arxiv.org/abs/1106.1573}{{\ttfamily
  arXiv:1106.1573 [hep-ph]}}.

\bibitem{Gao:2011mz}
X.~Gao, ``{Conserved cosmological perturbation in Galileon models},''
  \href{http://dx.doi.org/10.1088/1475-7516/2011/10/021}{{\em JCAP} {\bfseries
  10} (2011) 021}, \href{http://arxiv.org/abs/1106.0292}{{\ttfamily
  arXiv:1106.0292 [astro-ph.CO]}}.

\bibitem{Babichev:2011iz}
E.~Babichev, C.~Deffayet, and G.~Esposito-Farese, ``{Constraints on
  Shift-Symmetric Scalar-Tensor Theories with a Vainshtein Mechanism from
  Bounds on the Time Variation of G},''
  \href{http://dx.doi.org/10.1103/PhysRevLett.107.251102}{{\em Phys. Rev.
  Lett.} {\bfseries 107} (2011) 251102},
  \href{http://arxiv.org/abs/1107.1569}{{\ttfamily arXiv:1107.1569 [gr-qc]}}.

\bibitem{DeFelice:2011hq}
A.~De~Felice, T.~Kobayashi, and S.~Tsujikawa, ``{Effective gravitational
  couplings for cosmological perturbations in the most general scalar-tensor
  theories with second-order field equations},''
  \href{http://dx.doi.org/10.1016/j.physletb.2011.11.028}{{\em Phys. Lett. B}
  {\bfseries 706} (2011) 123--133},
  \href{http://arxiv.org/abs/1108.4242}{{\ttfamily arXiv:1108.4242 [gr-qc]}}.

\bibitem{Khoury:2011ay}
J.~Khoury, G.~E.~J. Miller, and A.~J. Tolley, ``{Spatially Covariant Theories
  of a Transverse, Traceless Graviton, Part I: Formalism},''
  \href{http://dx.doi.org/10.1103/PhysRevD.85.084002}{{\em Phys. Rev. D}
  {\bfseries 85} (2012) 084002},
  \href{http://arxiv.org/abs/1108.1397}{{\ttfamily arXiv:1108.1397 [hep-th]}}.

\bibitem{Qiu:2011cy}
T.~Qiu, J.~Evslin, Y.-F. Cai, M.~Li, and X.~Zhang, ``{Bouncing Galileon
  Cosmologies},'' \href{http://dx.doi.org/10.1088/1475-7516/2011/10/036}{{\em
  JCAP} {\bfseries 10} (2011) 036},
  \href{http://arxiv.org/abs/1108.0593}{{\ttfamily arXiv:1108.0593 [hep-th]}}.

\bibitem{Renaux-Petel:2011rmu}
S.~Renaux-Petel, S.~Mizuno, and K.~Koyama, ``{Primordial fluctuations and
  non-Gaussianities from multifield DBI Galileon inflation},''
  \href{http://dx.doi.org/10.1088/1475-7516/2011/11/042}{{\em JCAP} {\bfseries
  11} (2011) 042}, \href{http://arxiv.org/abs/1108.0305}{{\ttfamily
  arXiv:1108.0305 [astro-ph.CO]}}.

\bibitem{DeFelice:2011bh}
A.~De~Felice and S.~Tsujikawa, ``{Conditions for the cosmological viability of
  the most general scalar-tensor theories and their applications to extended
  Galileon dark energy models},''
  \href{http://dx.doi.org/10.1088/1475-7516/2012/02/007}{{\em JCAP} {\bfseries
  02} (2012) 007}, \href{http://arxiv.org/abs/1110.3878}{{\ttfamily
  arXiv:1110.3878 [gr-qc]}}.

\bibitem{Kimura:2011td}
R.~Kimura, T.~Kobayashi, and K.~Yamamoto, ``{Observational Constraints on
  Kinetic Gravity Braiding from the Integrated Sachs-Wolfe Effect},''
  \href{http://dx.doi.org/10.1103/PhysRevD.85.123503}{{\em Phys. Rev. D}
  {\bfseries 85} (2012) 123503},
  \href{http://arxiv.org/abs/1110.3598}{{\ttfamily arXiv:1110.3598
  [astro-ph.CO]}}.

\bibitem{Wang:2011dt}
H.~Wang, T.~Qiu, and Y.-S. Piao, ``{G-Curvaton},''
  \href{http://dx.doi.org/10.1016/j.physletb.2011.12.016}{{\em Phys. Lett. B}
  {\bfseries 707} (2012) 11--21},
  \href{http://arxiv.org/abs/1110.1795}{{\ttfamily arXiv:1110.1795 [hep-ph]}}.

\bibitem{Kimura:2011dc}
R.~Kimura, T.~Kobayashi, and K.~Yamamoto, ``{Vainshtein screening in a
  cosmological background in the most general second-order scalar-tensor
  theory},'' \href{http://dx.doi.org/10.1103/PhysRevD.85.024023}{{\em Phys.
  Rev. D} {\bfseries 85} (2012) 024023},
  \href{http://arxiv.org/abs/1111.6749}{{\ttfamily arXiv:1111.6749
  [astro-ph.CO]}}.

\bibitem{DeFelice:2011th}
A.~De~Felice, R.~Kase, and S.~Tsujikawa, ``{Vainshtein mechanism in
  second-order scalar-tensor theories},''
  \href{http://dx.doi.org/10.1103/PhysRevD.85.044059}{{\em Phys. Rev. D}
  {\bfseries 85} (2012) 044059},
  \href{http://arxiv.org/abs/1111.5090}{{\ttfamily arXiv:1111.5090 [gr-qc]}}.

\bibitem{Appleby:2011aa}
S.~Appleby and E.~V. Linder, ``{The Paths of Gravity in Galileon Cosmology},''
  \href{http://dx.doi.org/10.1088/1475-7516/2012/03/043}{{\em JCAP} {\bfseries
  03} (2012) 043}, \href{http://arxiv.org/abs/1112.1981}{{\ttfamily
  arXiv:1112.1981 [astro-ph.CO]}}.

\bibitem{DeFelice:2011aa}
A.~De~Felice and S.~Tsujikawa, ``{Cosmological constraints on extended Galileon
  models},'' \href{http://dx.doi.org/10.1088/1475-7516/2012/03/025}{{\em JCAP}
  {\bfseries 03} (2012) 025}, \href{http://arxiv.org/abs/1112.1774}{{\ttfamily
  arXiv:1112.1774 [astro-ph.CO]}}.

\bibitem{Zhou:2011ix}
S.-Y. Zhou and E.~J. Copeland, ``{Galileons with Gauge Symmetries},''
  \href{http://dx.doi.org/10.1103/PhysRevD.85.065002}{{\em Phys. Rev. D}
  {\bfseries 85} (2012) 065002},
  \href{http://arxiv.org/abs/1112.0968}{{\ttfamily arXiv:1112.0968 [hep-th]}}.

\bibitem{Goon:2012mu}
G.~Goon, K.~Hinterbichler, A.~Joyce, and M.~Trodden, ``{Gauged Galileons From
  Branes},'' \href{http://dx.doi.org/10.1016/j.physletb.2012.06.065}{{\em Phys.
  Lett. B} {\bfseries 714} (2012) 115--119},
  \href{http://arxiv.org/abs/1201.0015}{{\ttfamily arXiv:1201.0015 [hep-th]}}.

\bibitem{Shirai:2012iw}
N.~Shirai, K.~Bamba, S.~Kumekawa, J.~Matsumoto, and S.~Nojiri, ``{Generalized
  Galileon Model: Cosmological reconstruction and the Vainshtein mechanism},''
  \href{http://dx.doi.org/10.1103/PhysRevD.86.043006}{{\em Phys. Rev. D}
  {\bfseries 86} (2012) 043006},
  \href{http://arxiv.org/abs/1203.4962}{{\ttfamily arXiv:1203.4962 [hep-th]}}.

\bibitem{Goon:2012dy}
G.~Goon, K.~Hinterbichler, A.~Joyce, and M.~Trodden, ``{Galileons as
  Wess-Zumino Terms},'' \href{http://dx.doi.org/10.1007/JHEP06(2012)004}{{\em
  JHEP} {\bfseries 06} (2012) 004},
  \href{http://arxiv.org/abs/1203.3191}{{\ttfamily arXiv:1203.3191 [hep-th]}}.

\bibitem{Ali:2012cv}
A.~Ali, R.~Gannouji, M.~W. Hossain, and M.~Sami, ``{Light mass galileons:
  Cosmological dynamics, mass screening and observational constraints},''
  \href{http://dx.doi.org/10.1016/j.physletb.2012.10.009}{{\em Phys. Lett. B}
  {\bfseries 718} (2012) 5--14},
  \href{http://arxiv.org/abs/1207.3959}{{\ttfamily arXiv:1207.3959 [gr-qc]}}.

\bibitem{Liu:2012ww}
Z.-G. Liu and Y.-S. Piao, ``{A Galileon Design of Slow Expansion: Emergent
  universe},'' \href{http://dx.doi.org/10.1016/j.physletb.2012.11.068}{{\em
  Phys. Lett. B} {\bfseries 718} (2013) 734--739},
  \href{http://arxiv.org/abs/1207.2568}{{\ttfamily arXiv:1207.2568 [gr-qc]}}.

\bibitem{Barreira:2012kk}
A.~Barreira, B.~Li, C.~M. Baugh, and S.~Pascoli, ``{Linear perturbations in
  Galileon gravity models},''
  \href{http://dx.doi.org/10.1103/PhysRevD.86.124016}{{\em Phys. Rev. D}
  {\bfseries 86} (2012) 124016},
  \href{http://arxiv.org/abs/1208.0600}{{\ttfamily arXiv:1208.0600
  [astro-ph.CO]}}.

\bibitem{Gubitosi:2012hu}
G.~Gubitosi, F.~Piazza, and F.~Vernizzi, ``{The Effective Field Theory of Dark
  Energy},'' \href{http://dx.doi.org/10.1088/1475-7516/2013/02/032}{{\em JCAP}
  {\bfseries 02} (2013) 032}, \href{http://arxiv.org/abs/1210.0201}{{\ttfamily
  arXiv:1210.0201 [hep-th]}}.

\bibitem{Barreira:2013jma}
A.~Barreira, B.~Li, A.~Sanchez, C.~M. Baugh, and S.~Pascoli, ``{Parameter space
  in Galileon gravity models},''
  \href{http://dx.doi.org/10.1103/PhysRevD.87.103511}{{\em Phys. Rev. D}
  {\bfseries 87} (2013) 103511},
  \href{http://arxiv.org/abs/1302.6241}{{\ttfamily arXiv:1302.6241
  [astro-ph.CO]}}.

\bibitem{deFromont:2013iwa}
P.~de~Fromont, C.~de~Rham, L.~Heisenberg, and A.~Matas, ``{Superluminality in
  the Bi- and Multi- Galileon},''
  \href{http://dx.doi.org/10.1007/JHEP07(2013)067}{{\em JHEP} {\bfseries 07}
  (2013) 067}, \href{http://arxiv.org/abs/1303.0274}{{\ttfamily arXiv:1303.0274
  [hep-th]}}.

\bibitem{Deffayet:2013lga}
C.~Deffayet and D.~A. Steer, ``{A formal introduction to Horndeski and Galileon
  theories and their generalizations},''
  \href{http://dx.doi.org/10.1088/0264-9381/30/21/214006}{{\em Class. Quant.
  Grav.} {\bfseries 30} (2013) 214006},
  \href{http://arxiv.org/abs/1307.2450}{{\ttfamily arXiv:1307.2450 [hep-th]}}.

\bibitem{Li:2013tda}
B.~Li, A.~Barreira, C.~M. Baugh, W.~A. Hellwing, K.~Koyama, S.~Pascoli, and
  G.-B. Zhao, ``{Simulating the quartic Galileon gravity model on adaptively
  refined meshes},''
  \href{http://dx.doi.org/10.1088/1475-7516/2013/11/012}{{\em JCAP} {\bfseries
  11} (2013) 012}, \href{http://arxiv.org/abs/1308.3491}{{\ttfamily
  arXiv:1308.3491 [astro-ph.CO]}}.

\bibitem{Sami:2013ssa}
M.~Sami and R.~Myrzakulov, ``{Late time cosmic acceleration: ABCD of dark
  energy and modified theories of gravity},''
  \href{http://dx.doi.org/10.1142/S0218271816300317}{{\em Int. J. Mod. Phys. D}
  {\bfseries 25} no.~12, (2016) 1630031},
  \href{http://arxiv.org/abs/1309.4188}{{\ttfamily arXiv:1309.4188 [hep-th]}}.

\bibitem{Khoury:2013tda}
J.~Khoury, ``{Les Houches Lectures on Physics Beyond the Standard Model of
  Cosmology},'' \href{http://arxiv.org/abs/1312.2006}{{\ttfamily
  arXiv:1312.2006 [astro-ph.CO]}}.

\bibitem{Burrage:2015lla}
C.~Burrage, D.~Parkinson, and D.~Seery, ``{Beyond the growth rate of cosmic
  structure: Testing modified gravity models with an extra degree of
  freedom},'' \href{http://dx.doi.org/10.1103/PhysRevD.96.043509}{{\em Phys.
  Rev. D} {\bfseries 96} no.~4, (2017) 043509},
  \href{http://arxiv.org/abs/1502.03710}{{\ttfamily arXiv:1502.03710
  [astro-ph.CO]}}.

\bibitem{Koyama:2015vza}
K.~Koyama, ``{Cosmological Tests of Modified Gravity},''
  \href{http://dx.doi.org/10.1088/0034-4885/79/4/046902}{{\em Rept. Prog.
  Phys.} {\bfseries 79} no.~4, (2016) 046902},
  \href{http://arxiv.org/abs/1504.04623}{{\ttfamily arXiv:1504.04623
  [astro-ph.CO]}}.

\bibitem{Brax:2015dma}
P.~Brax, C.~Burrage, and A.-C. Davis, ``{The Speed of Galileon Gravity},''
  \href{http://dx.doi.org/10.1088/1475-7516/2016/03/004}{{\em JCAP} {\bfseries
  03} (2016) 004}, \href{http://arxiv.org/abs/1510.03701}{{\ttfamily
  arXiv:1510.03701 [gr-qc]}}.

\bibitem{Saltas:2016nkg}
I.~D. Saltas and V.~Vitagliano, ``{Covariantly Quantum Galileon},''
  \href{http://dx.doi.org/10.1103/PhysRevD.95.105002}{{\em Phys. Rev. D}
  {\bfseries 95} no.~10, (2017) 105002},
  \href{http://arxiv.org/abs/1611.07984}{{\ttfamily arXiv:1611.07984
  [hep-th]}}.

\bibitem{Ishak:2018his}
M.~Ishak, ``{Testing General Relativity in Cosmology},''
  \href{http://dx.doi.org/10.1007/s41114-018-0017-4}{{\em Living Rev. Rel.}
  {\bfseries 22} no.~1, (2019) 1},
  \href{http://arxiv.org/abs/1806.10122}{{\ttfamily arXiv:1806.10122
  [astro-ph.CO]}}.

\bibitem{Nicolis:2008in}
A.~Nicolis, R.~Rattazzi, and E.~Trincherini, ``{The Galileon as a local
  modification of gravity},''
  \href{http://dx.doi.org/10.1103/PhysRevD.79.064036}{{\em Phys. Rev. D}
  {\bfseries 79} (2009) 064036},
  \href{http://arxiv.org/abs/0811.2197}{{\ttfamily arXiv:0811.2197 [hep-th]}}.

\bibitem{Deffayet:2009wt}
C.~Deffayet, G.~Esposito-Farese, and A.~Vikman, ``{Covariant Galileon},''
  \href{http://dx.doi.org/10.1103/PhysRevD.79.084003}{{\em Phys. Rev. D}
  {\bfseries 79} (2009) 084003},
  \href{http://arxiv.org/abs/0901.1314}{{\ttfamily arXiv:0901.1314 [hep-th]}}.

\bibitem{Babich:2004gb}
D.~Babich, P.~Creminelli, and M.~Zaldarriaga, ``{The Shape of
  non-Gaussianities},''
  \href{http://dx.doi.org/10.1088/1475-7516/2004/08/009}{{\em JCAP} {\bfseries
  08} (2004) 009}, \href{http://arxiv.org/abs/astro-ph/0405356}{{\ttfamily
  arXiv:astro-ph/0405356}}.

\bibitem{Mooij:2015yka}
S.~Mooij and G.~A. Palma, ``{Consistently violating the non-Gaussian
  consistency relation},''
  \href{http://dx.doi.org/10.1088/1475-7516/2015/11/025}{{\em JCAP} {\bfseries
  11} (2015) 025}, \href{http://arxiv.org/abs/1502.03458}{{\ttfamily
  arXiv:1502.03458 [astro-ph.CO]}}.

\bibitem{Namjoo:2012aa}
M.~H. Namjoo, H.~Firouzjahi, and M.~Sasaki, ``{Violation of non-Gaussianity
  consistency relation in a single field inflationary model},''
  \href{http://dx.doi.org/10.1209/0295-5075/101/39001}{{\em EPL} {\bfseries
  101} no.~3, (2013) 39001}, \href{http://arxiv.org/abs/1210.3692}{{\ttfamily
  arXiv:1210.3692 [astro-ph.CO]}}.

\bibitem{Martin:2012pe}
J.~Martin, H.~Motohashi, and T.~Suyama, ``{Ultra Slow-Roll Inflation and the
  non-Gaussianity Consistency Relation},''
  \href{http://dx.doi.org/10.1103/PhysRevD.87.023514}{{\em Phys. Rev. D}
  {\bfseries 87} no.~2, (2013) 023514},
  \href{http://arxiv.org/abs/1211.0083}{{\ttfamily arXiv:1211.0083
  [astro-ph.CO]}}.

\bibitem{Cai:2018dkf}
Y.-F. Cai, X.~Chen, M.~H. Namjoo, M.~Sasaki, D.-G. Wang, and Z.~Wang,
  ``{Revisiting non-Gaussianity from non-attractor inflation models},''
  \href{http://dx.doi.org/10.1088/1475-7516/2018/05/012}{{\em JCAP} {\bfseries
  05} (2018) 012}, \href{http://arxiv.org/abs/1712.09998}{{\ttfamily
  arXiv:1712.09998 [astro-ph.CO]}}.

\bibitem{NANOGrav:2023gor}
{\bfseries NANOGrav} Collaboration, G.~Agazie {\em et~al.}, ``{The NANOGrav 15
  yr Data Set: Evidence for a Gravitational-wave Background},''
  \href{http://dx.doi.org/10.3847/2041-8213/acdac6}{{\em Astrophys. J. Lett.}
  {\bfseries 951} no.~1, (2023) L8},
  \href{http://arxiv.org/abs/2306.16213}{{\ttfamily arXiv:2306.16213
  [astro-ph.HE]}}.

\bibitem{Ferrante:2022mui}
G.~Ferrante, G.~Franciolini, A.~Iovino, Junior., and A.~Urbano, ``{Primordial
  non-Gaussianity up to all orders: Theoretical aspects and implications for
  primordial black hole models},''
  \href{http://dx.doi.org/10.1103/PhysRevD.107.043520}{{\em Phys. Rev. D}
  {\bfseries 107} no.~4, (2023) 043520},
  \href{http://arxiv.org/abs/2211.01728}{{\ttfamily arXiv:2211.01728
  [astro-ph.CO]}}.

\bibitem{franciolini:2023pbf}
G.~Franciolini, A.~Iovino, Junior., V.~Vaskonen, and H.~Veermae, ``{The recent
  gravitational wave observation by pulsar timing arrays and primordial black
  holes: the importance of non-gaussianities},''
  \href{http://arxiv.org/abs/2306.17149}{{\ttfamily arXiv:2306.17149
  [astro-ph.CO]}}.

\bibitem{Gorji:2023sil}
M.~A. Gorji, M.~Sasaki, and T.~Suyama, ``{Extra-tensor-induced origin for the
  PTA signal: No primordial black hole production},''
  \href{http://dx.doi.org/10.1016/j.physletb.2023.138214}{{\em Phys. Lett. B}
  {\bfseries 846} (2023) 138214},
  \href{http://arxiv.org/abs/2307.13109}{{\ttfamily arXiv:2307.13109
  [astro-ph.CO]}}.

\end{thebibliography}\endgroup
\bibliographystyle{utphys}

\end{document}